\documentclass[twocolumn]{aastex701}
\usepackage{graphicx} 

\usepackage{bm}
\usepackage{comment}
\usepackage{amsmath}
\usepackage{rotating}
\usepackage{layouts}
\usepackage{tasks}
\usepackage{enumitem}
\usepackage{placeins}
\usepackage{lineno}
\usepackage{amsmath} 
\usepackage{gensymb}
\usepackage{float}
\usepackage{booktabs}
\usepackage{longtable}
\usepackage{wrapfig}
\usepackage{xspace}
\linenumbers

\def\prt{\texttt{pRT}\xspace}

\def\water{H$_{2}$O\xspace}
\def\methane{CH$_{4}$\xspace}
\def\ammonia{NH$_{3}$\xspace}
\def\cotwo{CO$_{2}$\xspace}

\begin{document}

\newcommand{\nopar}{{\parfillskip=0pt \par}}

\newcommand{\Mjup}{\rm{M}_{\rm{Jup}}\xspace}
\newcommand{\Rjup}{\rm{R}_{\rm{Jup}}\xspace}

\newcommand{\todo}[1]{\textcolor{red}{[#1]}}

\shorttitle{Spectroscopy of HD~13724~B with the NIRSpec IFU}
\shortauthors{Hoch et al.}

\title{\emph{The Dyn-Atmo Survey}: High-contrast Imaging Spectroscopy of the Substellar Companion HD~13724~B with the JWST NIRSpec IFU}

\author[0000-0002-9803-8255]{Kielan K. W. Hoch}
\affiliation{Space Telescope Science Institute, 3700 San Martin Dr, Baltimore, MD 21218, USA}
\email{khoch@stsci.edu}

\author[0000-0001-7443-6550]{Alex Madurowicz}
\affiliation{Space Telescope Science Institute, 3700 San Martin Dr, Baltimore, MD 21218, USA}
\email{amadurowicz@stsci.edu}

\author[0000-0002-9792-3121]{Evert Nasedkin}
\affiliation{School of Physics, Trinity College Dublin, The University of Dublin, Dublin 2, Ireland}
\email{nasedkin@tcd.ie}

\author[0000-0003-4203-9715]{Emily Rickman}
\affiliation{European Space Agency (ESA), ESA Office, Space Telescope Science Institute, 3700 San Martin Dr, Baltimore, MD 21218, USA}
\email{erickman@stsci.edu}

\author[0000-0001-8170-7072]{Daniella C. Bardalez Gagliuffi}
\affiliation{Amherst College, Department of Physics and Astronomy, 25 East Drive, Amherst, MA 01002, USA}
\affiliation{Department of Astrophysics, American Museum of Natural History, Central Park West at 79th Street, NY 10024, USA}
\email{dbardalezgagliuffi@amherst.edu}

\author[0000-0001-5864-9599]{James Mang}
\altaffiliation{NSF Graduate Research Fellow}
\affiliation{Department of Astronomy, University of Texas at Austin, 2515 Speedway, Austin, TX 78712, USA}
\email{j_mang@utexas.edu}

\author[0000-0002-4404-0456]{Caroline V. Morley}
\affiliation{Department of Astronomy, University of Texas at Austin, 2515 Speedway, Austin, TX 78712, USA}
\email{cmorley@utexas.edu}

\author[0000-0002-9807-5435]{Christopher A. Theissen}
\affiliation{Department of Astronomy \& Astrophysics, University of California, San Diego, La Jolla, CA 92093, USA}
\email{ctheissen@ucsd.edu}

\author[0000-0001-6396-8439]{William O. Balmer}
\altaffiliation{51 Pegasi b Fellow}
\affiliation{Center for Interdisciplinary Exploration and Research in Astrophysics (CIERA), Northwestern University, 1800 Sherman Avenue, Evanston, IL 60201, USA}
\email{wbalmer1@jhu.edu}

\author[0000-0003-2649-2288]{Brendan P. Bowler}
\affiliation{Department of Physics, University of California, Santa Barbara, Santa Barbara, CA 93106, USA}
\email{bpbowler@ucsb.edu}

\author[0000-0002-2682-0790]{Emily Calamari}
\affiliation{Department of Astrophysics, American Museum of Natural History, Central Park West at 79th Street, NY 10024, USA}
\email{ecalamari@amnh.org}

\author[0009-0001-4688-6949]{Henry Dennen}
\affiliation{Amherst College, Department of Physics and Astronomy, 25 East Drive, Amherst, MA 01002, USA}
\email{hdennen26@amherst.edu}

\author[0000-0001-6251-0573]{Jacqueline K. Faherty}
\affiliation{Department of Astrophysics, American Museum of Natural History, Central Park West at 79th Street, NY 10024, USA}
\email{jfaherty@amnh.org}

\author[0000-0003-4557-414X]{Kyle Franson}
\altaffiliation{NHFP Sagan Fellow}
\affiliation{Department of Astronomy \& Astrophysics, University of California, Santa Cruz, CA 95064, USA}
\email{kfranson@ucsc.edu}

\author[0000-0002-0078-5288]{Mark R. Giovinazzi}
\affiliation{Amherst College, Department of Physics and Astronomy, 25 East Drive, Amherst, MA 01002, USA}
\email{mgiovinazzi@amherst.edu}

\author[0009-0007-5027-3129]{Lisseth Gonzales Quevedo}
\affiliation{Department of Astronomy, University of Massachusetts Amherst, 710 North Pleasant Street, Amherst, MA 01003-9305, USA}
\email{lgonzalesuque@umass.edu}

\author[0000-0003-0192-6887]{Elena Manjavacas}
\affiliation{AURA for the European Space Agency (ESA), ESA Office, Space Telescope Science Institute, 3700 San Martin Drive, Baltimore, MD, 21218 USA}
\affiliation{ Department of Physics \& Astronomy, Johns Hopkins University, 3400 N. Charles Street, Baltimore, MD 21218, USA}
\email{emanjavacas@stsci.edu}

\author{Piper Lentz}
\affiliation{Amherst College, Department of Physics and Astronomy, 25 East Drive, Amherst, MA 01002, USA}
\email{piperlentz@gmail.com}

\author[0009-0007-3572-0664]{Klara Matuszewska}
\affiliation{Amherst College, Department of Physics and Astronomy, 25 East Drive, Amherst, MA 01002, USA}
\email{kmatuszewska26@amherst.edu}

\author[0009-0002-4970-3930]{Ashley Messier}
\affiliation{ Department of Physics \& Astronomy, Johns Hopkins University, 3400 N. Charles Street, Baltimore, MD 21218, USA}
\email{amessie2@jh.edu}

\author[0000-0002-3191-8151]{Marshall Perrin}
\affiliation{Space Telescope Science Institute, 3700 San Martin Dr, Baltimore, MD 21218, USA}
\email{mperrin@stsci.edu}

\author{Laurent Pueyo}
\affiliation{Space Telescope Science Institute, 3700 San Martin Dr, Baltimore, MD 21218, USA}
\affiliation{ Department of Physics \& Astronomy, Johns Hopkins University, 3400 N. Charles Street, Baltimore, MD 21218, USA}
\email{pueyo@stsci.edu}

\author[0000-0003-2233-4821]{Jean-Baptiste Ruffio}
\affiliation{Department of Astronomy \& Astrophysics,  University of California, San Diego, La Jolla, CA 92093, USA}
\email{jruffio@ucsd.edu}

\author[0000-0002-4479-8291]{Kimberly Ward-Duong}
\affiliation{Department of Astronomy, Smith College, Northampton, MA 01063, USA}
\email{kwardduong@smith.edu}

\author[0000-0002-6618-1137]{Jerry W. Xuan}
\altaffiliation{51 Pegasi b Fellow}
\affiliation{Department of Earth, Planetary, and Space Sciences, University of California, Los Angeles, CA 90095, USA}
\email{wxuan@caltech.edu}

\correspondingauthor{Kielan K. W. Hoch}
\email{khoch@stsci.edu}

\collaboration{all}{The Dyn-Atmo collaboration}


\begin{abstract}


We present the first spectral analysis from the JWST Cycle 3 GO Program \#6362, the \textit{Dyn-Atmo} Survey of dynamical-atmospheric benchmarks, characterizing the atmospheres of directly-imaged companions with dynamical masses and measured orbits. 
This program has obtained moderate-resolution spectra (R$\sim$2700, 3-5$\mu$m) using the NIRSpec/G395H IFU.
We observe the T4-dwarf companion, HD~13724~B, companion to a G-type star at a projected angular separation of $0\farcs2$. 
Updated orbital monitoring constrains the mass to $38.3\pm0.6~M_{\rm{Jup}}$, which we use as a Gaussian prior when inferring the atmospheric state. 
We fit the continuum-subtracted spectrum using the BT-SETTL, Sonora Diamondback, Sonora Flame Skimmer, and NEWERA-PHOENIX self-consistent model grids, together with \texttt{petitRADTRANS} atmospheric retrievals.
We adopt bulk parameter values of $T_\mathrm{eff}$=$1150\pm$195 K, $\log g$=5.1$\pm0.3$ dex, [M/H]=0.4$\pm$0.4, $\log_{10}(K_{zz})$=6.33$_{-2.22}^{+2.59}$, and C/O=$0.50\pm0.13$ using a bayesian average across model grids. 
We find degeneracies in the models between surface gravity, metallicity, and $K_{zz}$ for Sonora Diamondback and Flame Skimmer. 
The Flame Skimmer and NEWERA-PHOENIX vertical mixing strengths are incompatible, with Flame Skimmer finding a systematically lower value.
The retrievals favor an atmosphere in chemical equilibrium, with enhanced metallicity and a solar C/O ratio, but with no detection of disequilibrium chemistry, and no detection of trace gases other than CH$_4$, CO$_{2}$, $^{12}$CO and H$_{2}$O. 
Ultimately, we identify critical challenges in forward modeling continuum-subtracted G395H spectra in the low signal-to-noise regime. We find that including independent dynamical mass measurements better constrains atmospheric parameters, but degeneracies between self-consistent model parameters limit accurate atmospheric parameter inferences.

\end{abstract}

\section{Introduction}\label{intro}

Substellar companions serve as an important test bed to derive fundamental parameters and atmospheric properties but often rely on evolutionary models to infer their masses. Mass determines the size, rate of nuclear fusion, luminosity, temperature, surface gravity, and overall evolution of brown dwarfs, but their inherent inability to sustain hydrogen fusion causes them to cool over time, leading to a degeneracy between age, mass, and luminosity which hinders characterization. Additionally, the cool atmospheres of brown dwarfs can harbor silicate clouds, producing variable weather and adding another layer of complexity. Recent studies have revealed discrepancies of up to 30\% in dynamical and model-derived masses for cold brown dwarf companions \citep{brandt2020,dupuy2014,sahlmann2020}, and unexpectedly high C/O potentially due to oxygen sinks unaccounted in cloud formation \cite{calamari2022}, underscoring the paramount importance of calibrating both evolutionary and spectral substellar models.

Brown dwarfs in binary systems with a main-sequence star have the advantage of being
coeval and cospatial with their primary. Moreover, if the orbital orientation is such that the primary star exhibits tangential acceleration in its proper motion between the Hipparcos and Gaia epochs \citep[e.g.,][]{brandtHGCA2021}, as well as radial velocity variability, the model-independent, or dynamical, mass of the brown dwarf secondary can be measured along with orbital parameters.

We can leverage this direct mass measurement of substellar companions to improve our understanding of atmospheres probed by high-contrast spectroscopy, in a new spectral regime with JWST. Dynamical mass constraints allow for an independent prior on the companion mass and can decrease uncertainties when measuring atmospheric parameters \citep[e.g.,][]{2023ApJ...956...99B,2023AJ....165...39F,rickman2019,rickman2020,rickman2024}. Additionally, dynamical mass measurements coupled with precise age measurements are fundamental for testing evolutionary models and constraining the mass-luminosity-age relations of brown dwarfs \citep{bildsten1997,marley2007,marleau2014,xuan2024c, 2025arXiv251206083L}. NIRSpec direct spectroscopy of companions with dynamical mass measurements will improve atmospheric characterization by incorporating constraints on fundamental physical parameters. 

High-contrast imaging spectroscopy of substellar companions with the NIRSpec IFU on the James Webb Space Telescope (JWST) is revolutionizing our understanding of these objects' atmospheres and has opened the door to new challenges and questions about atmospheric, chemical, and formation processes. JWST provides high signal to noise spectra of these companions at wavelengths beyond 3 microns revealing disequilibrium chemistry effects \citep{hoch2024,mukherjee2025}, molecular detections at unprecedented detail \citep{hoch2024,lew2024,luhman2023,ZJ2025,xuan2024,ruffio2026,xuan2026}, as well as the ability to observe fainter and cooler objects inaccessible with ground-based observatories \citep{hoch2025,madurowicz2025,bardalez-gagliuffi2025,matthews2024}. 


JWST spectra beyond 3 microns are challenging to fit using self-consistent atmospheric model grids. The power of applying this technique was demonstrated by \cite{hoch2024} by conducting a high-contrast direct spectral analysis on the T-type substellar companion HD~19467~B, which has a measured dynamical mass and extensive spectral analyses from ground-based observations. By extending the spectrum to a new wavelength regime, \cite{hoch2024} highlighted the need to update disequilibrium chemistry models, and \cite{wogan2025} adjusted the CO$_2$ abundances in Sonora models to be consistent with disequilibrium CO abundances seen in substellar companions and isolated brown dwarfs \citep{beiler2024,wogan2025}. \cite{petrus2024} illustrated the limits of current self-consistent atmospheric model grids on VHS~1256~b at high signal-to-noise ($S/N$) and noted that each parameter's estimate is significantly influenced by the wavelength range considered and the atmospheric model grid chosen. This is attributed to the lack of understanding of systematic errors from the assumptions used in calculating forward model grids and the challenges in replicating complex atmospheric structures. Further modeling efforts on VHS 1256 b are incorporating more complex cloud parameterization to self-consistent models and atmospheric retrievals are being conducted on broader wavelength ranges to improve the spectral fits in the wavelengths covered by JWST \citep[e.g.,][]{miles2023, radcliffe2026, deregt_vhs1256_2026,whiteford2026}.

Here, we present the first results and observations from JWST Cycle 3 GO program 6362 (PI: E. Rickman, co-PI: D. Bardalez Gagliuffi). The goal of the \emph{Dyn-Atmo} survey is to create a sample of dynamical and atmospheric benchmarks to anchor atmospheric properties in well constrained fundamental parameters and break the age-mass-luminosity degeneracy of substellar objects. This degeneracy has hampered progress in understanding fundamental properties of brown dwarfs and their evolutionary models as they are tied to self-consistent atmospheric models. We now have the opportunity to both observe their atmospheres and independently measured masses and robustly test current evolutionary models. To conduct this analysis, this program will homogeneously acquire near-infrared spectra between $2.9-5.3\,\mu$m with NIRSpec/IFU at R$\sim$2700. The sample includes substellar companions with measured dynamical masses and that were resolved with ground-based or space-based near-infrared imaging. All 10 targets have masses in the deuterium-burning range ($\sim10-80~M_{\rm{Jup}}$), effective temperatures of 650--2700 K, ages from 30 Myr--10 Gyr, and orbit Solar-type stars. Therefore, the diversity of the sample is representative across spectral types, masses, temperatures,
and ages with the goal to verify atmospheric and evolutionary models of brown dwarfs. 


The first target of the \emph{Dyn-Atmo} survey, HD~13724~B, is the closest on-sky directly-imaged companion with the NIRSpec IFU to date ($\sim$230 mas, $\sim$14 au). HD~13724~B was first detected with the CORALIE echelle spectrograph \citep{rickman2019} as part of the CORALIE radial velocity search for extrasolar planets. \citep{rickman2019} determined HD~13724~B to have a minimum mass, $m\sin i$, of 26.77$_{-2.2}^{+4.4}M_{\rm{Jup}}$ around a G3/G5V dwarf star \citep{kharchenko2001} at 43.48~$\pm$~0.06~pc. Spectroscopic and imaging follow-up with SPHERE detected the companion at an angular separation of 175.6$\pm$4.5~mas \citep{rickman2020} and obtained low-resolution spectra allowing for the estimation of a T4 or T4.5 spectral type and a dynamical mass measurement of 50.5$_{-3.5}^{+3.3}M_{\rm{Jup}}$ \citep{rickman2020}. The mass was later updated in \cite{brandt2021} to be 36.2$_{-1.5}^{+1.6}M_{\rm{Jup}}$ using calibrated Hipparcos-Gaia EDR3 accelerations. More recently, the mass of HD~13724~B has been refined with additional precise astrometry from VLTI/GRAVITY observations, yielding a dynamical mass of $38.34^{+0.62}_{-0.61}~M_{\rm{Jup}}$ (Rickman et al. in prep.). The age of the HD~13724 system has a range of estimates from 0.47 to 5.51~Gyr, inferred from different methodologies which are discussed in \cite{brandt2021}.




This paper is presented as follows. In Section \ref{obs} we summarize the observations taken with the NIRSpec IFU. In Section \ref{data} we describe the uniform method we use to reduce the IFU data and extract spectra for each target in the survey, which expands upon the methods in \cite{ruffio2024}. 
We step through our forward modeling framework in Section \ref{fmodel} using two public atmospheric model grids and two JWST era model grids from the Sonora family of models and the PHOENIX family of models to explore disequilibrium chemistry and the additional dynamical mass prior. 
We summarize our retrieval methods and results in Section \ref{retrievals}. In Section \ref{discussion} we discuss the impact of using the dynamical mass prior, degeneracies found in the models themselves, and the differences in parameterization of vertical mixing among model families and their implications for future works from the \emph{Dyn-Atmo} survey.  
Section \ref{conclusion} summarizes the major results and conclusions of this work.






\section{Observations}\label{obs}
HD~13724~B was observed with the JWST NIRSpec IFU on September 9, 2024 (UT) as part of program GO~6362 (PI: E. Rickman, co-PI: D. Bardalez Gagliuffi). The F290LP filter was chosen for the wavelength range 2.9--5.3 $\mu$m, combined with the grating G395H to achieve the highest available resolution (R$\sim$2700). This setup was based on the observations from the GTO program 1414 in \cite{ruffio2024} and \cite{hoch2024}. 
The star remains in the FOV for this set of observations, with the companion separated by only $\sim$200 mas. The orientation of the detector is determined by the V3PA. We calculated the optimal V3PA by computing an updated orbit fit of the system from archival astrometry to determine the position of the companion relative to the host star at the time of scheduled observation. Using this information, we selected a V3PA angle that would ensure that HD~13724~B would (i) not fall in the same IFU slice as the host star due to saturation and (ii) not fall under a diffraction spike from the host star.

We did not observe a reference star, as the reduction method demonstrated in \cite{ruffio2024,hoch2024} is sufficient for detection of the companion.
A 9-point small cycling dither pattern was chosen to help with the spatial undersampling of the NIRSpec FOV, as recommended in \cite{ruffio2024}.
We used the NRSIRS2RAPID readout pattern, with observations limited to 4 groups per integration to avoid saturating the detector as the host star is bright (V=7.87), G-type star \citep{hog2000}.



\begin{deluxetable}{lcrcc} 
\centering
\tabletypesize{\scriptsize} 
\tablewidth{0pt} 
\tablecaption{Summary of NIRSpec IFU G395H Observations.}
\label{tab:obs}
\tablehead{ 
  \colhead{Target} &  \colhead{Groups} & DIT/exp & Dithers & DIT} 
\startdata
HD 13724 & 4 &  1313 s & 9 & 197 min. \\
\enddata
 \tablenotetext{~}{All exposures use NRSIRS2RAPID readout pattern, 1 integration per dither.}
\end{deluxetable}

\section{Data Reduction}\label{data}



The data are analyzed using the JWST pipeline \citep{Rauscher_2024} and the Broad Repository for Exoplanet Analysis, Detection, and Spectroscopy (BREADS) \citep{Agrawal_2023} in accordance with the methodology laid out in \cite{Ruffio2023}\footnote{The code repository is available at the following url: \url{https://github.com/jruffio/breads}.} including a modification to the spline nodes which was first demonstrated in \citet{RuffioXuan2026}. By modeling data directly in the IFU detector plane we perform a computational starlight subtraction to remove the stellar contamination. This leaves part of the planetary signal as correlations in the continuum subtracted residuals due to its spectrally resolved molecular absorption bands. 

The stellar model is composed of a multiplicative product between a spline-based continuum unique to each row of pixels on the detector and a continuum-normalized spectrum of the star derived from the science data itself (excluding the position of the planet) for the high spectral frequency component of the starlight. The spline model is adapted from \cite{RuffioXuan2026}, who demonstrated that non-uniform node spacing could improve the significance of detections by minimizing spline residuals near the edges of the spectral traces where the curvature is large. This has the consequence that the empirical covariance matrix must be estimated in multiple different spectral subsets due to the different correlation properties resulting from imperfections in the spline model. More details about the computation of the covariance matrices can be found in Appendix \ref{ref:covariance_section}.

\begin{figure*}
    \centering
    \includegraphics[width=0.9\textwidth]{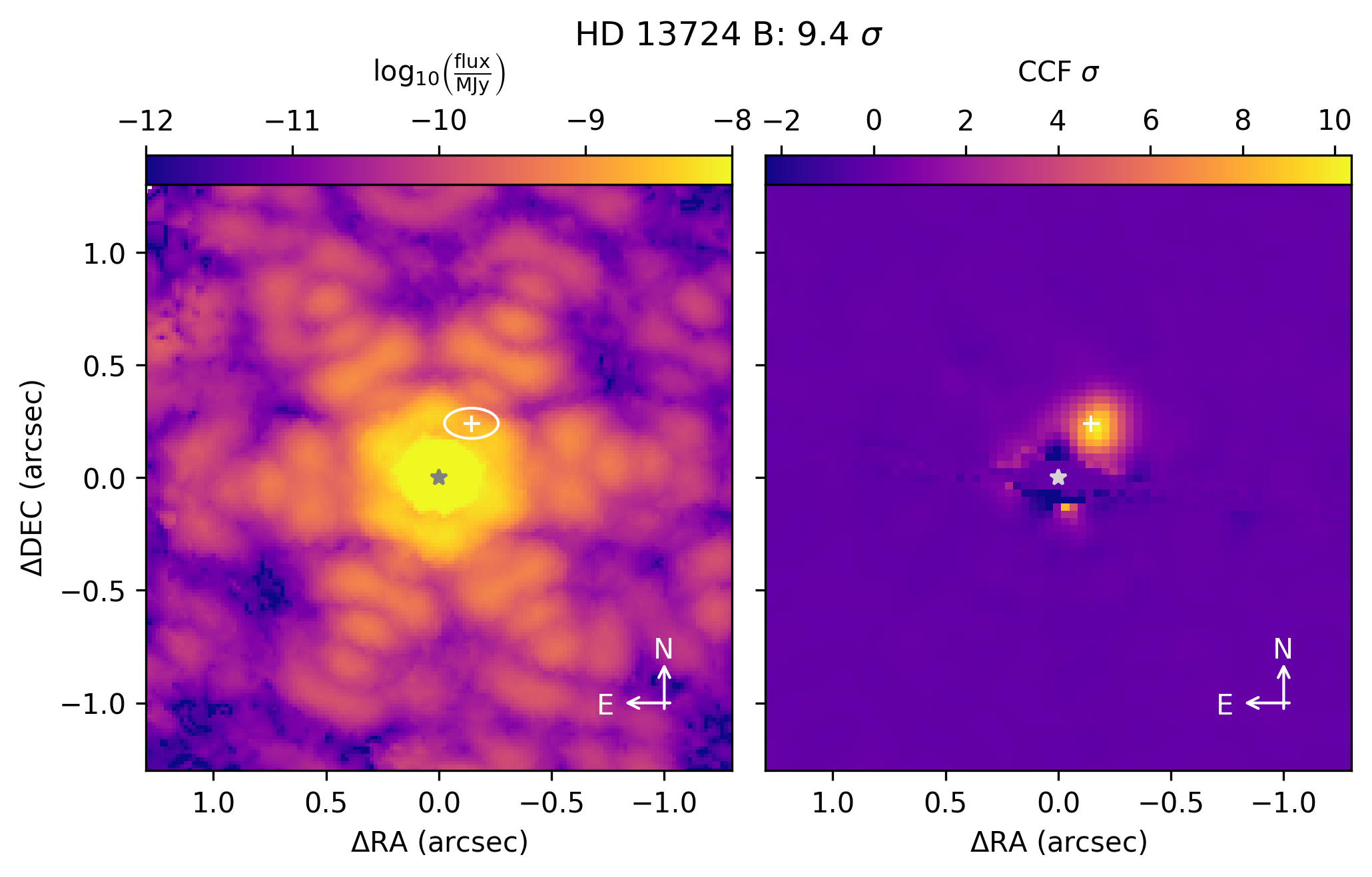}
    \caption{(Left) Interpolation of a slice through the non-star-subtracted point cloud at $\lambda = 4.68$ $\mu$m showing the flux pattern from the stellar primary in the shape of the JWST PSF. The position of the planet on the first Airy ring is indicated with a cross and the uncertainty on that position is shown with an ellipse scaled by a factor of 20$\times$ for visibility. (Right) Detection of HD 13724 B in sky coordinates at 9.4$\sigma$ computed with template-based cross-correlation at the expected position.}
    \label{fig:sep-det}
\end{figure*}

The observation sequence contains nine dither positions to improve the spatial sampling, and we compute the starlight subtraction on each dither independently. However, the entire sequence is combined in the final step into a single point cloud. The point cloud refers the abstract four-dimensional object ($\alpha$, $\delta$, $\lambda$, F), where $\alpha$, is the right ascension, $\delta$ is the declination, $\lambda$ is the wavelength, and F is the stage 2 calibrated flux in units of MJy. The residuals in point-cloud space of the starlight subtraction are fit at each wavelength and position within the FOV with a synthetic WebbPSF (now STPSF) \cite{Perrin2012,Perrin2014} to extract continuum-subtracted spectral flux cubes. The planet can be detected in the residual data by performing cross-correlation with a suitable model template (analogous to \cite{Ruffio2019} Eqn. 5 with a suitable normalization). To estimate the signal-to-noise of the planet, we cross correlate a Sonora Elf-Owl model template \cite{mukherjee_2023} with the following parameters:  $T_\mathrm{eff}$ = 1100 K, $\log g$ = 4.5, $[\mathrm{M}/\mathrm{H}]$ = 0.5, $\textrm{C}/\textrm{O}$ = 0.458, $\log K_\mathrm{zz}$ = 2. Similarly to the data, the model was continuum-subtracted using the spline filter utility module in BREADS. We compute the CCF-based detection map in IFU coordinates and the planet appears as a peak in Figure~\ref{fig:sep-det} at 9.4$\sigma$ in its expected position.

\subsection{Spectral Features}\label{spec-feat}

\begin{figure*}
\centering
\includegraphics[width=\textwidth]{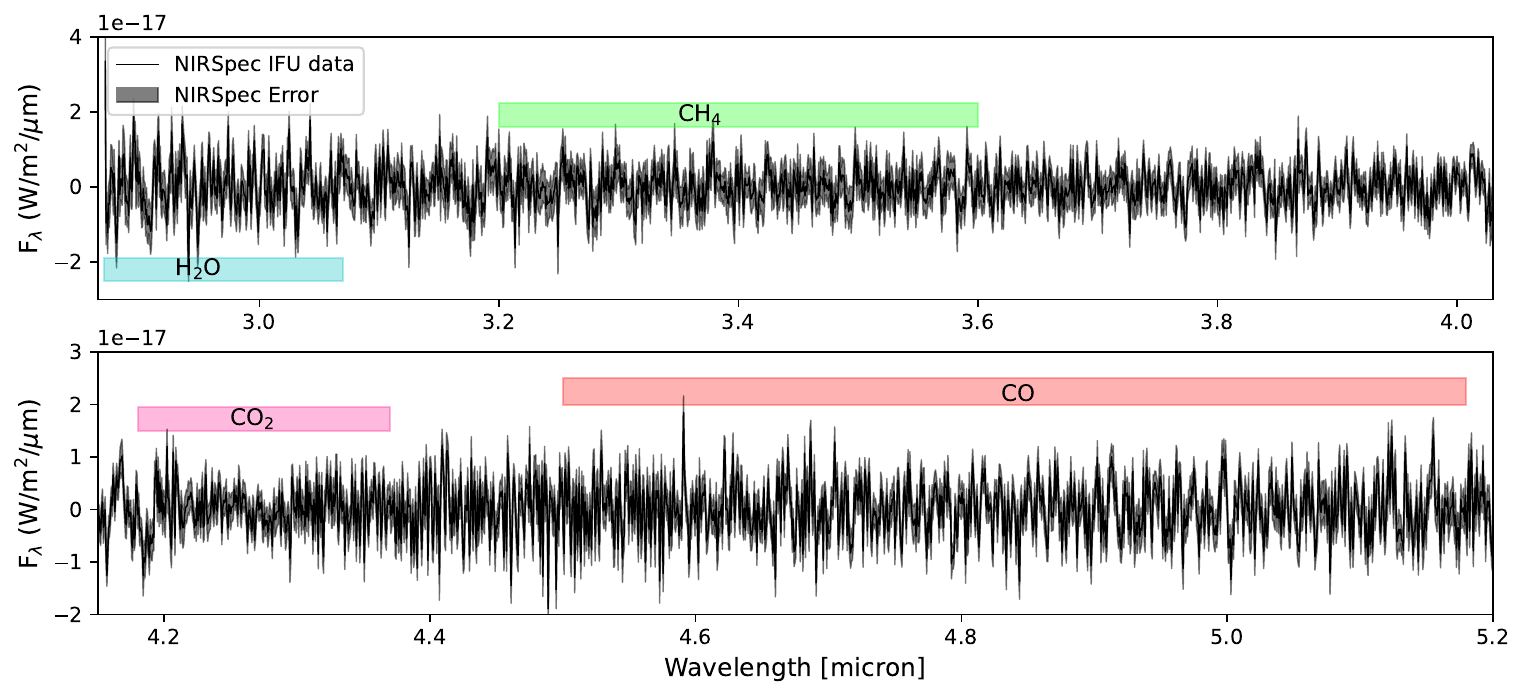}
\caption{JWST NIRSpec IFU G395H/F290LP R$\sim$2700 continuum subtracted spectrum of HD 13724 B spanning 2.9--5.3 $\mu$m. The data processing developed to extract this high-contrast spectrum are summarized in Section \ref{data}. The NIRSpec spectrum reveals the presence of H$_2$O (blue), CH$_4$ (green), CO (red), and CO$_2$ (pink), which are labeled in their corresponding wavelength ranges. The spectrum is in black, the error is in light gray. The top panel focuses in on the first detector wavelength range of NIRSpec (2.87--4.03 $\mu$m), and the bottom panel shows the second detector wavelength range of NIRSpec (4.18--5.35 $\mu$m).}
\label{fig:spec}
\end{figure*}

The spectrum of HD~13724~B is rich in detail, with many molecular lines detected at NIRSpec's $R\sim 2700$ resolution despite the strong stellar contribution at the separation of the companion. Before examining the spectrum in detail via forward model comparisons, we provide a general overview describing the features observed in the spectrum. The extracted NIRSpec spectra is shown in Figure \ref{fig:spec} with molecular features  and their regions labeled in violet rectangles. We see evidence for absorption from H$_2$O from about 2.9--3 $\mu$m, faint CH$_4$ molecular features appear around 3.3~$\mu$m, CO$_2$ absorption appears around 4.2~$\mu$m, and CO absorption can be seen from 4.3--5 $\mu$m. These species are expected from T-Type objects such as HD~13724~B \citep{hoch2024,ruffio2024,morley2012}. 

\section{Forward Modeling Using Self-Consistent Atmospheric Models}\label{fmodel}

Due to the high-contrast flux ratio between HD 13724 B and its host star and their close separation on sky, the post-processing required to extract the companion spectrum does not retain the spectral continuum information \citep{ruffio2024}. However, continuum spectra and photometry are available from the ground with VLT/GRAVITY \citep{kammerer2025} and SPHERE IFS (Rickman et al. in prep) for the wavelength range 0.95–1.35$\mu$m. We fit the suite of data using chi-squared minimization and affine-invariant ensemble samplers \citep{blake2010}, and a log-likelihood function described in \cite{burgasser2016}, \cite{hsu2021}, and \cite{theissen2022} to determine the best-fit model from each grid. The effective temperature ($T_\mathrm{eff}$) and surface gravity ($\log g$) are estimated using a Markov Chain Monte Carlo (MCMC) method built using the \texttt{emcee} package, which uses an implementation of the affine-invariant ensemble sampler \citep{goodman2010,foreman-mackey2013}. Our MCMC runs used 50 walkers, 20000 steps, and a burn-in of 400 steps to ensure parameters were well-mixed. The details of the MCMC calculations follow those described in \cite{wilcomb2020}. 


To determine robust best-fit atmospheric models from each grid, we fit each available spectra individually, we then conducted a joint-fit on all spectra reported in Table \ref{tab:atm_param}, and then conducted the same analysis with a dynamical mass prior reported in Table \ref{tab:atm_param_dyn}. The spectra fit were the NIRSpec continuum-subtracted data presented in this paper, and ground-based data from VLTI/GRAVITY \citep{kammerer2025} and SPHERE IFS (Rickman et al., in prep.). To provide the tightest constraints on our analysis we use the dynamical mass Gaussian prior using the mass 38.34$^{+0.62}_{-0.61}M_{\mathrm{Jup}}$ from Rickman et al., in prep.

\subsection{BT-SETTL}
BT-SETTL is a self-consistent atmosphere model grid that estimates the abundance and size distributions of dust grains using timescales of condensation, coalescence, mixing, and gravitational settling for 55 different solids in separate atmospheric layers \citep{allard2012}. Radiative transfer in the atmosphere is calculated following the PHOENIX code \citep{hauschildt1997,allard2001}. Abundances of molecules such as CO, CH$_4$, CO$_2$, N$_2$, and NH$_3$, are calculated in non-equilibrium chemistry with CIFIST 2015. Vertical mixing is accounted for via mixing length theory under hydrostatic and chemical equilibrium. The grid used spans effective temperatures from 800--3000 K and surface gravities from 2.5--5.5 dex. 

\begin{figure*}
\centering
\includegraphics[width=\textwidth]{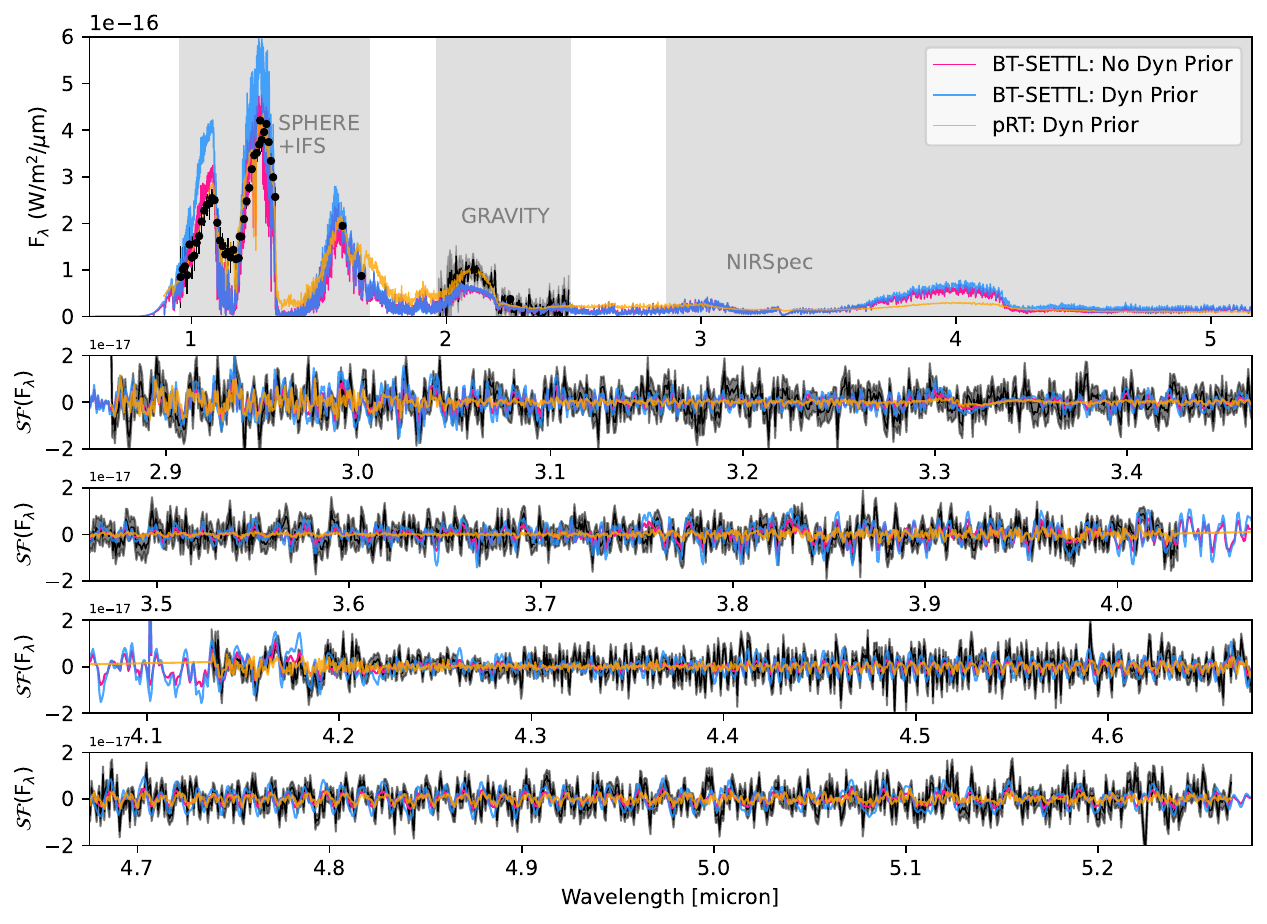}
\caption{Our atmospheric modeling results showing the best fit forward model without and with the dynamical mass prior (pink and blue respectively), and our best fit retrieval in orange on the entire dataset. The top panel shows the full wavelength range of the three datasets with labels in gray and data in black behind the three models. The bottom four panels are the same models continuum subtracted in the same method as the NIRSpec G395H data stretched across 2.87--5.35 $\mu$m against the NIRSpec spectra in black. The self-consistent models do a poor job fitting the GRAVITY spectra and the shorter wavelengths in the SPHERE IFS data.}
\label{fig:spectral_plots}
\end{figure*}

\begin{figure*}
\centering
\includegraphics[width=\textwidth]{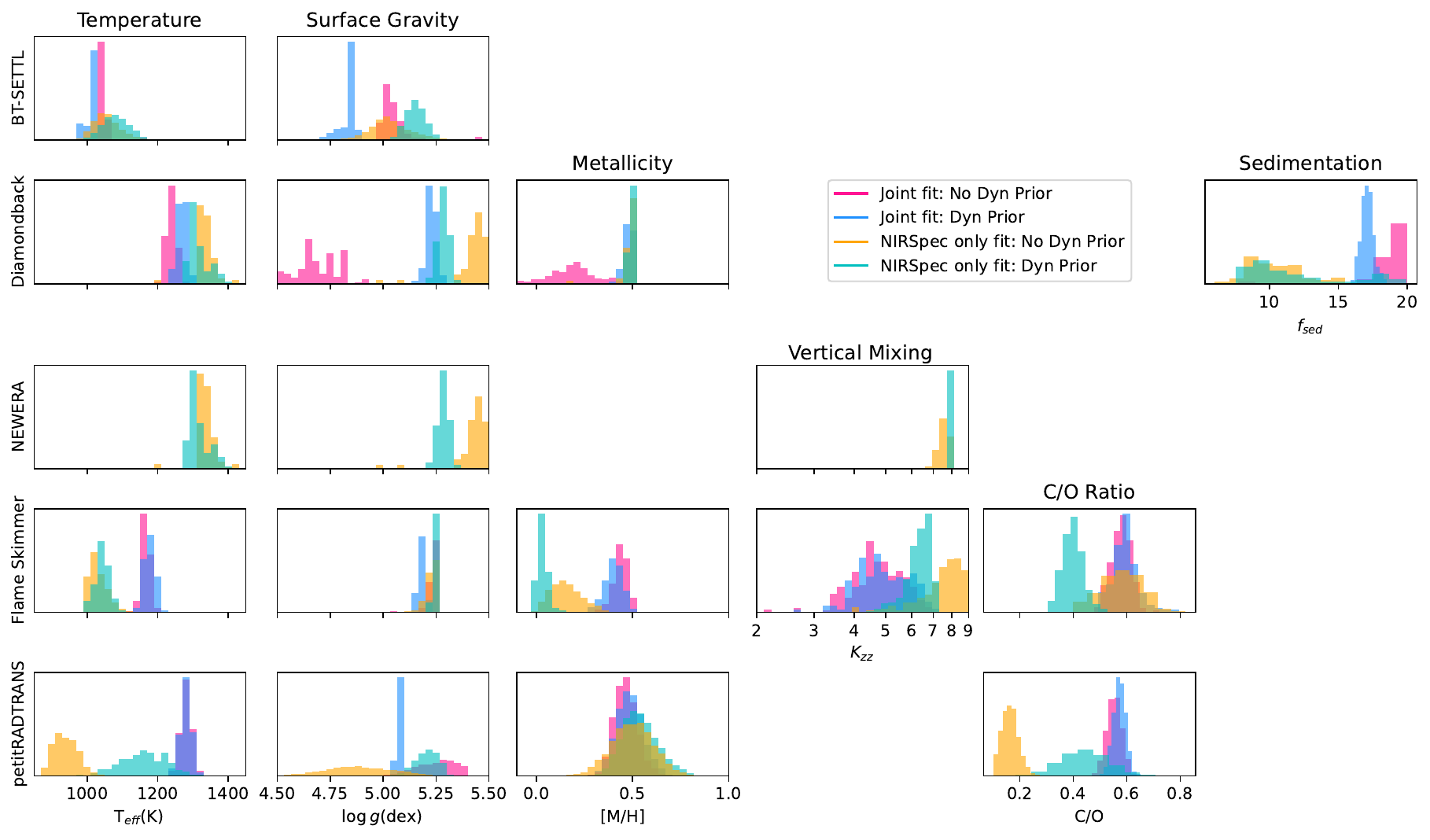}
\caption{Forward modeling and retrieval results for both the joint fits with and without the dynamical mass prior and NIRSpec only fits with and without the dynamical mass prior and the retrieval results. Each model grid, as well as the retrieval, span different parameter spaces. Each row is a different model and retrieval from BT-SETTL--Sonora Flame Skimmer and each column represents the parameter space allowed from each self-consistent model grid and retrieval. NEWERA-PHOENIX is the only grid that only spans NIRSpec G395H wavelengths, so a joint fit was not conducted with this grid.}
\label{fig:all_hist}
\end{figure*}

BT-SETTL was the model grid that had the overall best fit to our data with parameters of T$_{\mathrm{eff}}=1045^{+6}_{-5}$ K and $\log$g=5.00$^{+0.05}_{-0.03}$ dex without the dynamical mass prior, and T$_{\mathrm{eff}}=1014^{+5}_{-10}$\,K and $\log$g=4.9$^{+0.01}_{-0.03}$ dex with a dynamical mass prior. The resulting parameter values from the joint fit are consistent with the only-NIRSpec fit shown in Figure \ref{fig:spectral_plots}. When the Gaussian mass prior is included, the effective temperature and surface gravity decrease and the reduced $\chi^{2}$ is smaller by only 0.49. The distributions for the joint fit and NIRSpec only fits with and without the Gaussian mass prior are shown in Figure \ref{fig:all_hist}. 

\subsection{Sonora Diamondback}
Sonora Diamondback is a self-consistent atmosphere model grid within the Sonora family of models (Sonora Bobcat: \citealt{marley2021}; Sonora Cholla: \citealt{karadili2021}; Sonora Diamondback: \citealt{morley2024}; Sonora Elf Owl \citealt{mukherjee_2023}; Sonora Red Diamondback \citealt{davis2025}) used to explore cloudy atmospheres near the L-T transition. This grid uses radiative-convective equilibrium described in \cite{marley1999} with clouds parameterized following \cite{ackerman2001}. The opacities for molecules and atoms, and collision-induced opacity of hydrogen, helium, and solar abundances from \cite{lodders2010} are included in the models. The vertical mixing for the cloud model is based on mixing length theory and chemical equilibrium is assumed throughout the atmosphere. The grid spans effective temperatures from 900--2400 K, surface gravities from 3.5--5.5 dex, [M/H] from -0.5 to 0.5 dex with respect to Solar, and cloud sedimentation efficiency parameter f$_{\mathrm{sed}}$ ranges from 1--8 and cloud free in the publicly available grid. In order to mathemetically include the cloud-free models in our fit, we consider cloud-free models to have fsed=20 and linearly interpolate between fsed=8 and the cloud-free models. The modeling results across datasets and the joint with and without the gaussian mass prior are shown in Tables \ref{tab:atm_param}--\ref{tab:atm_param_dyn}.

The metallicity fit by Sonora Diamondback is inconsistent among the three datasets when fit individually (Table \ref{tab:atm_param}). However when we include the dynamical mass prior the metallicity measurements are more consistent with one another except for GRAVITY (Table \ref{tab:atm_param_dyn}). We find that the f$_{\mathrm{sed}}$ parameter is particularly sensitive to the SPHERE-IFS only-fit. We attribute this to the sensitivity to clouds in the shorter wavelengths covered by SPHERE, and a high f$_{\mathrm{sed}}$ value indicates a preference for an almost cloudless atmosphere consistent with HD 13724 B's spectral type. 




For the combined dataset, we find a best fit effective temperature of $T_\mathrm{eff}$=$1240^{+13}_{-13}$ K, surface gravity of $\log g$=4.2$^{+0.2}_{-0.3}$ dex, [M/H]=0.20$^{+0.01}_{-0.10}$, and f$_{\mathrm{sed}}$=19$\pm1$. 
With the dynamical mass prior, we find a best fit effective temperature of $T_\mathrm{eff}$=$1269^{+13}_{-14}$ K, surface gravity of $\log g$=5.20$\pm0.02$ dex, [M/H]=0.5$^{+0.01}_{-0.03}$, and f$_{\mathrm{sed}}$=17$\pm1$. Including all available spectral data lowers the effective temperature and increases the f$_{\mathrm{sed}}$ towards a cloudless atmosphere. With the additional inclusion of the dynamical mass prior, the surface gravity and metallicity increase and become more consistent with the disequilibrium grids. This illustrates the need for continuum information in the shorter wavelengths provided by ground-based spectra to constrain cloud properties and metallicity that are not well fit when only modeling NIRSpec continuum-subtracted spectra.

\begin{deluxetable*}{lcccccc}[!ht] 
\tabletypesize{\normalsize} 
\tablewidth{0pt} 
\tablecaption{Summary of atmospheric parameters derived from MCMC fits with an uninformative mass prior.}
\label{tab:atm_param}
\tablehead{ 
  \colhead{HD 13724 B} & \colhead{${T}_\mathrm{{eff}}$ (K)} & \colhead{$\log g$} & \colhead{[M/H]} & \colhead{$\log_{10}(K_{zz}$)} & \colhead{f$_{\mathrm{sed}}$} & \colhead{$\chi^{2}/\nu$}
}
\startdata 
\multicolumn{7}{c}{\textbf{BT-SETTL}} \\
\hline
NIRSpec G395H & $1053^{+42}_{-33}$ & $5.0^\pm0.1$ & n/a & n/a & n/a & 2.968 \\
SPHERE IFS & $962^{+11}_{-13}$ & 3.6$\pm0.1$ & n/a & n/a & n/a & 6.672 \\
GRAVITY & $943^{+23}_{-27}$ & 4.3$^{+0.5}_{-0.3}$ & n/a & n/a & n/a & 1.676 \\
All data combined & $1045^{+6}_{-5}$ & 5.0$\pm0.1$ & n/a & n/a & n/a & 232.55 \\
\hline
\multicolumn{7}{c}{\textbf{Sonora Diamondback}} \\
\hline
NIRSpec G395H & $1303^{+18}_{-24}$ & 5.4$\pm0.1$  & 0.5$^{+0.02}_{-0.04}$  & n/a & 9.7$^{+2.9}_{-1.6}$ & 2.963 \\
SPHERE IFS & $1233^{+33}_{-35}$ & 3.7$^{+0.4}_{-0.1}$ & 0.0$\pm0.1$  & n/a & 17.6$^{+1.5}_{-3.3}$ & 2.391 \\
GRAVITY & $1309^{+31}_{-36}$ & 5.0$^{+0.4}_{-0.6}$  & -0.1$\pm0.1$  & n/a & 10.8$^{+6.7}_{-7.1}$ & 1.724 \\
All data combined & $1240^{+14}_{-13}$ & 4.2$^{+0.2}_{-0.3}$  & 0.2$^{+0.01}_{-0.1}$  & n/a & 19$\pm1$ & 237.26 \\
\hline
\multicolumn{7}{c}{\textbf{Sonora Flame Skimmer}} \\
\hline
  \colhead{HD 13724 B} & \colhead{${T}_\mathrm{{eff}}$ (K)} & \colhead{$\log g$} & \colhead{[M/H]} & \colhead{$\log_{10}(K_{zz}$)} & \colhead{C/O} & \colhead{$\chi^{2}/\nu$} \\
\hline
NIRSpec G395H & $1027^{+26}_{-17}$ & $5.2^{+0.01}_{-0.03}$  &  0.2$^{+0.08}_{-0.09}$  & 7.89$^{+0.75}_{-1.04}$ & 0.59$\pm0.1$ & 3.417 \\ 
SPHERE IFS & 1292$^{+5}_{-12}$ & 4.8$^{+0.06}_{-0.03}$  &  0.4$^{+0.07}_{-0.01}$  & unconstrained & 0.45$^{+0.07}_{-0.08}$ & 9.452 \\ 
GRAVITY & 1095$^{+28}_{-39}$ & 5.1$\pm0.1$  &  unconstrained  & unconstrained & unconstrained & 1.803 \\
All data combined & $1167^{+16}_{-13}$ & $5.2^{+0.01}_{-0.02}$  &  0.4$^{+0.03}_{-0.04}$  & 5.04$^{+0.78}_{-1.04}$ & 0.59$\pm0.4$ & 235.74 \\ 
\hline
\multicolumn{7}{c}{\textbf{NEWERA-PHOENIX}} \\
\hline
NIRSpec G395H & 1125$^{+45}_{-26}$ & 4.5$\pm0.2$ & n/a &  7.8$\pm0.4$  & n/a & 2.918 \\
\hline
\hline
\multicolumn{7}{c}{\textbf{Adopted values}} \\
\hline
NIRSpec G395H & 1127 & 5.0 & 0.4 & 7.85 & 0.59 & -- \\
All data combined & 1151 & 4.8 & 0.3 & 7.04 & 0.59 & -- \\
Allowed Range of Values & 1010--1321 & 4.0--5.5 & 0.1--0.5 & 4.0--8.2 & 0.50--0.68 & -- \\
\enddata
\end{deluxetable*}


\begin{deluxetable*}{lcccccc}[!ht] 
\tabletypesize{\normalsize} 
\tablewidth{0pt} 
\tablecaption{Summary of atmospheric parameters derived from MCMC fits with Dynamical Mass Prior.}
\label{tab:atm_param_dyn}
  \tablehead{\colhead{HD 13724 B} & \colhead{${T}_\mathrm{{eff}}$ (K)} & \colhead{$\log g$} & \colhead{[M/H]} & \colhead{$\log_{10}$$(K_{zz}$)} & \colhead{f$_{\mathrm{sed}}$} & \colhead{$\chi^{2}\nu$}
}
\startdata 
\multicolumn{7}{c}{\textbf{BT-SETTL}} \\
\hline
NIRSpec G395H & $1080^{+38}_{-35}$ & $5.2\pm0.1$ & n/a & n/a & n/a & 2.983 \\
SPHERE IFS & $1062^{+11}_{-14}$ & 4.7$\pm0.03$ & n/a & n/a & n/a & 9.106 \\
GRAVITY & $962^{+22}_{-20}$ & 4.7$\pm0.1$ & n/a & n/a & n/a & 1.669 \\
All data combined & $1014^{+5}_{-10}$ & 4.9$^{+0.01}_{-0.03}$ & n/a & n/a & n/a & 232.06 \\
\hline
\multicolumn{7}{c}{\textbf{Sonora Diamondback}} \\
\hline
NIRSpec G395H & $1305^{+40}_{-12}$ & 5.3$\pm$0.02  & 0.3$^{+0.001}_{-0.013}$  & n/a & 10.2$^{+6.5}_{-1.7}$ & 2.998 \\
SPHERE IFS & $1212^{+41}_{-54}$ & 5.1$\pm0.1$ & 0.4$^{+0.1}_{-0.2}$  & n/a & 15.4$^{+1.5}_{-1.6}$ & 3.321 \\
GRAVITY & $1316^{+31}_{-34}$ & 5.2$\pm0.1$  & 0.0$\pm0.1$  & n/a & 9.5$^{+6.8}_{-5.8}$ & 1.717 \\
All data combined & $1269^{+13}_{-14}$ & 5.2$\pm0.02$  & 0.5$^{+0.01}_{-0.03}$  & n/a & 17$\pm1$ & 236.50 \\
\hline
\multicolumn{7}{c}{\textbf{Sonora Flame Skimmer}} \\
\hline
  \colhead{HD 13724 B} & \colhead{${T}_\mathrm{{eff}}$ (K)} & \colhead{$\log g$} & \colhead{[M/H]} & \colhead{$\log_{10}(K_{zz}$)} & \colhead{C/O} & \colhead{$\chi^{2}/\nu$} \\
\hline
NIRSpec G395H & $1043\pm23$ & $5.2^{+0.004}_{-0.01}$  &  0.0$^{+0.03}_{-0.02}$  & 6.49$^{+0.36}_{-0.57}$ & 0.41$\pm0.4$ & 3.300 \\ 
SPHERE IFS & 1273$^{+15}_{-16}$ & 5.2$^{+0.01}_{-0.02}$  &  0.4$^\pm0.1$  & 8.66$^{+0.26}_{-0.62}$ & 0.41$^{+0.09}_{-0.08}$ & 8.829 \\ 
GRAVITY & 1100$^{+29}_{-26}$ & 4.9$^{+0.05}_{-0.04}$  &  0.0$^{+0.06}_{-0.03}$  & 8.70$^{+0.20}_{-0.40}$ & 0.72$^{+0.29}_{-0.32}$ & 1.748 \\ 
All data combined & $1178^{+14}_{-16}$ & $5.2^{+0.01}_{-0.02}$  &  0.4$^{+0.04}_{-0.05}$  & 5.0$^{+1.2}_{-0.89}$ & 0.9$^\pm0.04$ & 235.26 \\ 
\hline
\multicolumn{7}{c}{\textbf{NEWERA-PHOENIX}} \\
\hline
NIRSpec G395H & 1151$^{+6}_{-3}$ & 5.10$^{+0.02}_{-0.02}$ & n/a &  7.95$^{+0.32}_{-0.87}$  & n/a & 2.990 \\
\hline
\hline
\multicolumn{7}{c}{\textbf{Adopted values}} \\
\hline
NIRSpec G395H & 1145 & 5.2 & 0.3 & 7.66 & 0.41 \\
All data combined & 1154 & 5.1 & 0.4 & 5.00 & 0.59 \\
Allowed Range of Values & 1004--1345 & 4.8--5.3 & 0.0--0.5 & 4.11--8.92 & 0.37--0.62 & -- \\
\enddata
\end{deluxetable*}






\subsection{Sonora Flame Skimmer}
JWST NIRSpec data of substellar objects has shown non-equilibrium chemistry is essential to consider when modeling their atmospheres \citep{hoch2024,madurowicz2025,balmer2025,lew2024,xuan2026}. 
Therefore, to continue exploring the complex chemistry accessible with NIRSpec, we utilize the upcoming Sonora family atmosphere grid, Flame Skimmer, which includes disequilibrium chemistry processes \citep{mang2026}. 

Sonora Flame Skimmer extends the Sonora Bobcat (equilibrium) \citep{marley2021} and Sonora Elf Owl (disequilibrium) \citep{Mukherjee2024, wogan2025} models to colder temperatures and lower surface gravities. In addition to expanding this parameter space, Sonora Flame Skimmer incorporates updated opacities and chemistry. Key improvements include the rainout of volatile species such as H$_2$O, even for atmospheres in chemical disequilibrium. Similar to Sonora Elf Owl v2 \citep{wogan2025}, the CO$_2$ abundances are updated and treated self-consistently with the improved chemistry prescription in Sonora Flame Skimmer.

The subset of Sonora Flame Skimmer used in this work spans effective temperatures 1000--1300 K, vertical mixing parameterized by $K_{\rm zz}$ from 10$^2$--10$^9$, C/O of 0.5--2.5 times solar ($C/O_{\odot}$ = 0.45), [M/H] from 0--0.5, and surface gravity $\log g$ of 4.75--5.25 dex. The parameters for this subset were chosen to encompass the resultant ranges from the public grids reported in Tables \ref{tab:atm_param}--\ref{tab:atm_param_dyn}. We use the same forward modeling approach to our smaller grid but increase the number of walkers to 100 as we are now fitting for 8 parameters available with Sonora Flame Skimmer (see Tables \ref{tab:atm_param}--\ref{tab:atm_param_dyn}).

For the GRAVITY spectra, Sonora Flame Skimmer was unable to constrain the parameters [M/H], $K_{\rm zz}$ and C/O without the dynamical mass prior, so we do not quote these values in Table \ref{tab:atm_param}. With the dynamical mass prior, we were able to constrain all parameters and find a best fit effective temperature of $T_\mathrm{eff}$=1100$^{+29}_{-26}$ K, surface gravity of $\log g$=4.9$^{+0.05}_{-0.04}$ dex, [M/H]=0.0$^{+0.06}_{-0.03}$, a C/O of 0.72$^{+0.29}_{-0.32}$, and a $\log_{10}(K_{\rm zz})$=8.70$^{+0.20}_{-0.40}$. 



When fitting the SPHERE IFS spectra, we were unable to constrain $K_{\rm zz}$ without the dynamical mass prior. When we included the dynamical mass prior, we were able to constrain all parameters in the grid finding a best fit effective temperature of $T_\mathrm{eff}$=1273$^{+15}_{-16}$ K, surface gravity of $\log g$=5.20$^{+0.01}_{-0.02}$ dex, [M/H]=0.4$\pm0.1$, a C/O of 0.41$^{+0.09}_{-0.08}$, and a $\log_{10}(K_{\rm zz})$=8.66$^{+0.26}_{-0.62}$. The best-fit surface gravity was higher when implementing the dynamical mass prior on the independent SPHERE IFS run, which is to be expected due to a constrained range of radii from the dynamical mass prior.

Next, we fit all of the available spectra jointly using the Sonora Flame Skimmer grid. The effective temperature increased, as well as the metallicity when including all available spectra; however the surface gravity and C/O remained the same. Including the ground-based data improved the constraints on $\log_{10}(K_{\rm zz})$ and lowered the best-fit value to 5.0$^{+1.2}_{-0.89}$ from 8.66$^{+0.26}_{-0.62}$ when only considering our NIRSpec G395H data. When including the dynamical mass prior on our joint fits, the effective temperature, surface gravity, metallicity, and C/O are consistent without the prior. However, $K_{\rm zz}$ best-fit values have a broader range and extend to higher values. 




\subsection{NEWERA-PHOENIX}
We also use NEWERA-PHOENIX models from \cite{hauschildt2025} and Barman et al., in prep., which include disequilibrium chemistry induced by vertical mixing, as demonstrated in \cite{hoch2024,Mukherjee2024}. The grid of synthetic spectra was calculated with the PHOENIX model atmosphere code \citep{hauschildt1999} adopting solar atomic abundances \citep{asplund2009}. The current grid spans ${T}_\mathrm{{eff}}$ = 800--1350 K and $\log g$ = 4--5.5 dex, which spans the range of values from Section \ref{tab:atm_param}, and is only a subset of the much larger NEWERA-PHOENIX grid that will be made publicly available. Chemical equilibrium is initially assumed for all species, including condensates. The models are in hydrostatic and radiative-convective equilibrium. To simulate cloud free conditions, the condensate opacities are not included in the radiative transfer calculation (similar to the “COND” models of \citealp{allard2001}). Mixing ratios of certain molecules (H$_2$O, CO, CO$_2$, CH$_4$, N$_2$, NH$_3$) are modified to account for disequilibrium brought about by vertical mixing, following the procedure in \cite{barman2011} and parameterized by the eddy diffusion coefficient ($K_{zz}$, with grid values of 10$^2$, 10$^4$, 10$^6$, 10$^8$). The molecular line data are continuously updated in PHOENIX and includes the latest recommended data from ExoMol \citep{tennyson2020}, HITRAN \citep{tan2022} and HITEMP \citep{hargreaves2020}. The line data for the molecules that have prominent absorption features across our NIRSpec data are from the following sources: H$_2$O \citep{polyansky2018}, CO \citep{hitemp2010}, CO$_2$ \citep{hitemp2010,li2015}, and CH$_4$ \citep{Yurchenko2014,Yurchenko2017}. The details about the latest line list selection process and the complete list of line data included with PHOENIX is reported in \cite{hauschildt2025}.  

We only report measurements for our NIRSpec G395H spectra, as the NEWERA-PHOENIX grid only covers 2.87--5.35 $\mu$m. We follow the same forward modeling process as above without the prior and with the mass prior (Tab. \ref{tab:atm_param}--\ref{tab:atm_param_dyn}). With the dynamical mass prior, we measure $T_\mathrm{eff}$=1151$^{+6}_{-3}$ K, surface gravity of $\log g$=5.1$^{+0.02}_{-0.02}$ dex, and a $\log_{10}(K_{zz})$=7.95$^{+0.32}_{-0.87}$. The Gaussian mass prior increases the surface gravity, the effective temperature and $K_{zz}$ when using the NEWERA-PHOENIX grid. The increase in surface gravity will in turn cause a decrease in radius with the dynamical mass prior constraining the range of radius values.


\subsection{Adopted bulk parameters from forward models}
Our lowest reduced $\chi^2$ value for the dynamical mass prior when including all available spectral data was from the BT-SETTL model grid (see last column of Table \ref{tab:atm_param_dyn}). However, to encompass the wide variety of parameters we fit for across the model grids used, we chose to adopt values based on the average best fit values from NIRSpec G395H and all data combined and quote a range of allowed values. Our adopted values and their ranges for only NIRSpec G395H and all data combined are reported at the end of Table \ref{tab:atm_param} and are an effective temperature $T_\mathrm{eff}$=$1139\pm$182\,K, a surface gravity of $\log g$=4.9$\pm0.9$ dex, a metallicity of [M/H]=0.33$\pm$0.23, $\log_{10}(K_{zz})$=7.46$_{-3.46}^{+0.74}$, and a C/O=$0.59\pm0.1$. We do the same process for the fits with the dynamical mass prior, reported at the end of Table \ref{tab:atm_param_dyn}, and are an effective temperature $T_\mathrm{eff}$=$1150\pm$195 K, a surface gravity of $\log g$=5.1$\pm0.3$ dex, a metallicity of [M/H]=0.4$\pm$0.4, $\log_{10}(K_{zz})$=6.33$_{-2.22}^{+2.59}$, and a C/O=$0.50\pm0.13$. When the Gaussian dynamical mass prior is used, the resultant effective temperature and surface gravity values increase and the [M/H], $K_{zz}$, and C/O values decrease. 

\subsection{Dynamical Mass Prior Importance on Forward Model Metallicity Inferences}\label{dyn_mass_imp}

\begin{figure} 
    \centering
        \centering
        \includegraphics[width=\linewidth]{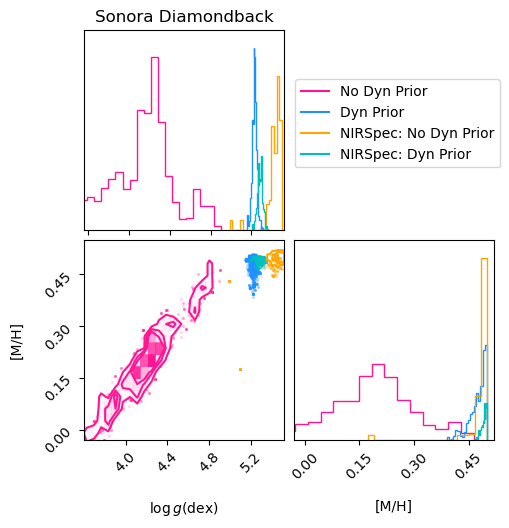}
    \hfill 
        \centering
        \includegraphics[width=\linewidth]{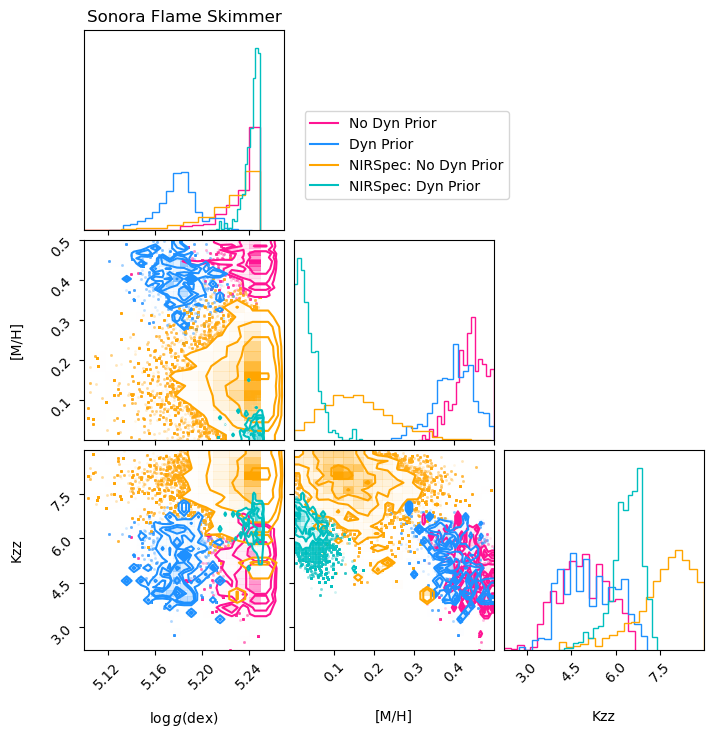}
    \caption{Posterior distributions for Sonora Diamondback (top) and Sonora Flame Skimmer (bottom) for metallicity, surface gravity, and Kzz (only for Flame Skimmer). We show the posteriors for the joint fit without the dynamical mass prior (pink), the joint fit with the dynamical mass prior (blue), the NIRSpec G395H only fit without the dynamical mass prior (orange), and the NIRSpec G395H only fit with the dynamical mass prior (green). For Sonora diamondback, the dynamical mass prior on the joint fit constrains both parameters to higher values, however the NIRSpec only fits with the dynamical mass prefer a lower surface gravity and a higher metallicity. For Flame Skimmer, the dataset fit and the dynamical mass prior show inconsistent measurements of metallicity and show that there is a degeneracy between all three parameters.}
    \label{fig:corner_mh}
\end{figure}

Dynamical mass measurements provide an additional constraint when forward modeling spectral data to better infer atmospheric parameters by adding an additional constraint to one of the parameters, surface gravity. This additional constraint impacts the precision of our fits, but has shown to impact metallicity inferences\citep{zafar2026}. Only two of our grids vary metallicity, Sonora Diamondback and Flame Skimmer. Figure \ref{fig:corner_mh} shows the posterior distributions between metallicity and surface gravity for our joint and NIRSpec-only fits with and without the dynamical mass prior. For Diamondback, the dynamical mass prior helps narrow the posteriors for surface gravity and metallicity in our joint fit, and found higher values for both metallicity and surface gravity for all datasets. The dynamical mass prior helps constrain the surface gravity and metallicity measurements, however, their peak values are inconsistent with the NIRSpec-only fit. \cite{zafar2026} finds that SPHEREx spectra of mid T-type nearby brown dwarfs are not well fit by Sonora Diamondback when comparing CO$_2$ indices against CH$_4$ indices for different f$_{\mathrm{sed}}$ up to cloudless. The connection to clouds and chemistry in mid T-type brown dwarfs is not well represented by forward models and results in discrepancies when attempting to fit metallicity and surface gravity.


Sonora Flame Skimmer allows metallicity, surface gravity, and $K_{zz}$ to be probed as well. The bottom plot of Figure \ref{fig:corner_mh} shows our joint fit and NIRSpec only fit with and without the dynamical mass prior. Sonora Flame Skimmer is a cloudless grid with chemical disequilibrium, but shows metallicity measurements that span [Fe/H] = 0--0.5 dex even with the dynamical mass prior constraining surface gravity. $K_{zz}$ is also not well constrained when introducing a dynamical mass prior indicating a possible degeneracy between metallicity, surface gravity, and $K_{zz}$, which is also seen in SPHEREx spectral fitting \citep{zafar2026}. 



\subsection{Molecular Abundances}
To continue to probe the carbon chemistry in disequilibrium forward model grids, we turn to the NEWERA-PHOENIX grid used in \cite{hoch2024} for an old (9.4$\pm1.0$ Gyr) mid T-type brown dwarf companion, HD 19467 B. The NEWERA-PHOENIX models contain atomic abundances of the solar composition from \cite{asplund2009}. We utilize the mini NEWERA-PHOENIX model grid that varies the mixing ratios of H$_2$O, CO, CO$_2$, and CH$_4$. The mixing ratios are scaled by factors ranging from 0.1 to 10 one at a time by holding the rest of the mixing ratios at solar. Synthetic spectra that varied H$_2$O and CH$_4$ span the 2.87--4.03 $\mu$m range (NRS1), and synthetic spectra that varied CH$_4$ and CO span the 4.15--5.35 $\mu$m range (NRS2). The bulk atmospheric parameters for this grid are $T_\mathrm{eff}$=1100 K, surface gravity of $\log g$=4.5 dex, and a $K_{zz}=5\times10^{5}$, which are encompassed within our adopted values in Table \ref{tab:atm_param_dyn} and within the estimated $K_{zz}$ for Flame Skimmer in PHOENIX parameter space shown in Figure \ref{fig:kzz_newera_fs}.

\begin{figure} 
    \centering
        \centering
        \includegraphics[width=\linewidth]{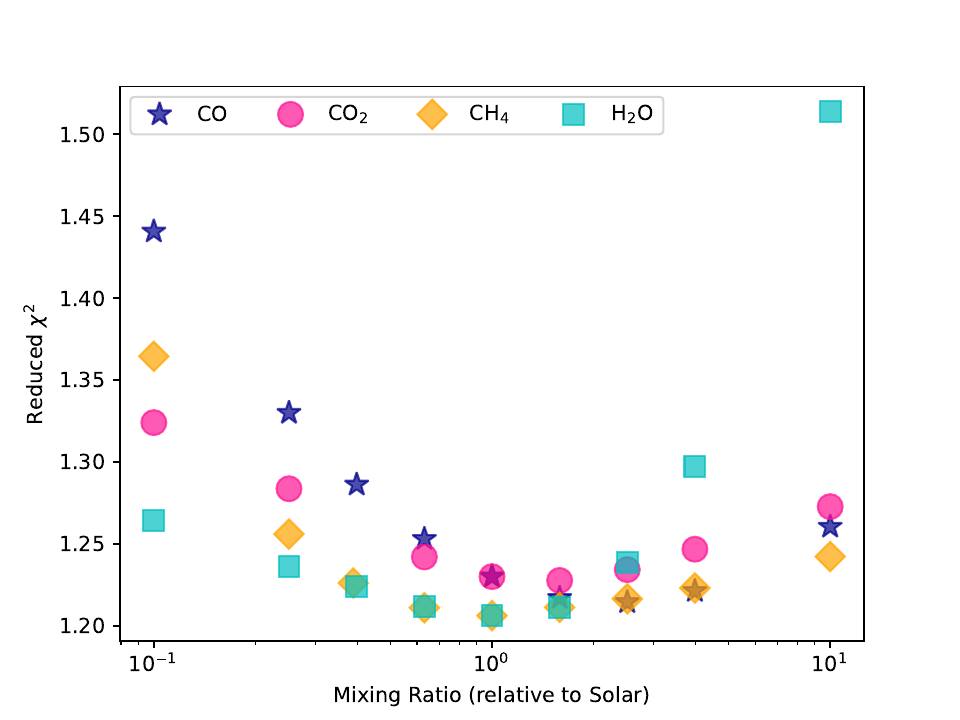}
        \caption{NEWERA-PHOENIX model $\chi^{2}$ results with varying abundances for H$_2$O, CO, CO$_2$, and CH$_4$ to our continuum-subtracted NIRSpec spectrum. The abundances are given in units relative to the ratio in the Sun, such that a value of one implies the solar value. The scalings of H$_2$O (shown in lightblue diamonds) and CH$_4$ (shown in orange diamonds) prefer solar. The scalings of CO prefer 2.512 (shown in dark blue stars) and the scalings of CO$_2$ prefer 1.585 (shown in pink stars). This could suggest enhanced carbon abundances within the atmosphere of HD 13724 B.}
    \label{fig:molecules}
\end{figure}

For H$_2$O and CH$_4$, we used the spectra from NRS1 (2.87--4.05 $\mu$m)  to compare and calculate $\chi^{2}$ values divided by the degrees of freedom for each mixing ratio. We start by holding all other molecules at solar, and only use the models that vary H$_2$O. The resultant $\chi^{2}$ values are plotted in blue for H$_2$O in the left-hand panel of Figure \ref{fig:molecules}. We do the same calculation for CH$_4$ and plot the values in violet in the left-hand panel of Figure \ref{fig:molecules}. 

For CO$_2$ and CO, we used the spectra from NRS2 (4.15--5.35 $\mu$m) and compared it to the models that vary the mixing ratios of the two molecules. We did the same analysis as we did with H$_2$O and CH$_4$ in NRS1. These values for CO$_2$ are plotted in blue in the right-hand panel of Figure \ref{fig:molecules}. We conduct the same analysis on the models that vary CO and obtain the curve plotted in pink in Figure \ref{fig:molecules}.

The $\chi^{2}$ values for both CO and CO$_2$ were minimized when the mixing ratio scale factor was enhanced relative to the solar value. We calculate the 1 - $\sigma$ uncertainties by taking the values from models within $\pm$ 1 of our lowest $\chi^{2}$. For CO, we find a best fit mixing ratio of 2.512$^{+1.47}_{-0.927}$ scaled to solar. For CO$_2$, we find a best fit mixing ratio of 1.585$^{+0.927}_{-0.585}$ scaled to solar. CH$_4$ showed a shallow curve making its mixing ratio difficult to constrain (see Fig. \ref{fig:molecules}). 

For H$_2$O and CH$_4$, the $\chi^{2}$ was minimized when the mixing ratio scale factor was 1.0 resulting in best fit values of 1.000$^{+0.585}_{-0.369}$. However, the $\chi^{2}$ curves for both molecules were broad, indicating that the scaling of the mixing ratios did not have a large impact on the $\chi^{2}$ values and is considered consistent with no scaling at all (Fig. \ref{fig:molecules}). Two carbon-bearing molecules measuring enhanced mixing ratios and Sonora Flame Skimmer providing a best fit C/O that is enhanced relative to solar suggest that the atmosphere of HD 13724 B has enhanced carbon abundance.

\section{Atmospheric Retrievals}\label{retrievals}

\subsection{Retrieval setup}
To complement our self-consistent  forward model fits, we carry out atmospheric retrievals on the spectra of HD 13724 B. We used \texttt{petitRADTRANS} (\prt, v3.3.3) \citep{molliere_petitradtrans_2019,nasedkin_atmospheric_2024} to infer the bulk properties of the object and its atmosphere and determine how these inferences depend on the combination of observations used and the choice of model parameterization. We used the built-in forward models from \prt to calculate the emission spectrum given a chosen parameterization of the temperature profile, the atmospheric chemistry, and the cloud properties. This forward model was coupled to the \texttt{MultiNest} \citep{feroz_multinest_2009,feroz_importance_2013} implementation of the nested sampling algorithm \citep{skilling_nestedsampling_2004} via \texttt{pyMultiNest} \citep{buchner_x-ray_2014} to calculate the posterior probability distributions and estimate the Bayesian evidence. We used 400 live points and a sampling efficiency of 0.8 (without using constant sampling efficiency), which allowed us to explore the parameter space in a computationally feasible runtime: the 32 retrievals required a total of about $5\times10^5$ CPU hours. To perform model comparison we use the estimate of the Bayesian evidence provided by \texttt{MultiNest}, but also calculate the BIC and BPICS heuristics to ensure that the model comparisons are robust to the changing prior volume \citep{thorngren2026}. The table of model comparison statistics for the primary retrievals used in this study is found in Table \ref{tab:ret_model_comparison}, with the complete Table available in the appendices in table \ref{tab:complete_comparison}.

We ran retrievals on combinations of the NIRSpec/G395H, VLTI/GRAVITY \citep{kammerer2025} and VLT/SPHERE \citep{rickman2019} spectroscopic observations, together with photometric measurements from VLT/SPHERE \citep{rickman2019}. This enabled an exploration of how each dataset affected the inferred parameter distributions. The GRAVITY spectrum was allowed to float using an additive offset to help account for uncertainties in the absolute flux calibration. 


The temperature profile was parameterised by inferring the log temperature gradient $d\log T/d\log P$ at different pressures within the atmosphere, and integrating and interpolating to obtain the temperature profile at each of the 154 discretized pressure layers, following the procedure from \cite{zhang_elpis_2023}. However, we also explored varying the number of points at which the temperature gradient is measured (5 or 10), as well as the width of the prior distributions used to constrain the temperature profile to a physically plausible regime. 

Equilibrium, disequilibrium, and freely retrieved chemical abundances were all tested. Disequilibrium chemistry was parameterized via a quench pressure for the CO, CH$_{4}$ and H$_{2}$O network, where the abundances of these species at pressures lower than the quench pressure were fixed to the equilibrium abundance at the quench pressure. In total we included gas-phase opacities for 
\water \citep{ExoMol_H2O}, 
$^{12}$CO \citep{rothman_hitemp_2010},
$^{13}$CO \citep{rothman_hitemp_2010},
\cotwo \citep{ExoMol_CO2},
\methane \citep{exomol_ch4MM_2024},
HCN \citep{ExoMol_HCN},
FeH \citep{wende_feh_2010},
H$_{2}$S \citep{ExoMol_H2S},
\ammonia \citep{ExoMol_NH3},
PH$_{3}$ \citep{exomol_ph3},
Na \citep{allard_new_2019},
and K \citep{allard_k-h_2016}. To fit the NIRSpec/G395H observations, we calculated the models using line-by-line opacities defined at a spectral resolving power of $10^6$, downsampled to $10^5$. For the GRAVITY observations we used correlated-k opacity tables at $R=1000$, while for SPHERE we binned the correlated-k tables to $R=100$ for the spectroscopic observations and $R=40$ for the photometry.

While it was soon recognized that a clear atmosphere provided a sufficient fit to the observed spectra, for completeness we also tested including clouds. In the simplest case we included an opaque cloud deck, inferring the cloud top pressure. A patchy silicate cloud was also tested, fitting for the cloud coverage fraction, the cloud base pressure, the mean particle radius, the mass fraction at the cloud base and the power law by which the abundance decreased with altitude. Neither the grey cloud nor the more complex silicate cloud improved the goodness of fit, and as they were not favored by any evidence metric they will not be discussed further.

In order to compare the models to the data, we calculated a likelihood value for each model-data pair and summed these contributions. Prior to the likelihood calculation, the model was convolved with the instrumental line spread function. For NIRSpec/G395H this resolving power was allowed to vary linearly as a function of wavelength, and was fit for as in \cite{xuan2026}. For GRAVITY a fixed resolving power of 500 was used \citep{kammerer2021}, while for SPHERE a resolving power of 39 was used, matching the native resolution of each instrument. For photometry, the model spectrum was used to calculate synthetic photometric points using \texttt{species}, which accounts for the filter transmission profiles from the SVO \citep{SVO}. Following the convolution, the model spectra are rebinned to the wavelength bins of the data. As the NIRSpec/G395H observations produce a continuum-subtracted spectrum, we apply an identical continuum subtraction procedure to the rebinned model spectrum. We fit a spline to the model, using the same node spacing as for the data, and subtract this spline from the model. For the NIRSpec/G395H and the GRAVITY data a full covariance matrix was available to be used in the likelihood calculation, while for the SPHERE spectrum and photometry the measured Gaussian uncertainties were used.

A key test of the \textit{Dyn-Atmo} survey is to understand how having a dynamical mass prior aids in inferring atmospheric parameters. With a well-constrained mass of $38.34\pm0.6$ M$_{\rm Jup}$, we were also able to test how applying a Gaussian prior on the mass impacted the remaining parameters, as compared to a uniform prior from 13 to 80 M$_{\rm Jup}$.

\begin{table}[t]
\centering
\caption{Priors and median values for run R04 of the atmospheric retrievals.}
\label{tab:v31priors}
\begin{tabular}{lll}
\toprule
Parameter & Prior & Median $\pm$ std \\
\midrule
$M_{\rm pl}$ [$M_{\rm Jup}$] & $\mathcal{N}(38.23, 0.6)$ & $38.3 \pm 0.6$ \\
$R_{\rm pl}$ [$R_{\rm Jup}$] & $\mathcal{U}(0.7, 2.0)$ & $0.707 \pm 0.008$ \\
T$_{\rm bot}$ & $\mathcal{U}(200, 7200)$ & $2375 \pm 48$ \\
$\nabla \log T_{1}$ & $\mathcal{N}(0.15, 0.01)$ & $0.152 \pm 0.009$ \\
$\nabla \log T_{2}$ & $\mathcal{N}(0.18, 0.04)$ & $0.14 \pm 0.01$ \\
$\nabla \log T_{3}$& $\mathcal{N}(0.21, 0.05)$ & $0.20 \pm 0.01$ \\
$\nabla \log T_{4}$& $\mathcal{N}(0.16, 0.06)$ & $0.07 \pm 0.01$ \\
$\nabla \log T_{5}$ & $\mathcal{N}(0.08, 0.025)$ & $0.05 \pm 0.02$ \\
$\nabla \log T_{6}$ & $\mathcal{N}(0.06, 0.02)$ & $0.05 \pm 0.02$ \\
$\nabla \log T_{7}$ & $\mathcal{N}(0.0, 0.1)$ & $-0.02 \pm 0.07$ \\
$\nabla \log T_{8}$ & $\mathcal{N}(0.0, 0.1)$ & $0.03 \pm 0.08$ \\
$\nabla \log T_{9}$& $\mathcal{N}(0.0, 0.1)$ & $0.01 \pm 0.09$ \\
$\nabla \log T_{10}$ & $\mathcal{N}(0.0, 0.1)$ & $0.00 \pm 0.09$ \\
$\left[\mathrm{M/H}\right]$ & $\mathcal{U}(-1.000, 1.000)$ & $0.63 \pm 0.08$ \\
C/O & $\mathcal{U}(0.1000, 1.000)$ & $0.57 \pm 0.02$ \\
$^{13}$C/$^{12}$C & $\mathcal{U}(10^{-4}, 10^{1})$ & $0.01 \pm 0.02$ \\
RV [km s$^{-1}$] & $\mathcal{U}(-1000, 1000)$ & $-6.4 \pm 0.8$ \\
GRAV offset & $\mathcal{U}(-1, 1)\times10^{-16}$ & $-2.9\pm0.6\times10^{-17}$ \\
G395H R$_{\rm slope}$ & $\mathcal{U}(400.0, 1600.0)$ & $979 \pm 394$ \\
G395H R$_{0}$ & $\mathcal{U}(-800.0, 400.0)$ & $-195 \pm 390$ \\
\bottomrule
\end{tabular}
\end{table}

\begin{table*}[ht]
    \centering
    \caption{Abridged model-comparison summary for the full HD\,13724\,B retrieval suite, the full comparison is available in table \ref{tab:complete_comparison}. Data shorthand: `NS' NIRSpec, `P' photometry, `G' GRAVITY, `S' SPHERE. $\Delta\log\mathcal{Z}$, $\Delta$BIC and $\Delta\mathrm{BPIC}_{\mathrm{S}}$ are normalised to the best-evidence run R04. }
    \label{tab:ret_model_comparison}
    \begin{tabular}{llccccrrrrc}
    \toprule
    Run & PT & Data & Mass & Chem. & Cloud & $n_{\rm par}$ & $\Delta\log\mathcal{Z}$ & $\Delta$BIC & $\Delta\mathrm{BPIC}_{\mathrm{S}}$ & $\chi^2_\nu$ \\
    \midrule
    R01 & 10 PT & NS & Free & Eq. & Clear & 19 & -4376.1 & 20143.3 & 20143.9 & 0.505 \\
    R02 & 10 PT & NS & Dyn. & Eq. & Clear & 19 & -4372.3 & 20142.3 & 20142.2 & 0.505 \\
    R03 & 10 PT & NS+G+S & Free. & Eq. & Clear & 20 & -1.0 & -3.5 & -2.8 & 0.549 \\
    R04 & 10 PT & NS+G+S & Dyn. & Eq. & Clear & 20 & 0.0 & 0.0 & 0.0 & 0.550 \\
    \bottomrule
    \end{tabular}
    
\end{table*}

\begin{table*}[ht]
    \centering
    \caption{Abridged table of inferred and derived parameters for the HD\,13724\,B retrieval suite. The full comparison is available in table \ref{tab:complete_params}.}
    \label{tab:allparams}
    
    \begin{tabular}{lcccccc}
    \toprule
    Run & $M$ [$M_{\rm Jup}$] & $R$ [$R_{\rm Jup}$] & $T_{\rm eff}$ [K] & $\log g$ [cgs] & {[M/H]} & C/O \\
    \midrule
    R01 & $32\pm12$ & $1.05\pm0.08$ & $939_{-33}^{+35}$ & $4.9\pm0.2$ & $0.5_{-0.1}^{+0.1}$ & $0.16_{-0.03}^{+0.03}$ \\
    R02 & $38.2\pm0.5$ & $0.78\pm0.05$ & $1160_{-68}^{+63}$ & $5.21\pm0.06$ & $0.53_{-0.09}^{+0.09}$ & $0.44_{-0.09}^{+0.09}$ \\
    R03 & $66.\pm12$ & $0.708\pm0.009$ & $1331_{-15}^{+14}$ & $5.53\pm0.08$ & $0.59_{-0.07}^{+0.07}$ & $0.54_{-0.03}^{+0.03}$ \\
    R04 & $38.3\pm0.6$ & $0.707\pm0.008$ & $1322_{-8}^{+9}$ & $5.30\pm0.01$ & $0.63_{-0.08}^{+0.08}$ & $0.57_{-0.02}^{+0.02}$ \\
    
    \bottomrule
    \end{tabular}
\end{table*}

\subsection{Atmospheric retrieval results}
The results of the atmospheric retrievals broadly reflect those of the forward models. The inferred parameter values depend strongly on the choice of datasets included in the retrieval and the model parameterization. Many retrievals, particularly those relying solely on the NIRSpec/G395H data, find unphysical solutions for the chemical composition of the atmosphere. As the additional datasets are added, the measurements approach those of the forward models, and the overall evidence in support of the model increases.  The SPHERE photometry stands out as a difficult-to-fit dataset, with very small uncertainties, and a low H-band flux that drives the retrievals to an implausibly high metallicity solution. As these photometric data were not included in the forward model fits, we do not include them in further analysis of the retrieval results. 
In order to perform a reasonable comparison, we select a set of retrievals which enable robust comparisons to the forward model grids.
These are R01 through R04, which use equilibrium chemistry, and compare the use of a dynamical mass prior (R02 and R04) and use of NIRSpec only (R01, R02) and the combined NIRSpec, GRAVITY and SPHERE spectra (R03 and R04).
The bulk parameters inferred from these retrievals are included in table \ref{tab:allparams}, with the complete set of results available in table \ref{tab:complete_params}. 
Complete parameter measurements for R04 are included in table \ref{tab:v31priors}.
The best fit models all share similar reduced $\chi^{2}$ values of ${\sim0.5}$, with the best fit model of R04 shown in Figure \ref{fig:spectral_plots}.



Selecting from the baseline retrievals, run R04 is favoured by the Bayes factor (though not the BPIC$_{\rm S}$ or BIC, which favor the free mass prior over the dynamical mass), and provides physically plausible parameters when fitting the full spectrum of HD~13724~B.
In this retrieval, equilibrium chemistry is assumed, and the temperature profile was calculated as in  \cite{zhang_elpis_2023}, using the 10PT spacing and tight prior variation as described in table \ref{tab:v31priors}.  
A Gaussian prior centered on the known dynamical mass was used.
In general the retrievals find a radius consistent with the lower bound of the prior, pushing towards 0.7 R$_{\rm Jup}$. 
We find broadly similar properties to those found using the self-consistent models: the radius is $0.707\pm0.007$ R$_{\rm Jup}$ giving a log surface gravity of $5.30\pm0.01$ in cgs units.
The atmosphere is enriched in metals, with a metallicity of [M/H]=$0.63\pm0.08$, or $2.5\pm0.5\times$ the stellar value, and a C/O ratio of $0.57\pm0.02$, compatible with the solar value of 0.55.

These parameters are representative of the entire suite of retrievals, though the diversity between models is much greater than the posterior uncertainty on the parameters from any single retrieval. In general, a small radius is preferred, suggesting that these fits are suffering from the known `small-radius problem' \citep[e.g.][]{burningham2017,balmer2025}. 
The metallicity is enhanced in all but 5 pathological cases, out of 32 total retrievals. The free retrievals struggle to find physically plausible solutions, finding CO abundances of $>10$\% of the atmosphere by mass. The low $S/N$ of the NIRSpec/G395H data means the observed CO lines have inconsistent depths, as the noise scatters the data by about the same amount as the line depths themselves. This makes inferring the chemical abundances challenging. These implausible chemical abundances force the retrieval to compensate by using the PT profile to fit the low resolution near-infrared spectra, which results in large variations in the measured effective temperature, from 700 K up to 1600 K. However, all evidence metrics favour retrievals with more physically plausible chemical compositions and temperatures: the true value of the metallicity likely lies between 0.4 and 0.71, while the effective temperature is between 906 and 1350 K, with a C/O ratio consistent with Solar. This preferred family of retrievals required assumptions of equilibrium chemistry and tight constraints on the PT profile, essentially forcing the retrieval to reproduce the self consistent model. This is validated by examining the model spectra themselves in Fig. \ref{fig:spectral_plots}: while the SPHERE H-band photometry lacks constraining power due to the large uncertainties, the equilibrium chemistry models are much more consistent in shape and amplitude with the self-consistent models in the H-band than the free chemistry models. Further H-band spectroscopic observations are likely necessary to provide reliable, data-driven measurements of the atmospheric chemical abundances. While the retrievals provides a useful crosscheck to the self-consistent models, these results highlight the continued importance of the physically motivated models to measure atmospheric parameters in the low-S/N range. Unlike many similar JWST observations, such as those of VHS 1256 b \citep{miles2023}, this dataset is limited by the observational uncertainties rather than the model flexibility.

\subsubsection{Temperature profile}

\begin{figure}
    \centering
    \includegraphics[width=0.95\linewidth]{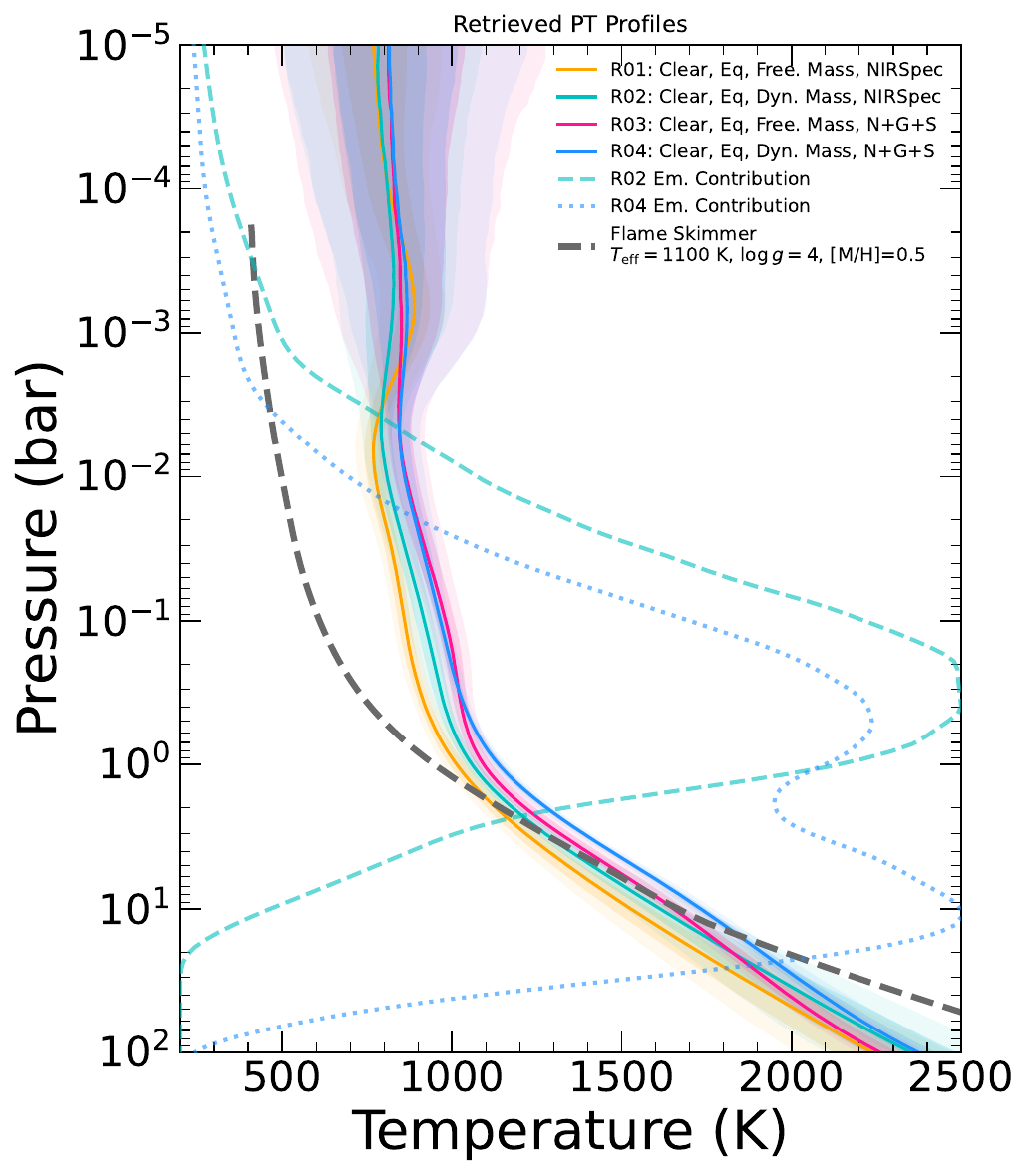}
    \caption{Retrieved temperature profiles for equilibrium chemistry retrievals, comparing NIRSpec only retrievals to retrievals on the combined spectra, as well as between the free and dynamical mass priors. The shaded regions indicate the 1$\sigma$ and 2$\sigma$ confidence intervals.  The dashed grey line indicates a Flame Skimmer temperature profile from the model closest to the best fit, showing that the retrievals consistently identify a warmer upper atmosphere and a shallower temperature gradient in the photosphere. The dashed and dotted blue and lines denote the wavelength averaged emission contribution function, highlighting where the retrieval is sensitive to the emission. The addition of the SPHERE and GRAVITY data provides additional sensitivity at pressures greater than 1 bar.}
    \label{fig:retrieval_PT}
\end{figure}

The retrieved temperature profiles for the baseline retrievals are shown in Fig. \ref{fig:retrieval_PT}. 
We compared the use of 5 or 10 nodes to fit the temperature gradient, finding that the more flexible 10 point profile was preferred (R01 is preferred over R06).
The temperature profiles are generally consistent with each other, with R03 and R04 being consistently warmer than the NIRSpec only retrievals, R01 and R02.
The inclusion of the SPHERE and GRAVITY data provide additional sensitivity deep in the atmosphere, reducing the uncertainty on the temperature profile between 1 and 10 bar.
Both R03 and R04 deviate slightly from an adiabat at around 10 bar, which is often thought to indicate the presence of a cloud. However, clear atmospheres are always favoured by the Bayesian evidence.
The retrieved temperature profile is warmer in the upper atmosphere (at pressures less than 1 bar) than the profile from the best-fit Sonora Diamondback model. 
This heating is consistent regardless of the combination of datasets used. At 14 au separation, it is unlikely that this is due to heating from the star, as the equilibrium temperature of the companion is less than 100 K. Instead, it may reflect auroral heating processes  seen in other T-dwarfs, such as WISE-1935 \cite{faherty_methane_2024} and SIMP-0136 \citep{nasedkin2025}, which has been attributed to electron precipitation in the upper atmosphere \citep{smith_heating_2026}. 
However, confirmation of such a mechanism would require a detection of pulsed radio emission to identify the presence of a high-energy electron beam. 
Likewise, HD 13724 B does not show the inverted temperature structure characteristic of this localised heating, suggesting that there may be other physical processes at play.  

By calculating low resolution models drawn from the posterior parameter distribution and integrating the flux from 0.5 to 30 micron, we calculate the effective temperature to be T$_{\rm eff}=1322\pm9$ K, placing this retrieval towards the warmer end of the self-consistent model fits. Similar to the Flame Skimmer grid, the temperature increases when including all of the data, compared to the NIRSpec-only fits.

\subsubsection{Chemistry}

{\scriptsize
\begin{table}[htbp]
\centering
\caption{Leave-one-out molecular Bayes factors for the NIRSpec-only family. The baseline is the full free-chemistry retrieval, R05. Positive values favour keeping the molecule in the model, species in bold have statistically significant detections. }
\label{tab:moleculebayes}
\begin{tabular}{lrr}
\toprule
Species & $\Delta\log_{10} \mathcal{Z}$ & $\Delta \mathrm{BPICs}$ \\
\midrule
\textbf{CH$_\mathbf{4}$} & $14.9$ & $161.0$ \\
\textbf{CO$_\mathbf{2}$} & $10.4$ & $119.6$ \\
\textbf{$^{\mathbf{12}}$CO} & $7.9$ & $88.1$ \\
\textbf{H$_\mathbf{2}$O} & $6.3$ & $55.4$ \\
NH$_3$ & $0.02$ & $-0.1$ \\
$^{13}$CO & $-0.01$ & $-0.2$ \\
Na & $-0.03$ & $-1.8$ \\
HCN & $-0.07$ & $-2.9$ \\
H$_2$S & $-0.1$ & $-2.2$ \\
PH$_3$ & $-0.2$ & $-2.7$ \\
\bottomrule
\end{tabular}
\end{table}
}
\begin{figure}
    \centering
    \includegraphics[width=\linewidth]{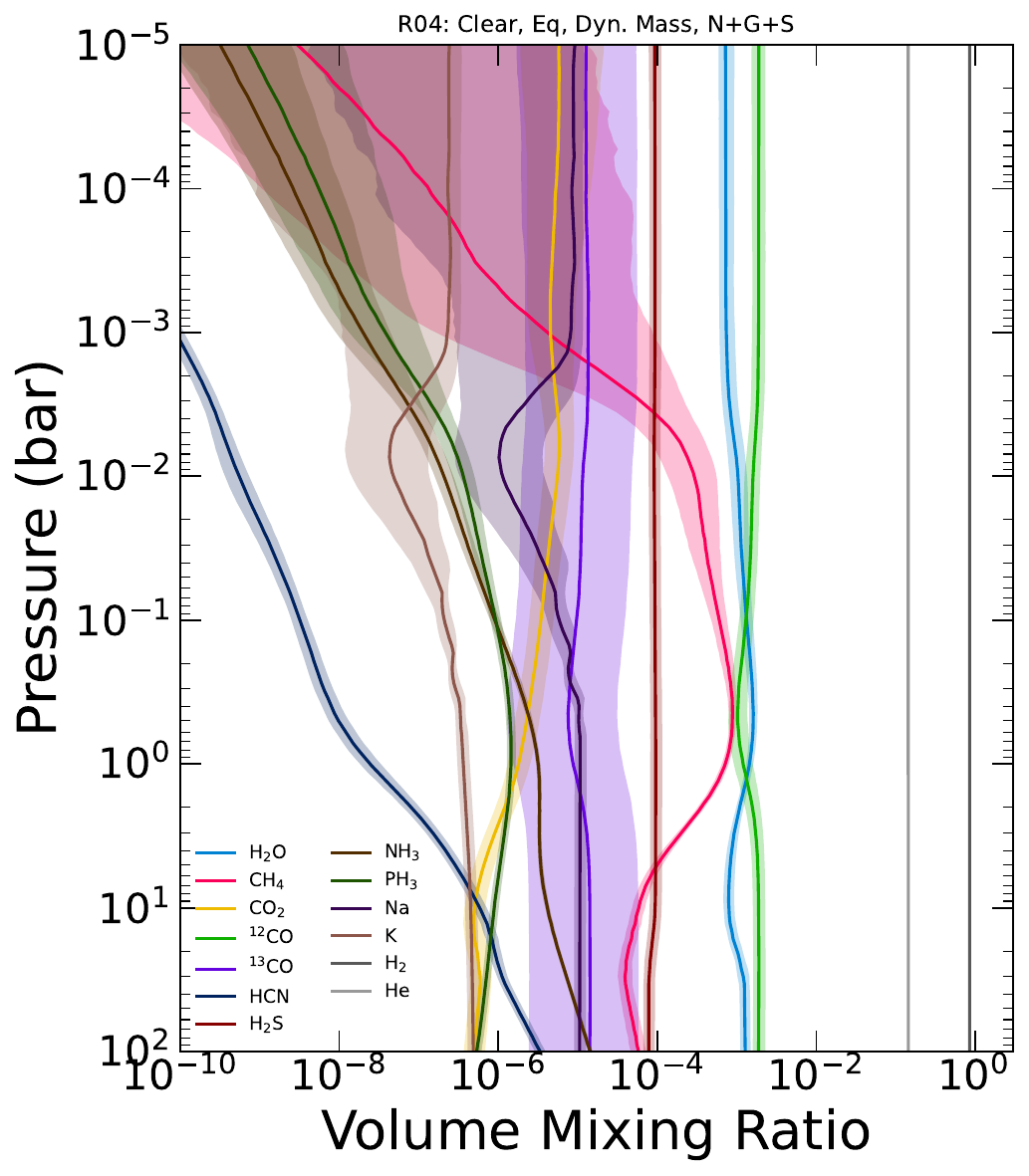}
    \caption{Volume mixing ratios for the best fit equilibrium atmosphere from R04, with [M/H]=$0.63\pm0.08$ and C/O=$0.57\pm0.02$. The shaded region indicates the 1$\sigma$ confidence interval.}
    \label{fig:retrieval_chemistry}
\end{figure}

The molecular abundances for R04 are shown in Fig. \ref{fig:retrieval_chemistry}.
The retrievals reveal that while free chemistry (R05) is the preferred model, equilibrium chemistry (R06) is required to infer physically plausible results. Furthermore, equilibrium (R06) chemistry is preferred over disequilibrium chemistry (R07) for both our NIRSpec only runs and our full data runs. The P$_{\mathrm{quench}}$ upper limit is found to be at 0.1 bar, and there is no statistically significant detection of disequilibrium chemistry, implying relatively weak vertical mixing in the atmosphere. 
All retrievals find an enriched atmosphere for HD 13724 B, even when compared to the stellar value of 0.23, with the best fit R04 retrieval finding a metallicity of $0.63\pm0.08$. 
This is similar in composition to the family of directly imaged exoplanets, which are known to have metal rich compositions \citep[e.g.][]{RuffioXuan2026,balmer_aflepgrav_2025,nowak_betapic_2020}. 

We test for the presence of individual trace species using leave-one-out retrievals.
Following the recommendations of \cite{thorngren2026}, a $\log_{10}$ Bayes factor of 4.2 is approximately equivalent to a p-value of 0.05, and we take this as the threshold for detection. 
The Bayes factors testing the detection significance of different molecular species and isotopologues are shown in table \ref{tab:moleculebayes}. CH$_4$ has the strongest detection, followed by CO$_2$, $^{12}$CO and H$_2$O. No other species are significantly detected.

\subsubsection{Dependence on dynamical mass}
The dynamical mass provides critical constraining power for the retrievals. When allowed to freely vary, the mass estimates from the retrieval range from 17 M$_{\rm Jup}$ to 72 M$_{\rm Jup}$, resulting in log surface gravities ranging from 4.3 to 5.6. 
The available wavelength coverage and $S/N$, together with the lack of absolute flux information in the G395H data mean that the observations are only weakly sensitive to the surface gravity. 
The H-band spectral shape and the alkali line depth in the J bands typically provide the greatest sensitivity to the surface gravity, and the lack of data here strongly limits the reliability of measurements. 

However, an independent mass estimate can be used as a prior to restrict the available parameter space. 
Adding this prior consistently led to higher surface gravity measurements, which is reflected in the forward model analysis. 
Even with a precise dynamical mass the radius remains difficult to measure using retrievals. The radius is consistently pushed to the lower limit of the prior, and if allowed will shrink to unphysically values of 0.7 R$_{\rm Jup}$. 
Most retrievals for which the radii was constrained (i.e. not limited by the prior) suffered from unphysical atmospheric compositions, with metallicities greater than 10$\times$ that of the host star. 
This may be due to very high inferred CO abundances, likely driven by the scatter in the data near the 4.5$\mu$m CO fundamental band, as the residuals in this region show systematic correlations with the CO opacity.
Notably, none of the retrievals or forward model grids were able reproduce the amplitude of the CO lines between 4.4 and 5.0 micron. 
For the baseline set of retrievals T$_{\rm eff}$ and [M/H] do not show strong dependency on the use of the dynamical mass prior. 
In the NIRSpec only case the use of the dynamical mass increased the C/O ratio from an implausible $0.16\pm0.03$ to a more physically likely value of $0.44\pm0.09$.


\subsubsection{Dependence on data inclusion}
The inferred parameters depend strongly on which datasets are included in the retrieval.
Using only the NIRSpec/G395H data, the retrieved temperature profile is marginally more uncertain than when adding the SPHERE and GRAVITY data, particularly at higher pressures, and is slightly cooler at all pressures. 
The GRAVITY data, with higher $S/N$ and resolving power than the SPHERE data provides the strongest additional constraints, increasing the precision on the inferred temperature profile and chemical parameters. 
Under the assumption of equilibrium chemistry and using only the NIRSpec data with a dynamical mass prior we measured a radius, metallicity, and effective temperature compatible at the 1$\sigma$ level with the full spectrum inference, with larger uncertainties on all parameters, as presented in table \ref{tab:allparams}. 
The inferred C/O ratio is not compatible, finding a mild relative depletion of carbon with a C/O ratio $0.44\pm0.09$, compared to the solar-compatible value of $0.57\pm0.02$ found when using the full dataset. 



%
%

\section{Discussion}\label{discussion}


Here we presented the closest on-sky substellar companion ($\sim$230 mas), HD 13724 B, detected with the NIRSpec IFU G395H filter as a part of the \textit{Dyn-Atmo} Survey (Fig. \ref{fig:sep-det}). The technique utilized in this survey enables direct detections and spectroscopy of close separation, high contrast companions. However, the continuum of the spectra cannot currently be recovered. Therefore, forward modeling and retrieval analyses are being conducted on continuum subtracted data. Not all objects observed in this mode will have continuum information such as photometric points from other instruments in JWST wavelength ranges. The challenge then becomes identifying the limitations when modeling NIRSpec continuum subtracted atmospheric spectra without photometry to provide additional continuum information. 


\subsection{Vertical Mixing and Disequilibrium Chemistry in JWST-Era models}\label{vertical_mix}

\begin{figure*}
\centering
\includegraphics[width=\textwidth]{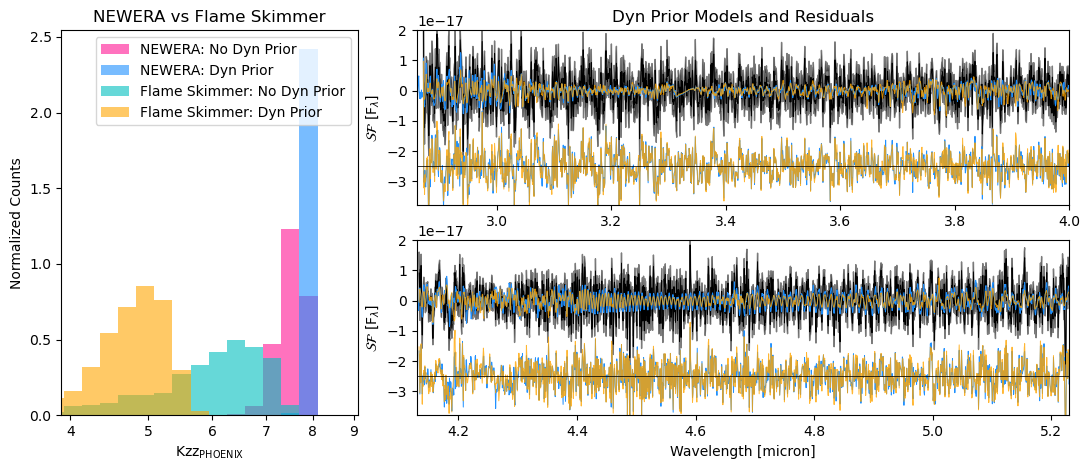}
\caption{We show a histogram plot of the best fit $K_{zz}$ values from Flame Skimmer and PHOENIX when only including the NIRSpec data on the left. We adjust the Flame Skimmer $K_{zz}$s by a factor of 0.025 to account for the different mixing timescale used in the PHOENIX grids. This reveals more discrepancies when exploring vertical mixing between newer disequilibrium chemistry atmospheric model grids. On the right hand side, we show the dynamical mass prior best fit models for Flame Skimmer (yellow) and PHOENIX (blue) against the data with the residuals of the fits plotted below in the respective colors with a black line at the offset zero point.}
\label{fig:kzz_newera_fs}
\end{figure*}

Vertical mixing is thought to impact substellar companion atmospheric spectra past 3$\mu$m from temperatures of $\sim$1200 K down to $\sim$250 K, and is parameterized in the Sonora Flame Skimmer and PHOENIX models via $K_{zz}$ \citep{zahnlemarley2014}. To measure the impact of vertical mixing on the spectrum, we compare the NIRSpec-only fits from NEWERA-PHOENIX and Flame Skimmer. Figure \ref{fig:kzz_newera_fs} shows the posterior histograms for Flame Skimmer and NEWERA with and without the dynamical mass prior against $K_{zz}$ in the first column. The NEWERA grid ends at $K_{zz}$ of 8, while Flame Skimmer extends to $K_{zz}$ of 9. With the dynamical mass prior, NEWERA prefers a higher $K_{zz}$ closer to 8, and Flame Skimmer prefers a lower $K_{zz}$ with a broader distribution of allowed values. The right panel of Figure \ref{fig:kzz_newera_fs} shows our NIRSpec spectrum split into NRS 1 and NRS 2 wavelengths with the best fit dynamical mass prior models for both Flame Skimmer and NEWERA with the residuals plotted below. Flame Skimmer over predicts the line depths in NRS1 compared to NEWERA near the 3.3$\mu$m CH$_4$ feature and under predicts the line depths in the CO region in NRS 2 when compared to NEWERA. The lower $K_{zz}$ value for Flame Skimmer increases the strength of CH$_4$ absorption and decreases the strength of the CO absorption which is also seen in \cite{zafar2026} and illustrates chemistry-sensitive carbon species and degenerate spaces in the grid. 

The Sonora and the PHOENIX families of grids have different base assumptions for their self-consistent models, and therefore $K_{zz}$ is not necessarily the same parameter across the two grids. For the PHOENIX model grid, the mixing time scale uses the effective length scale where the pressure scale height is multiplied by a factor ranging from 0.1--0.2 to better match that of Jupiter described in \cite{barman2011}, which is not done in Flame Skimmer. 

\begin{equation}
K_{zz,NE} \sim 0.025 \times K_{zz,FS}
\label{kzz_eq}
\end{equation}

Therefore, we take the average of those scale values and square them to multiply Flame Skimmer $K_{zz}$ values by 0.025 following section 4.2 in \cite{barman2011} and presented in equation \ref{kzz_eq}. This causes the $K_{zz}$ values of Flame Skimmer to be lower than those of NEWERA-PHOENIX, but will allow for a closer comparison of vertical mixing between the grid fits. \cite{zahnle2014} found that the PHOENIX models have lower CO mixing ratios and higher CH$_4$ mixing ratios following \cite{yung1988}, requiring higher $K_{zz}$ values to replicate the CO absorption. When considering these effects, we show a comparison of corrected posterior $K_{zz}$ from Flame Skimmer to NEWERA-PHOENIX and reveal Flame Skimmer prefers even lower $K_{zz}$ values when compared to NEWERA-PHOENIX shown in Figure \ref{fig:kzz_newera_fs}. More modeling work is needed to better fit the NIRSpec spectra in relation to disequilibrium chemistry, vertical mixing, and varying molecular abundances is required to improve atmospheric characterization of NIRSpec substellar companion spectra. 


\subsection{Implications for Forward Modeling Continuum Subtracted NIRSpec G395H Spectra}

In this work, we use four separate grids, but only two grids, Sonora Flame Skimmer and NEWERA-PHOENIX, treat disequilibrium chemistry which is important to consider in this wavelength range. We conducted individual fits on ground-based data and our NIRSpec continuum subtracted data and a joint fit on the entire spectral dataset. When only fitting the NIRSpec continuum subtracted spectra parameters such as $f_{sed}$, metallicity, $K_{zz}$, and C/O are not consistent across the individual fits and the joint fits even with the dynamical mass prior. To explore this phenomenon we look to the models themselves and the impacts of subtracting their continuum. 

\begin{figure*}
\centering
\includegraphics[width=\textwidth]{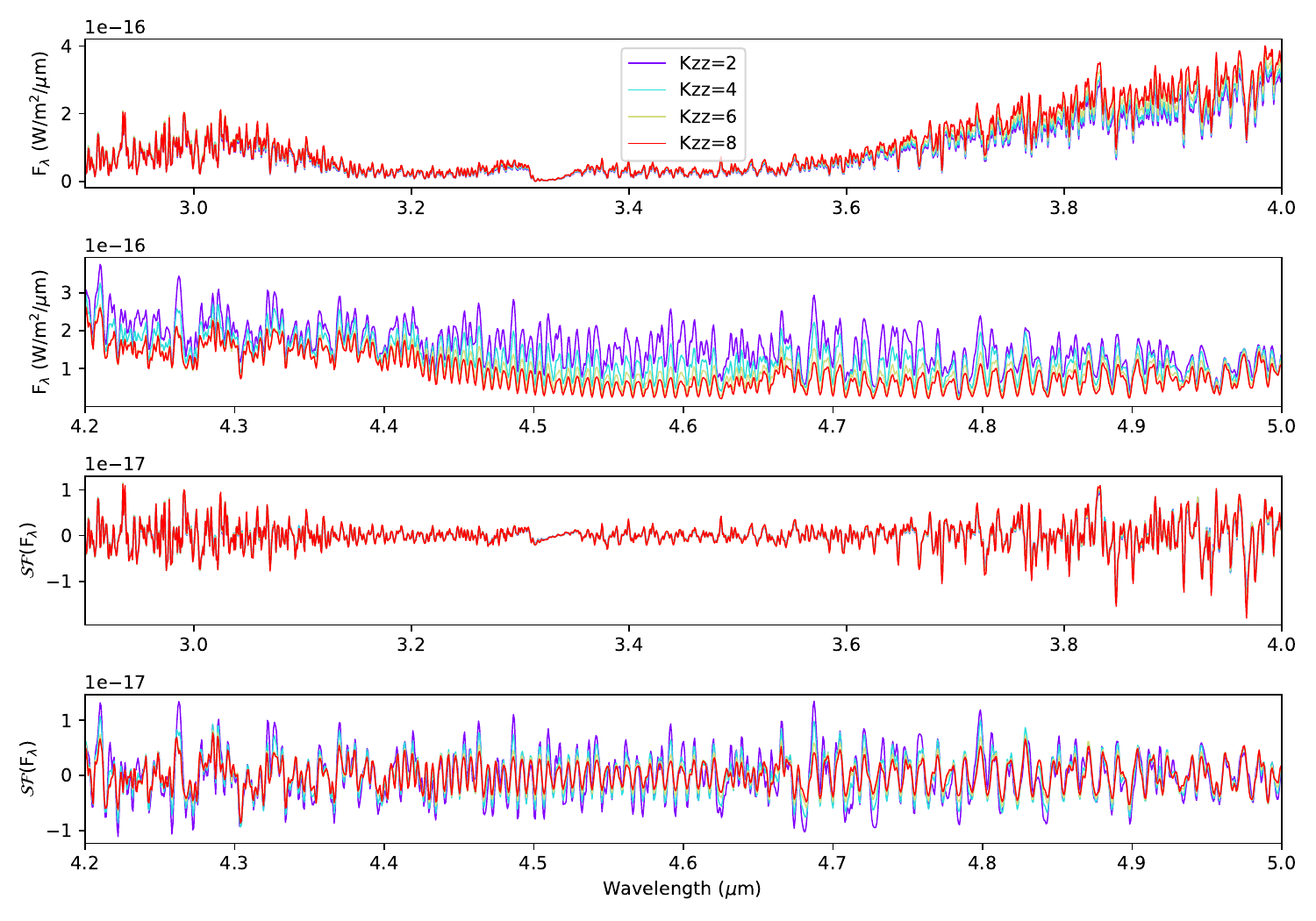}
\caption{Flux corrected and broadened to R$\sim$2700 NEWERA-PHOENIX models at an effective temperature of 1000 K, $\log g$=5.0 dex, and $K_{\rm zz}$ values from 10$^2$--10$^8$. The top two panels show the continuum included model spectra showing visible changes as $K_{\rm zz}$ changes. The bottom two panels show the continuum subtracted models using the same method as our NIRSpec data with the data error range below for the different $K_{\rm zz}$ values. The bottom two panels show the changes are not completely retained and become smaller when losing the continuum information in this wavelength range. However, the NEWERA-PHOENIX grid shows less variation in NRS1 when compared to Sonora Flame Skimmer.}
\label{fig:newera_kzz_comp}
\end{figure*}

To look at the spectral features and how they vary, we take NEWERA-PHOENIX models and vary $K_{zz}$ only. We match the resolution and wavelength spacing of the models to that of our NIRSpec data to retain the variation of resolution with wavelength. We plot models that are the same effective temperature (1000 K) and surface gravity (5.0 dex) and step through $K_{zz}$s of 10$^{2}$, 10$^{4}$, 10$^{6}$, and 10$^{8}$ in Figure \ref{fig:newera_kzz_comp}. We then use the same method to subtract the continuum of the model spectra as the data for a proper comparison. Detector NRS1 loses the differences seen past 3.6 $\mu$m when the continuum is subtracted, and NRS2 retains some of the spectral line differences. A similar effect is noted when this method is conducted on Flame Skimmer with C/O, $K_{zz}$, and surface gravity. Due to the collapse of spectral line variation when subtracting the continuum from forward models, high signal-to-noise continuum subtracted spectra are necessary for robust parameter measurements if no continuum information is available in this wavelength range. This is reflected in the wide variety of best fit values when including the full dataset including the ground-based data and when only fitting the continuum subtracted NIRSpec G395H data in Tables \ref{tab:atm_param}--\ref{tab:atm_param_dyn}. 

\subsection{Formation}
How these substellar companions form and how their formation pathways differ from exoplanets is still a major question in astronomy. There have been a variety of proposed formation tracers such as the atmospheric C/O ratios of the companion and host \citep{oberg2011,boss2026}, eccentricity of the companion's orbit \citep{bowler2020astrometry}, and atmospheric isopologue abundances and ratios in comparison to the local ISM \citep{zhang2021}. 

For the star HD 13724, we have spectral observations with HARPS and CORALIE and measured chemical abundances presented in Giovinazzi et al. in prep. to derive the C/O ratio. They find a C/O = 0.36 $\pm$ 0.09 by comparing absolute abundance values from the star. When compared with the Solar value (C/O$_{\odot}$ = 0.59$\pm$0.08; \citealt{asplund2021}), this ratio indicates the host star could be enhanced with oxygen. With this information from the host star's atmosphere we can compare with our atmospheric characterization of the companion presented here. For HD 13724 B, we conducted both forward model analysis and retrieval analysis. Both methods indicate that the atmospheric C/O of HD 13724 B is most likely close to Solar, which is indicative of carbon enhancement when compared to its host star. \cite{boss2026} shows companions formed via gas disk gravitational instability could have super-stellar C/O ratios as a result of their formation and orbital evolution. For an object with a measured dynamical mass of 38.34$^{+0.62}_{-0.61}$M$_{\mathrm{Jup}}$, we would expect formation closer to a star such as gravitational instability, and our results are in line with this assumption supported by \cite{boss2026}. 


HD 13724 B additionally has a measured eccentricity, which is high at 0.4121$^{+0.0090}_{-0.0092}$. \cite{bowler2020astrometry} found evidence for brown dwarf companions exhibiting higher eccentricities that resemble the orbital properties of wide stellar binaries. This suggests that brown dwarf companions with high eccentricities predominately form in a similar fashion such as gravitational instability. This is in agreement with our results with estimated super-stellar C/O ratio of the companion leading to possible formation via gas disk gravitational instability. Variation in planet metallicity with stellar C/O ratios could further complicate discerning between formation and evolution pathways of substellar companions and their protoplanetary disks \citep{bergin2024}. Understanding how bulk metallicity, surface gravity, and disequilibrium chemical processes in self-consistent models will enable more robust comparisons to theoretical formation models.

\subsection{Future Work from the \textit{Dyn-Atmo} Survey}
HD 13724 B is the first target of 10 analyzed in this survey. We use a combination of forward models and retrievals in this work to explore the new wavelength coverage and high-contrast now accessible with JWST. To obtain high-contrast spectra of all targets we lose the continuum information and we show here that obtaining high signal-to-noise is critical when modeling spectra without their continuum. Future results will follow a similar approach using a combination of datasets for each object using both retrievals and forward models which has been shown to be complementary to maximize the amount of information gained with high-contrast NIRSpec IFU spectroscopy \citep{RuffioXuan2026, xuan2026}.


\section{Conclusion}\label{conclusion}

Here we present the first atmospheric characterization and analysis from the \textit{Dyn-Atmo} survey (P.I.: Rickman, co-P.I.: Bardalez Gagliuffi) of the high-contrast direct spectra of the substellar companion, HD 13724 B, with the JWST NIRSpec IFU. This is the most closely-separated companion detected on-sky with the NIRSpec IFU to-date, resulting in a continuum subtracted, moderate resolution (R$\sim$2700) spectrum from 2.9--5.3$\mu$m (Figs. \ref{fig:sep-det}-\ref{fig:spec}). We described our data reduction methods in Section \ref{data}, which are built off of the breads framework first published in \cite{ruffio2024} and \cite{hoch2024}. 



To model and characterize the atmosphere of HD 13724 B we used a forward modeling framework utilizing four grids, two public grids (BT-SETTL and Sonora Diamondback), and two upcoming grids (Sonora Flame Skimmer and NEWERA-PHOENIX). To explore the complementarities of modeling different datasets, we fit ground-based SPHERE IFS and GRAVITY K-band spectra individually and the entire dataset jointly with our NIRSpec spectrum. We then implemented a dynamical mass prior, which allowed tighter constraints on parameters, however, we find that the metallicity and surface gravity measurements are inconsistent across Sonora Diamondback (top panel of fig. \ref{fig:corner_mh}). Our results are in line with the findings of \cite{zafar2026} that T-type brown dwarfs are not well represented with our current understanding of the connection of clouds and chemistry. 



We additionally use Sonora Flame Skimmer and NEWERA-PHOENIX to model HD 13724 B's spectra with a vertical mixing parameter $K_{\rm zz}$. The bottom panel of figure \ref{fig:corner_mh} shows another level of degeneracy with metallicity, surface gravity and $K_{\rm zz}$ in Sonora Flame Skimmer indicating that the connection between disequilibrium chemistry, surface gravity and bulk metallicity are also not well represented by current state-of-the-art models. We note that the NEWERA-PHOENIX models $K_{\rm zz}$ values in are inherently different than those in Sonora Flame Skimmer. In Figure \ref{fig:kzz_newera_fs} we account for this difference in scale height assumption and we find that the best-fit values are incompatible with NEWERA-PHOENIX inferring higher values while Sonora Flame Skimmer prefers lower values. This indicates that careful understanding of the assumed chemistry and mixing timescales of model grids used are critical when characterizing atmospheres using data past 3$\mu$m. 

We conducted both retrievals on both the NIRSpec G395H data and the combination of datasets. We find that the inclusion of the near-infrared data is important for constraining the temperature profile at pressures greater than 1 bar, and enables more precise constraints on the effective temperature and surface gravity than relying on NIRSpec alone. However, all of the retrievals struggle to measure physically plausible radii, generally finding a radius consistent with the lower limit of the prior (0.7 R$_{\rm Jup}$). The retrievals find an atmosphere in chemical equilibrium, with an enhanced metallicity between 0.5-0.7, depending on the choice of datasets and dynamical mass prior used. Free chemistry retrievals were unable to identify physically plausible parameters, and the most plausible retrievals required assumptions of equilibrium chemistry, a dynamical mass prior and a constrained temperature profile. This demonstrates the value of physically motivated and self-consistent models when fitting low $S/N$ data. As with the reduced $\chi^{2}$ analysis using the NEWERA-PHOENIX grid, we find a super-stellar C/O ratio when compared to the host star C/O value. Combining this with the fact that HD 13724 B has a high eccentricity value, we hypothesize that the mid T-type companion formed via gravitational instability \citep{bowler2020,boss2026}.

This work is only the first atmospheric characterization from the \textit{Dyn-Atmo} survey. This survey will reduce all of the JWST NIRSpec G395H data uniformly following our methodology reported in this work allowing for a homogeneous analysis across the companions with measured dynamical masses. Future efforts will conduct similar atmospheric characterization analyses using both forward model and retrieval frameworks on the entire sample of targets and will show how using a combination of both techniques in a multidisciplinary approach will improve our understanding of substellar atmospheric processes. 

\section{Acknowledgments}
\label{acknowledgements}

This paper reports work based on observations with the NASA/ESA/CSA JWST, associated with program GO-6362 (PI: Emily Rickman, co-PI: Daniella Bardalez Gagliuffi), obtained at the Space Telescope Science Institute, which is operated by AURA, Inc., under NASA contract NAS 5-03127. Funding is provided to the team by STScI through grant JWST-GO-06362.002-A. 

K.K.W.H acknowledges funding from the Giacconi Fellowship at the Space Telescope Science Institute as well as funding from the Space Telescope Science Institute's Director's Discretionary Research Funds grant number D0101.90370. K.K.W.H additionally acknowledges Travis Barman for creating the custom NEWERA-PHOENIX model grid used in this work, in addition to Dr. Sarah Moran and Dr. Brianna Lacy for helpful discussions supporting major findings presented.

J.M. acknowledges support from the National Science Foundation Graduate Research Fellowship Program under Grant No. DGE 2137420.
J.W.X is grateful for support from the Heising-Simons Foundation 51 Pegasi b Fellowship (grant \#2025-5887).
E.N. would like to thank Dr. Johanna Vos, and acknowledges support from Royal Society - Research Ireland University Research Fellowship (URF/1/221932, RF/ERE/221108).

The atmospheric retrievals were performed on the Luxembourg national supercomputer MeluXina.
The authors gratefully acknowledge the LuxProvide teams for their expert support.
This research made use of the SIMBAD database and the VizieR Catalogue access tool, both operated at the CDS, Strasbourg, France. The original descriptions of the SIMBAD and VizieR services were published in \citet{wenger2000} and \citet{ochsenbein2000}.

\appendix

\section{Computing Covariance Matrices}\label{ref:covariance_section}
\begin{figure}[H]
    \centering
    \includegraphics[width=0.95\linewidth]{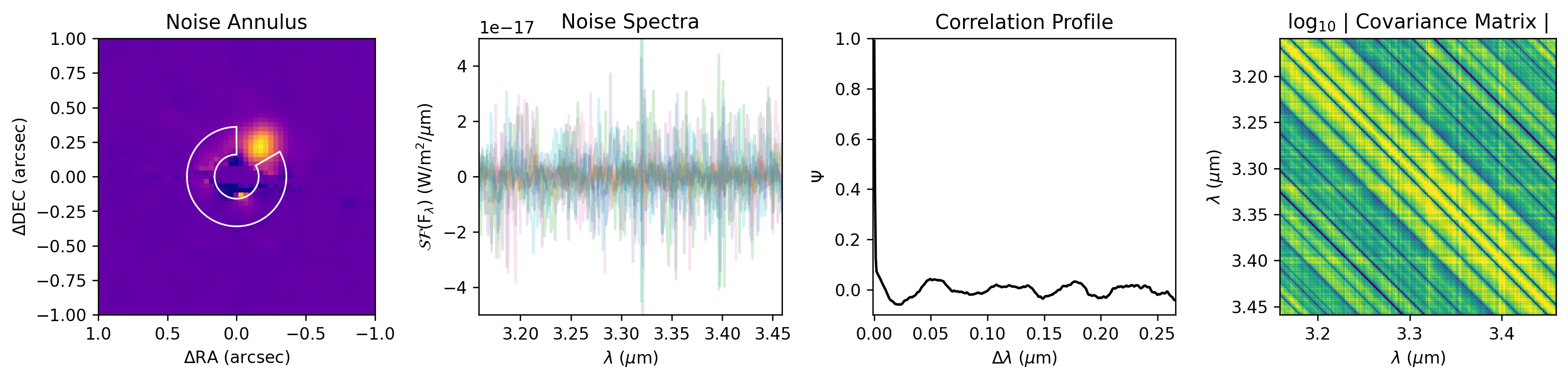}
    \caption{Example process of constructing the empirical error covariance matrix in the second of eight distinct wavelength regions: $\lambda \in [3.159, 3.459]$ $\mu$m. (Left) CCF detection map in IFU coordinates including the ``noise annulus" region which is used to extract the samples for estimating the empirical error covariance. (Middle Left) Subset of 11/239 noise samples extracted in this wavelength region at 20\% fractional opacity in a variety of colors for visual demonstration. (Middle Right) Auto-correlation profile $\Psi$ computed across the noise samples using Equation \ref{eq:Psi}. (Right) 2D Covariance matrix computed in this wavelength region as the element-wise product of the 2D correlation matrix with the outer product of the error bar vector.}
    \label{fig:annulus_correlation}
\end{figure}
\begin{figure}[H]
    \centering
    \includegraphics[width=0.95\textwidth]{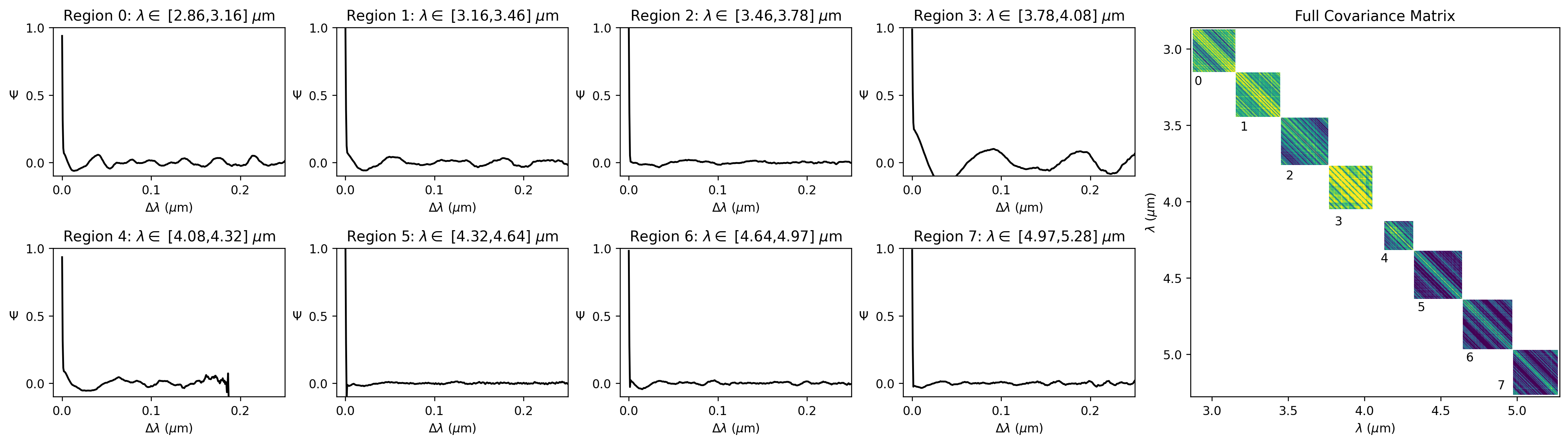}
    \caption{Auto-correlation profiles for each of the eight distinct wavelength regions with unique node spacing and the full covariance matrix with eight unique sub-blocks and zero elements in the off-block-diagonal regions.}
    \label{fig:eight_regions}
\end{figure}

Prior investigations \citep{Greco_2016, nasedkin_hcicovariance_2023} have demonstrated that obtaining unbiased parameter posteriors requires modeling of spectrally correlated noise when computing atmospheric model likelihoods. We outline an empirical approach to this issue and illustrate it in Figure \ref{fig:annulus_correlation}. 
The ``noise annulus" is defined as the region from 0.16" to 0.36" with a ~60 degree ``slice" centered at the position of the companion removed. The spectral samples extracted from this region are taken to represent the distribution of empirical noise residuals from which the correlations in the genuine spectra are estimated. The one-dimensional auto-correlation profile is assumed to be a simple function of $\Delta\lambda$, which is estimated with an average over positions and wavelengths of the residual samples $r$.
\begin{equation}\label{eq:Psi}
    \Psi(\Delta \lambda) = \langle r_{x,\lambda} r_{x,\lambda+\Delta \lambda} \rangle_{x,\lambda}
\end{equation}
The 1D correlation profile contains two notable features, a 2 pixel wide covariance originating from regular wavelength grid interpolation, and a larger oscillation originating from imperfections in the spline-based continuum subtraction. The first is identical to the previously demonstrated feature in \cite{ruffio2024} Appendix E, while the second is unique the continuum-subtracted data. This 1D profile is smoothed with a 20-pixel wide running mean and subsequently transformed into 2D covariance matrix by first applying the 1D function to a 2D difference matrix
\begin{equation}
    \Psi_{i,j} = \Psi(\Delta \lambda = | \lambda_i - \lambda_j |),
\end{equation}
and second by taking the element-wise product of the 2D correlation matrix with the outer product of the error bar vector $\sigma$
\begin{equation}
    C_{i,j} = \sigma_i\sigma_j\Psi_{i,j}.
\end{equation}
This task is performed eight times, once on each of the eight distinct wavelength regions that each have distinct spline node spacing, in line with the analysis in \cite{RuffioXuan2026} (see Methods / Supplementary Figure 3). The off-block-diagonal elements are assumed to be zero for simplicity and the non-trivial nature of estimating these much longer length scales. The variations in the 1D profiles across the different regions can be seen in Figure \ref{fig:eight_regions}. The exact implementation of this calculation in the code can be referenced at the following url: \url{https://breads.readthedocs.io/en/latest/tutorials/jwst/nirspec_2_analyzing_data_products_CCF_covariance.html#Computing-Covariance-Matrices}.

\section{Complete retrieval results}
{\scriptsize
\begin{longtable}{llcccrrrrc}
\caption{Model-comparison summary for the full HD\,13724\,B retrieval suite. Data shorthand: `NS' NIRSpec, `P' photometry, `G' GRAVITY, `S' SPHERE. . $\Delta\log\mathcal{Z}$, $\Delta$BIC and $\Delta\mathrm{BPIC}_{\mathrm{S}}$ are normalised to the best-evidence run R12. Log-evidence differences are only valid Bayes factors between runs sharing the same $n_{\rm data}$.}\label{tab:complete_comparison}\\
\toprule
Run & PT & Data & Chem. & Cloud & $n_{\rm par}$ & $\Delta\log\mathcal{Z}$ & $\Delta$BIC & $\Delta\mathrm{BPIC}_{\mathrm{S}}$ & $\chi^2_\nu$ \\
\midrule
\endfirsthead
\toprule
Run & PT & Data & Chem. & Cloud & $n_{\rm par}$ & $\Delta\log\mathcal{Z}$ & $\Delta$BIC & $\Delta\mathrm{BPIC}_{\mathrm{S}}$ & $\chi^2_\nu$ \\
\midrule
\endhead
\bottomrule
\endfoot
\multicolumn{10}{c}{Baseline}\\

R01 & 10 PT & NS & Eq. & Clear & 19 & -4471.991 & 20578.9 & 20585.2 & 0.505 \\
R02 & 10 PT & NS & Eq. & Clear & 19 & -4468.217 & 20577.9 & 20583.5 & 0.505 \\
R03 & 10 PT & NS+G+S & Eq. & Clear & 20 & -96.844 & 439.1 & 438.5 & 0.549 \\
R04 & 10 PT & NS+G+S & Eq. & Clear & 20 & -95.882 & 442.6 & 441.3 & 0.550 \\
\midrule
\multicolumn{10}{c}{Molecular detections}\\
R05 & 5 PT & NS & Free & Clear & 23 & -4438.487 & 20458.5 & 20442.9 & 0.461 \\
R05-CO & 5 PT & NS & Free & Clear & 21 & -4456.692 & 20536.3 & 20531.0 & 0.488 \\
R05-13CO & 5 PT & NS & Free & Clear & 21 & -4438.472 & 20447.3 & 20442.8 & 0.462 \\
R05-CH4 & 5 PT & NS & Free & Clear & 21 & -4472.752 & 20604.7 & 20605.9 & 0.508 \\
R05-CO2 & 5 PT & NS & Free & Clear & 21 & -4462.435 & 20569.5 & 20562.6 & 0.498 \\
R05-H2O & 5 PT & NS & Free & Clear & 21 & -4452.905 & 20502.1 & 20498.4 & 0.478 \\
R05-HCN & 5 PT & NS & Free & Clear & 20 & -4438.322 & 20438.8 & 20440.1 & 0.461 \\
R05-H2S & 5 PT & NS & Free & Clear & 20 & -4438.256 & 20437.8 & 20440.8 & 0.461 \\
R05-Na & 5 PT & NS & Free & Clear & 20 & -4438.417 & 20437.3 & 20441.2 & 0.461 \\
R05-PH3 & 5 PT & NS & Free & Clear & 20 & -4438.093 & 20438.7 & 20440.3 & 0.461 \\
\midrule
\multicolumn{10}{c}{Equilibrium chemistry}\\
R06 & 5 PT & NS & Eq. & Clear & 15 & -4458.895 & 20514.4 & 20542.9 & 0.495 \\
R07 & 5 PT & NS & Diseq. & Clear & 16 & -4458.424 & 20514.4 & 20538.2 & 0.493 \\
R08 & 10 PT & NS+G+S & Eq. & Clear & 20 & -95.882 & 442.6 & 441.3 & 0.550 \\
R09 & 10 PT & NS+G+S & Diseq. & Clear & 22 & -97.287 & 457.8 & 444.8 & 0.550 \\
R10 & 10 PT & NS+G+S & Free & Silicate & 34 & -97.422 & 493.8 & 405.3 & 0.535 \\
\midrule
\multicolumn{10}{c}{Including Photometry}\\
R11 & 10 PT & NS+P+G+S & Eq. & Clear & 20 & -13.990 & 54.9 & 54.3 & 0.565 \\
R12 & 10 PT & NS+P+G+S & Eq. & Clear & 20 & 0.000 & 0.0 & 0.0 & 0.550 \\
R13 & 10 PT & NS+P+G+S & Eq. & Clear & 20 & -100.962 & 444.3 & 446.9 & 0.678 \\
R14 & 10 PT & NS+P+G+S & Eq. & Clear & 21 & -0.194 & 9.7 & 2.0 & 0.551 \\
R15 & 10 PT & NS+P+G+S & Free & Clear & 29 & -73.078 & 365.3 & 308.7 & 0.638 \\
R16 & 10 PT & NS+P+G+S & Diseq. & Clear & 22 & -63.817 & 302.2 & 290.5 & 0.635 \\
R17 & 10 PT & NS+P+G+S & Free & Grey & 29 & -130.424 & 623.3 & 568.5 & 0.701 \\
R18 & 10 PT & NS+P+G+S & Free & Grey & 28 & -82.535 & 394.0 & 344.4 & 0.648 \\
R19 & 10 PT & NS+P+G+S & Free & Silicate & 34 & -0.239 & 60.6 & -25.3 & 0.538 \\
R20 & 10 PT & NS+P+G+S & Free & Silicate & 34 & -0.518 & 59.8 & -27.2 & 0.538 \\
R21 & 10 PT & NS+P+G+S & Free & Silicate & 34 & -0.545 & 61.2 & -26.8 & 0.538 \\
\midrule
\multicolumn{10}{c}{Data inclusion}\\
R22 & 5 PT & NS & Free & Grey & 25 & -4438.865 & 20476.6 & 20448.2 & 0.461 \\
R23 & 5 PT & NS+P & Free & Grey & 23 & -4350.749 & 20048.6 & 20034.3 & 0.471 \\
R24 & 5 PT & NS+P+G & Free & Grey & 24 & -608.315 & 2809.6 & 2782.5 & 0.521 \\
R25 & 5 PT & NS+P+G+S & Free & Grey & 24 & -98.849 & 372.0 & 354.2 & 0.650 \\
R26 & 5 PT & NS+P+S & Free & Grey & 23 & -3816.112 & 17587.3 & 17569.6 & 0.602 \\
R27 & 5 PT & NS+P & Free & Clear & 23 & -4349.960 & 20049.3 & 20033.7 & 0.471 \\
R28 & 5 PT & NS+P+G & Free & Clear & 23 & -608.630 & 2801.1 & 2780.1 & 0.521 \\
R29 & 5 PT & NS+P+G+S & Free & Clear & 23 & -107.286 & 427.9 & 414.5 & 0.668 \\
R30 & 5 PT & NS+P+G+S & Free & Clear & 23 & -92.339 & 395.5 & 376.4 & 0.659 \\
R31 & 5 PT & NS+P+S & Free & Clear & 22 & -3815.948 & 17579.1 & 17567.6 & 0.594 \\
R32 & 5 PT & NS+P+G+S & Free & Silicate & 29 & -54.776 & 243.1 & 199.4 & 0.605 \\
R33 & 10 PT & NS & Free & Clear & 28 & -4438.963 & 20496.3 & 20447.3 & 0.460 \\
\end{longtable}
}
{\scriptsize
\begin{longtable}{lcccccc}
\caption{Complete inferred and derived parameters for the HD\,13724\,B retrieval suite. For free-chemistry runs [M/H] and C/O are the derived from the inferred abundances; for (dis)equilibrium runs they are the sampled [M/H] and C/O parameters.}
\label{tab:complete_params}\\
\toprule
Run & $M$ [$M_{\rm Jup}$] & $R$ [$R_{\rm Jup}$] & $T_{\rm eff}$ [K] & $\log g$ [cgs] & {[M/H]} & C/O \\
\midrule
\endfirsthead
\toprule
Run & $M$ [$M_{\rm Jup}$] & $R$ [$R_{\rm Jup}$] & $T_{\rm eff}$ [K] & $\log g$ [cgs] & {[M/H]} & C/O \\
\midrule
\endhead
\bottomrule
\endfoot
R01 & $32.00\pm12.00$ & $1.0490\pm0.0760$ & $939.3_{-32.7}^{+35.4}$ & $4.880\pm0.164$ & $0.490_{-0.110}^{+0.110}$ & $0.162_{-0.031}^{+0.031}$ \\
R02 & $38.15\pm0.45$ & $0.7820\pm0.0510$ & $1160.6_{-68.4}^{+63.2}$ & $5.209\pm0.056$ & $0.531_{-0.093}^{+0.093}$ & $0.438_{-0.090}^{+0.090}$ \\
R03 & $66.00\pm12.00$ & $0.7076\pm0.0085$ & $1331.7_{-14.9}^{+14.1}$ & $5.531\pm0.080$ & $0.592_{-0.067}^{+0.067}$ & $0.543_{-0.025}^{+0.025}$ \\
R04 & $38.26\pm0.56$ & $0.7072\pm0.0074$ & $1322.6_{-8.2}^{+8.6}$ & $5.297\pm0.011$ & $0.634_{-0.076}^{+0.076}$ & $0.574_{-0.018}^{+0.018}$ \\
\midrule
R05 & $17.00\pm13.00$ & $1.9140\pm0.0920$ & $724.7_{-51.9}^{+555.5}$ & $4.098\pm0.322$ & $1.561_{-0.405}^{+0.289}$ & $0.969_{-0.024}^{+0.014}$ \\
R05-CO & $38.24\pm0.50$ & $1.8880\pm0.1000$ & $690.9_{-74.8}^{+652.4}$ & $4.444\pm0.046$ & $2.000_{-0.231}^{+0.451}$ & $0.872_{-0.073}^{+0.048}$ \\
R05-13CO & $38.22\pm0.53$ & $1.8800\pm0.1100$ & $744.9_{-53.8}^{+478.1}$ & $4.445\pm0.054$ & $1.711_{-0.234}^{+0.192}$ & $0.968_{-0.025}^{+0.016}$ \\
R05-CH4 & $38.21\pm0.53$ & $1.1300\pm0.1700$ & $864.9_{-67.6}^{+66.3}$ & $4.892\pm0.126$ & $0.919_{-0.815}^{+0.577}$ & $0.869_{-0.087}^{+0.055}$ \\
R05-CO2 & $38.25\pm0.55$ & $1.1000\pm0.1300$ & $895.5_{-71.1}^{+79.6}$ & $4.912\pm0.102$ & $0.834_{-0.362}^{+0.450}$ & $0.975_{-0.052}^{+0.016}$ \\
R05-H2O & $38.23\pm0.48$ & $1.4800\pm0.1500$ & $1094.9_{-191.3}^{+582.3}$ & $4.653\pm0.091$ & $1.837_{-0.107}^{+0.072}$ & $1.012_{-0.005}^{+0.005}$ \\
R05-HCN & $38.19\pm0.53$ & $1.8990\pm0.0990$ & $742.0_{-54.3}^{+561.8}$ & $4.439\pm0.046$ & $1.724_{-0.235}^{+0.174}$ & $0.968_{-0.023}^{+0.016}$ \\
R05-H2S & $38.24\pm0.56$ & $1.8970\pm0.0990$ & $723.8_{-46.3}^{+408.5}$ & $4.441\pm0.046$ & $1.723_{-0.256}^{+0.184}$ & $0.969_{-0.024}^{+0.016}$ \\
R05-Na & $38.21\pm0.53$ & $1.8900\pm0.1100$ & $1223.7_{-522.5}^{+775.5}$ & $4.443\pm0.052$ & $1.689_{-0.250}^{+0.193}$ & $0.967_{-0.028}^{+0.017}$ \\
R05-NH3 & $38.23\pm0.56$ & $1.8900\pm0.1000$ & $736.4_{-58.1}^{+496.8}$ & $4.442\pm0.049$ & $1.723_{-0.248}^{+0.185}$ & $0.968_{-0.025}^{+0.017}$ \\
R05-PH3 & $38.18\pm0.54$ & $1.9000\pm0.1000$ & $763.8_{-76.1}^{+735.7}$ & $4.439\pm0.047$ & $1.732_{-0.254}^{+0.175}$ & $0.968_{-0.026}^{+0.017}$ \\
\midrule
R06 & $38.32\pm0.50$ & $0.9230\pm0.0230$ & $1306.1_{-109.1}^{+94.4}$ & $5.066\pm0.022$ & $0.560_{-0.120}^{+0.120}$ & $0.663_{-0.056}^{+0.056}$ \\
R07 &$38.28\pm0.51$ & $0.9540\pm0.0460$ & $1260.1_{-71.8}^{+84.8}$ & $5.038\pm0.041$ & $0.380_{-0.100}^{+0.100}$ & $0.623_{-0.049}^{+0.049}$ \\
R09 & $38.23\pm0.56$ & $0.7071\pm0.0073$ & $1329.0_{-15.0}^{+14.2}$ & $5.296\pm0.011$ & $0.629_{-0.073}^{+0.073}$ & $0.576_{-0.018}^{+0.018}$ \\
R09 & $38.18\pm0.41$ & $0.9430\pm0.0290$ & $1260.8_{-13.9}^{+12.8}$ & $5.045\pm0.027$ & $0.601_{-0.172}^{+0.181}$ & $0.963_{-0.026}^{+0.017}$ \\
\midrule
R10 & $38.32\pm0.55$ & $0.9009\pm0.0010$ & $1197.3_{-5.0}^{+5.1}$ & $5.084\pm0.007$ & $0.936_{-0.054}^{+0.054}$ & $0.463_{-0.011}^{+0.011}$ \\
R11 & $38.46\pm0.53$ & $0.7052\pm0.0052$ & $1322.4_{-9.7}^{+8.1}$ & $5.295\pm0.012$ & $0.951_{-0.038}^{+0.038}$ & $0.528_{-0.009}^{+0.009}$ \\
R12 & $71.80\pm7.50$ & $0.9009\pm0.0009$ & $1202.0_{-3.4}^{+4.2}$ & $5.361\pm0.046$ & $0.941_{-0.058}^{+0.058}$ & $0.419_{-0.016}^{+0.016}$ \\
R13 & $38.44\pm0.55$ & $0.7055\pm0.0054$ & $1321.4_{-10.4}^{+7.9}$ & $5.295\pm0.011$ & $0.953_{-0.038}^{+0.038}$ & $0.528_{-0.009}^{+0.009}$ \\
R14 & $38.23\pm0.48$ & $0.9047\pm0.0044$ & $1216.4_{-27.3}^{+14.1}$ & $5.082\pm0.007$ & $0.916_{-0.125}^{+0.088}$ & $0.659_{-0.069}^{+0.052}$ \\
R15 & $38.46\pm0.54$ & $0.7057\pm0.0053$ & $1323.7_{-6.0}^{+5.4}$ & $5.301\pm0.009$ & $0.947_{-0.040}^{+0.040}$ & $0.528_{-0.009}^{+0.009}$ \\
R16 & $38.25\pm0.57$ & $0.9230\pm0.0180$ & $1384.7_{-13.4}^{+16.5}$ & $5.030\pm0.025$ & $-2.509_{-0.240}^{+0.139}$ & $1.108_{-0.650}^{+7.571}$ \\
R17 & $38.17\pm0.52$ & $0.9045\pm0.0044$ & $1176.6_{-13.2}^{+17.9}$ & $5.082\pm0.007$ & $0.948_{-0.121}^{+0.079}$ & $0.654_{-0.079}^{+0.063}$ \\
R18 & $38.36\pm0.51$ & $0.9076\pm0.0072$ & $1099.9_{-11.9}^{+10.2}$ & $5.081\pm0.009$ & $0.894_{-0.122}^{+0.093}$ & $1.030_{-0.020}^{+0.024}$ \\
R19 & $38.24\pm0.46$ & $0.9052\pm0.0047$ & $1220.4_{-17.0}^{+12.5}$ & $5.073\pm0.012$ & $0.579_{-0.231}^{+0.255}$ & $0.960_{-0.041}^{+0.021}$ \\
R20 & $38.19\pm0.48$ & $0.9044\pm0.0044$ & $1220.4_{-16.5}^{+13.5}$ & $5.075\pm0.011$ & $0.685_{-0.227}^{+0.182}$ & $0.965_{-0.029}^{+0.017}$ \\
\midrule
R21 & $38.21\pm0.52$ & $1.9000\pm0.1000$ & $746.1_{-62.8}^{+482.8}$ & $4.438\pm0.047$ & $1.720_{-0.253}^{+0.186}$ & $0.965_{-0.026}^{+0.018}$ \\
R22 & $37.60\pm0.03$ & $1.3330\pm0.0250$ & $959.5_{-63.1}^{+207.9}$ & $4.827\pm0.074$ & $1.384_{-0.291}^{+0.378}$ & $0.982_{-0.021}^{+0.013}$ \\
R23 & $37.83\pm0.29$ & $0.9160\pm0.0140$ & $1481.2_{-328.0}^{+1108.1}$ & $5.060\pm0.023$ & $1.530_{-0.225}^{+0.237}$ & $0.808_{-0.067}^{+0.063}$ \\
R24 & $38.16\pm0.30$ & $0.9223\pm0.0081$ & $1317.3_{-4.1}^{+5.9}$ & $5.066\pm0.009$ & $-0.950_{-0.073}^{+0.017}$ & $1477_{-167}^{+177}$ \\
R25 & $38.20\pm0.55$ & $0.9540\pm0.0450$ & $1300.9_{-36.9}^{+25.3}$ & $5.036\pm0.041$ & $-1.973_{-0.151}^{+0.158}$ & $0.500_{-0.000}^{+0.001}$ \\
R26 & $38.23\pm0.48$ & $1.2074\pm0.0994$ & $958.7_{-55.5}^{+145.5}$ & $4.832\pm0.071$ & $1.332_{-0.335}^{+0.380}$ & $0.986_{-0.020}^{+0.014}$ \\
R27 & $38.28\pm0.48$ & $0.9300\pm0.0250$ & $1582.4_{-460.3}^{+1076.3}$ & $5.059\pm0.023$ & $1.504_{-0.226}^{+0.227}$ & $0.806_{-0.063}^{+0.059}$ \\
R28 & $38.21\pm0.32$ & $0.9096\pm0.0053$ & $1111.0_{-14.6}^{+13.1}$ & $5.078\pm0.007$ & $0.626_{-0.047}^{+0.062}$ & $0.791_{-0.026}^{+0.031}$ \\
R29 & $66.70\pm6.70$ & $0.9026\pm0.0023$ & $1144.4_{-6.2}^{+5.1}$ & $5.327\pm0.044$ & $1.474_{-0.192}^{+0.189}$ & $0.738_{-0.080}^{+0.063}$ \\
R30 & $38.23\pm0.55$ & $1.9380\pm0.0560$ & $1299.1_{-39.8}^{+29.8}$ & $5.036\pm0.043$ & $-1.964_{-0.164}^{+0.155}$ & $0.500_{-0.000}^{+0.001}$ \\
R31 & $35.90\pm1.10$ & $0.9280\pm0.0420$ & $1236.7_{-46.9}^{+35.2}$ & $5.029\pm0.040$ & $-0.250_{-0.061}^{+0.061}$ & $4.492_{-1.081}^{+1.528}$ \\
R32 & $38.21\pm0.53$ & $1.8500\pm0.1300$ & $704.9_{-51.7}^{+614.4}$ & $4.464\pm0.065$ & $1.864_{-0.179}^{+0.139}$ & $0.901_{-0.052}^{+0.039}$ \\
\end{longtable}
}





\bibliography{bibliography,data_reduction}{}

@ARTICLE{bergin2024,
       author = {{Bergin}, Edwin A. and {Booth}, Richard A. and {Colmenares}, Maria Jose and {Ilee}, John D.},
        title = "{C/O Ratios and the Formation of Wide-separation Exoplanets}",
      journal = {\apjl},
         year = 2024,
        month = jul,
       volume = {969},
       number = {1},
          eid = {L21},
        pages = {L21},
          doi = {10.3847/2041-8213/ad5839},
archivePrefix = {arXiv},
       eprint = {2406.12037},
 primaryClass = {astro-ph.EP},
       adsurl = {https://ui.adsabs.harvard.edu/abs/2024ApJ...969L..21B}
}

@ARTICLE{zahnle2014,
       author = {{Zahnle}, Kevin J. and {Marley}, Mark S.},
        title = "{Methane, Carbon Monoxide, and Ammonia in Brown Dwarfs and Self-Luminous Giant Planets}",
      journal = {\apj},
         year = 2014,
        month = dec,
       volume = {797},
       number = {1},
          eid = {41},
        pages = {41},
          doi = {10.1088/0004-637X/797/1/41},
archivePrefix = {arXiv},
       eprint = {1408.6283},
 primaryClass = {astro-ph.EP},
       adsurl = {https://ui.adsabs.harvard.edu/abs/2014ApJ...797...41Z}
}

@ARTICLE{asplund2021,
       author = {{Asplund}, M. and {Amarsi}, A.~M. and {Grevesse}, N.},
        title = "{The chemical make-up of the Sun: A 2020 vision}",
      journal = {\aap},
         year = 2021,
        month = sep,
       volume = {653},
          eid = {A141},
        pages = {A141},
          doi = {10.1051/0004-6361/202140445},
archivePrefix = {arXiv},
       eprint = {2105.01661},
 primaryClass = {astro-ph.SR},
       adsurl = {https://ui.adsabs.harvard.edu/abs/2021A&A...653A.141A}
}

@ARTICLE{kharchenko2001,
       author = {{Kharchenko}, N.~V.},
        title = "{All-sky compiled catalogue of 2.5 million stars}",
      journal = {Kinematika i Fizika Nebesnykh Tel},
         year = 2001,
        month = oct,
       volume = {17},
       number = {5},
        pages = {409-423},
       adsurl = {https://ui.adsabs.harvard.edu/abs/2001KFNT...17..409K}
}

@ARTICLE{hog2000,
       author = {{H{\o}g}, E. and {Fabricius}, C. and {Makarov}, V.~V. and {Urban}, S. and {Corbin}, T. and {Wycoff}, G. and {Bastian}, U. and {Schwekendiek}, P. and {Wicenec}, A.},
        title = "{The Tycho-2 catalogue of the 2.5 million brightest stars}",
      journal = {\aap},
         year = 2000,
        month = mar,
       volume = {355},
        pages = {L27-L30},
       adsurl = {https://ui.adsabs.harvard.edu/abs/2000A&A...355L..27H}
}

@ARTICLE{mang2026,
       author = {{Mang}, James and {Chachan}, Yayaati and {Morley}, Caroline V. and {Batalha}, Natasha E. and {Wogan}, Nicholas F. and {Mukherjee}, Sagnick and {Fortney}, Jonathan J. and {Marley}, Mark S. and {Visscher}, Channon and {Gharib-Nezhad}, Ehsan},
        title = "{The Sonora Substellar Atmosphere Models. VII. Flame Skimmer: Cloud-free Atmospheric and Evolutionary Models for the Coldest Substellar Objects}",
      journal = {arXiv e-prints},
         year = 2026,
        month = aug,
          eid = {arXiv:2608.06454},
        pages = {arXiv:2608.06454},
          doi = {10.48550/arXiv.2608.06454},
archivePrefix = {arXiv},
       eprint = {2608.06454},
 primaryClass = {astro-ph.EP},
       adsurl = {https://ui.adsabs.harvard.edu/abs/2026arXiv260806454M}
}

@ARTICLE{whiteford2026,
       author = {{Whiteford}, Niall and {Faherty}, Jacqueline K. and {Burningham}, Ben and {Vos}, Johanna M. and {Petrus}, Simon and {Patapis}, Polychronis and {Biller}, Beth A. and {Skemer}, Andrew and {Hinkley}, Sasha and {Calamari}, Emily and {Su{\'a}rez}, Genaro and {Cruz}, Kelle L. and {Miles}, Brittany E. and {Carter}, Aarynn L. and {Martinez}, Francisco A. and {Rowland}, Melanie J. and {Absil}, Olivier and {Adams}, Arthur D. and {Balmer}, William O. and {Boccaletti}, Anthony and {Bonavita}, Mariangela and {Bonnefoy}, Micka{\"e}l and {Booth}, Mark and {Bowler}, Brendan P. and {Briesemeister}, Zackery W. and {Bryan}, Marta L. and {Calissendorff}, Per and {Cantalloube}, Faustine and {Charnay}, Benjamin and {Chauvin}, Ga{\"e}l and {Chen}, Christine H. and {Choquet}, Elodie and {Christiaens}, Valentin and {Cugno}, Gabriele and {Currie}, Thayne and {Danielski}, Camilla and {De Furio}, Matthew and {Dupuy}, Trent J. and {Factor}, Samuel M. and {Fitzgerald}, Michael P. and {Fortney}, Jonathan J. and {Franson}, Kyle and {Girard}, Julien H. and {Gonzales}, Eileen C. and {Grady}, Carol A. and {Henning}, Thomas and {Hines}, Dean C. and {Hood}, Callie E. and {Hoch}, Kielan K.~W. and {Howe}, Alex R. and {Janson}, Markus and {Kalas}, Paul and {Kammerer}, Jens and {Kennedy}, Grant M. and {Kervella}, Pierre and {Kim}, Minjae and {Kitzmann}, Daniel and {Kraus}, Adam L. and {Kuzuhara}, Masayuki and {Lagage}, Pierre-Olivier and {Lagrange}, Anne-Marie and {Lawson}, Kellen and {Lazzoni}, Cecilia and {Leisenring}, Jarron M. and {Lew}, Ben W.~P. and {Liu}, Michael C. and {Liu}, Pengyu and {Llop-Sayson}, Jorge and {Lloyd}, James P. and {Lueber}, Anna and {Macintosh}, Bruce and {M{\^a}lin}, Mathilde and {Manjavacas}, Elena and {Marino}, Sebasti{\'a}n and {Marley}, Mark S. and {Marois}, Christian and {Martinez}, Raquel A. and {Matthews}, Elisabeth C. and {Matthews}, Brenda C. and {Mawet}, Dimitri and {Mazoyer}, Johan and {McElwain}, Michael W. and {Metchev}, Stanimir and {Meyer}, Michael R. and {Millar-Blanchaer}, Maxwell A. and {Molli{\`e}re}, Paul and {Moran}, Sarah E. and {Morley}, Caroline V. and {Mukherjee}, Sagnick and {Palma-Bifani}, Paulina and {Pantin}, Eric and {Perrin}, Marshall D. and {Pueyo}, Laurent and {Quanz}, Sascha P. and {Quirrenbach}, Andreas and {Ray}, Shrishmoy and {Rebollido}, Isabel and {Adams Redai}, Jea and {Ren}, Bin B. and {Rickman}, Emily and {Sallum}, Steph and {Samland}, Matthias and {Sargent}, Benjamin and {Schlieder}, Joshua E. and {Stapelfeldt}, Karl R. and {Stone}, Jordan M. and {Sutlieff}, Ben J. and {Tamura}, Motohide and {Tan}, Xianyu and {Theissen}, Christopher A. and {Tremblin}, Pascal and {Uyama}, Taichi and {Vasist}, Malavika and {Vigan}, Arthur and {Wagner}, Kevin and {Wang}, Jason J. and {Ward-Duong}, Kimberly and {Wolff}, Schuyler G. and {Worthen}, Kadin and {Wyatt}, Mark C. and {Ygouf}, Marie and {Zurlo}, Alice and {Zhang}, Xi and {Zhang}, Keming and {Zhang}, Zhoujian and {Zhou}, Yifan},
        title = "{The JWST Early Release Science Program for Direct Observations of Exoplanetary Systems VIII: patchy forsterite and enstatite clouds in the atmosphere of VHS 1256 b, retrieval lessons learned and outlook to the future}",
      journal = {arXiv e-prints},
         year = 2026,
        month = aug,
          eid = {arXiv:2608.06583},
        pages = {arXiv:2608.06583},
          doi = {10.48550/arXiv.2608.06583},
archivePrefix = {arXiv},
       eprint = {2608.06583},
 primaryClass = {astro-ph.EP},
       adsurl = {https://ui.adsabs.harvard.edu/abs/2026arXiv260806583W}
}

@ARTICLE{beiler2024,
       author = {{Beiler}, Samuel A. and {Mukherjee}, Sagnick and {Cushing}, Michael C. and {Kirkpatrick}, J. Davy and {Schneider}, Adam C. and {Kothari}, Harshil and {Marley}, Mark S. and {Visscher}, Channon},
        title = "{A Tale of Two Molecules: The Underprediction of CO$_{2}$ and Overprediction of PH$_{3}$ in Late T and Y Dwarf Atmospheric Models}",
      journal = {\apj},
         year = 2024,
        month = sep,
       volume = {973},
       number = {1},
          eid = {60},
        pages = {60},
          doi = {10.3847/1538-4357/ad6759},
archivePrefix = {arXiv},
       eprint = {2407.15950},
 primaryClass = {astro-ph.EP},
       adsurl = {https://ui.adsabs.harvard.edu/abs/2024ApJ...973...60B}
}

@ARTICLE{radcliffe2026,
       author = {{Radcliffe}, Alice and {Charnay}, Benjamin and {Lagrange}, Anne-Marie and {Kiefer}, Flavien and {B{\'e}zard}, Bruno and {Petrus}, Simon and {Palma-Bifani}, Paulina and {Ravet}, Matthieu and {Leconte}, J{\'e}r{\'e}my and {Marleau}, Gabriel-Dominique},
        title = "{Next-generation Exo-REM atmospheric models Application to VHS 1256 b to emulate patchy clouds}",
      journal = {\aap},
         year = 2026,
        month = jul,
       volume = {711},
          eid = {A257},
        pages = {A257},
          doi = {10.1051/0004-6361/202659006},
archivePrefix = {arXiv},
       eprint = {2605.29070},
 primaryClass = {astro-ph.EP},
       adsurl = {https://ui.adsabs.harvard.edu/abs/2026A&A...711A.257R}
}

@ARTICLE{matthews2024,
       author = {{Matthews}, E.~C. and {Carter}, A.~L. and {Pathak}, P. and {Morley}, C.~V. and {Phillips}, M.~W. and {P.~M.}, S. Krishanth and {Feng}, F. and {Bonse}, M.~J. and {Boogaard}, L.~A. and {Burt}, J.~A. and {Crossfield}, I.~J.~M. and {Douglas}, E.~S. and {Henning}, Th. and {Hom}, J. and {Ko}, C.-L. and {Kasper}, M. and {Lagrange}, A.-M. and {Petit dit de la Roche}, D. and {Philipot}, F.},
        title = "{A temperate super-Jupiter imaged with JWST in the mid-infrared}",
      journal = {\nat},
         year = 2024,
        month = sep,
       volume = {633},
       number = {8031},
        pages = {789-792},
          doi = {10.1038/s41586-024-07837-8},
archivePrefix = {arXiv},
       eprint = {2503.01599},
 primaryClass = {astro-ph.EP},
       adsurl = {https://ui.adsabs.harvard.edu/abs/2024Natur.633..789M}
}

@ARTICLE{bardalez-gagliuffi2025,
       author = {{Bardalez Gagliuffi}, Daniella C. and {Balmer}, William O. and {Pueyo}, Laurent and {Brandt}, Timothy D. and {Giovinazzi}, Mark R. and {Millholland}, Sarah and {Black}, Brennen and {Lu}, Tiger and {Rice}, Malena and {Mang}, James and {Morley}, Caroline and {Lacy}, Brianna and {Girard}, Julien H. and {Matthews}, Elisabeth C. and {Carter}, Aarynn L. and {Bowler}, Brendan P. and {Faherty}, Jacqueline K. and {Fontanive}, Clemence and {Rickman}, Emily},
        title = "{JWST Coronagraphic Images of 14 Her c: A Cold Giant Planet in a Dynamically Hot Multiplanet System}",
      journal = {\apjl},
         year = 2025,
        month = jul,
       volume = {988},
       number = {1},
          eid = {L18},
        pages = {L18},
          doi = {10.3847/2041-8213/ade30f},
archivePrefix = {arXiv},
       eprint = {2506.09201},
 primaryClass = {astro-ph.EP},
       adsurl = {https://ui.adsabs.harvard.edu/abs/2025ApJ...988L..18B}
}

@ARTICLE{thorngren2026,
       author = {{Thorngren}, Daniel P. and {Sing}, David K. and {Mukherjee}, Sagnick},
        title = "{Bayesian Model Comparison and Significance: Widespread Errors and How to Correct Them}",
      journal = {\apjs},
         year = 2026,
        month = mar,
       volume = {283},
       number = {1},
          eid = {10},
        pages = {10},
          doi = {10.3847/1538-4365/ae0e71},
archivePrefix = {arXiv},
       eprint = {2510.00169},
 primaryClass = {astro-ph.EP},
       adsurl = {https://ui.adsabs.harvard.edu/abs/2026ApJS..283...10T}
}

@ARTICLE{bowler2020,
       author = {{Bowler}, Brendan P. and {Blunt}, Sarah C. and {Nielsen}, Eric L.},
        title = "{Population-level Eccentricity Distributions of Imaged Exoplanets and Brown Dwarf Companions: Dynamical Evidence for Distinct Formation Channels}",
      journal = {\aj},
         year = 2020,
        month = feb,
       volume = {159},
       number = {2},
          eid = {63},
        pages = {63},
          doi = {10.3847/1538-3881/ab5b11},
archivePrefix = {arXiv},
       eprint = {1911.10569},
 primaryClass = {astro-ph.EP},
       adsurl = {https://ui.adsabs.harvard.edu/abs/2020AJ....159...63B}
}

@ARTICLE{boss2026,
       author = {{Boss}, Alan P.},
        title = "{Gas Giants Formed by Gravitational Instability May Accrete Atmospheres with Superstellar Carbon-to-oxygen Ratios}",
      journal = {\apj},
         year = 2026,
        month = may,
       volume = {1002},
       number = {1},
          eid = {56},
        pages = {56},
          doi = {10.3847/1538-4357/ae5ba7},
       adsurl = {https://ui.adsabs.harvard.edu/abs/2026ApJ..1002...56B}
}

@ARTICLE{yung1988,
       author = {{Yung}, Yuk L. and {Drew}, William A. and {Pinto}, Joseph P. and {Friedl}, Randall R.},
        title = "{Estimation of the reaction rate for the formation of CH $_{3}$O from H + H $_{2}$CO: Implications for chemistry in the solar system}",
      journal = {\icarus},
         year = 1988,
        month = mar,
       volume = {73},
       number = {3},
        pages = {516-526},
          doi = {10.1016/0019-1035(88)90061-9},
       adsurl = {https://ui.adsabs.harvard.edu/abs/1988Icar...73..516Y}
}

@ARTICLE{zahnlemarley2014,
       author = {{Zahnle}, Kevin J. and {Marley}, Mark S.},
        title = "{Methane, Carbon Monoxide, and Ammonia in Brown Dwarfs and Self-Luminous Giant Planets}",
      journal = {\apj},
         year = 2014,
        month = dec,
       volume = {797},
       number = {1},
          eid = {41},
        pages = {41},
          doi = {10.1088/0004-637X/797/1/41},
archivePrefix = {arXiv},
       eprint = {1408.6283},
 primaryClass = {astro-ph.EP},
       adsurl = {https://ui.adsabs.harvard.edu/abs/2014ApJ...797...41Z}
}

@ARTICLE{zafar2026,
       author = {{Rustamkulov}, Zafar and {Kirkpatrick}, J. and {Akeson}, Rachel and {Werner}, Michael W and {Ashby}, Matthew and {Chang}, Tzu-Ching and {Chen}, Shuang-Shuang and {Cooray}, Asantha and {Crill}, Brendan and {Dore}, Olivier and {Dowell}, C. and {Faisst}, Andreas and {Hui}, Howard and {Jeong}, Woong-Seob and {Kang}, Miju and {Korngut}, Phil and {Lisse}, Carey and {Masters}, Daniel and {Melnick}, Gary and {Nguyen}, Chi and {Paladini}, Roberta and {Tolls}, Volker and {Yang}, Yujin and {Zemcov}, Michael},
        title = "{SPHEREx 0.75 to 5 $μ$m Spectra for a Sequence of Nearby Brown Dwarfs}",
      journal = {arXiv e-prints},
         year = 2026,
        month = jul,
          eid = {arXiv:2607.00543},
        pages = {arXiv:2607.00543},
          doi = {10.48550/arXiv.2607.00543},
archivePrefix = {arXiv},
       eprint = {2607.00543},
 primaryClass = {astro-ph.SR},
       adsurl = {https://ui.adsabs.harvard.edu/abs/2026arXiv260700543R}
}

@ARTICLE{hauschildt2025,
       author = {{Hauschildt}, P.~H. and {Barman}, T. and {Baron}, E. and {Aufdenberg}, J.~P. and {Schweitzer}, A.},
        title = "{The NewEra model grid}",
      journal = {\aap},
         year = 2025,
        month = jun,
       volume = {698},
          eid = {A47},
        pages = {A47},
          doi = {10.1051/0004-6361/202554171},
archivePrefix = {arXiv},
       eprint = {2504.17597},
 primaryClass = {astro-ph.SR},
       adsurl = {https://ui.adsabs.harvard.edu/abs/2025A&A...698A..47H}
}

@ARTICLE{ruffio2024,
       author = {{Ruffio}, Jean-Baptiste and {Perrin}, Marshall D. and {Hoch}, Kielan K.~W. and {Kammerer}, Jens and {Konopacky}, Quinn M. and {Pueyo}, Laurent and {Madurowicz}, Alex and {Rickman}, Emily and {Theissen}, Christopher A. and {Agrawal}, Shubh and {Greenbaum}, Alexandra Z. and {Miles}, Brittany E. and {Barman}, Travis S. and {Balmer}, William O. and {Llop-Sayson}, Jorge and {Girard}, Julien H. and {Rebollido}, Isabel and {Soummer}, R{\'e}mi and {Allen}, Natalie H. and {Anderson}, Jay and {Beichman}, Charles A. and {Bellini}, Andrea and {Bryden}, Geoffrey and {Espinoza}, N{\'e}stor and {Glidden}, Ana and {Huang}, Jingcheng and {Lewis}, Nikole K. and {Libralato}, Mattia and {Louie}, Dana R. and {Sohn}, Sangmo Tony and {Seager}, Sara and {van der Marel}, Roeland P. and {Wakeford}, Hannah R. and {Watkins}, Laura L. and {Ygouf}, Marie and {Mountain}, C. Matt},
        title = "{JWST-TST High Contrast: Achieving Direct Spectroscopy of Faint Substellar Companions Next to Bright Stars with the NIRSpec Integral Field Unit}",
      journal = {\aj},
         year = 2024,
        month = aug,
       volume = {168},
       number = {2},
          eid = {73},
        pages = {73},
          doi = {10.3847/1538-3881/ad5281},
archivePrefix = {arXiv},
       eprint = {2310.09902},
 primaryClass = {astro-ph.EP},
       adsurl = {https://ui.adsabs.harvard.edu/abs/2024AJ....168...73R}
}

@ARTICLE{xuan2024c,
       author = {{Xuan}, Jerry W. and {M{\'e}rand}, A. and {Thompson}, W. and {Zhang}, Y. and {Lacour}, S. and {Blakely}, D. and {Mawet}, D. and {Oppenheimer}, R. and {Kammerer}, J. and {Batygin}, K. and {Sanghi}, A. and {Wang}, J. and {Ruffio}, J.-B. and {Liu}, M.~C. and {Knutson}, H. and {Brandner}, W. and {Burgasser}, A. and {Rickman}, E. and {Bowens-Rubin}, R. and {Salama}, M. and {Balmer}, W. and {Blunt}, S. and {Bourdarot}, G. and {Caselli}, P. and {Chauvin}, G. and {Davies}, R. and {Drescher}, A. and {Eckart}, A. and {Eisenhauer}, F. and {Fabricius}, M. and {Feuchtgruber}, H. and {Finger}, G. and {F{\"o}rster Schreiber}, N.~M. and {Garcia}, P. and {Genzel}, R. and {Gillessen}, S. and {Grant}, S. and {Hartl}, M. and {Hau{\ss}mann}, F. and {Henning}, T. and {Hinkley}, S. and {H{\"o}nig}, S.~F. and {Horrobin}, M. and {Houll{\'e}}, M. and {Janson}, M. and {Kervella}, P. and {Kral}, Q. and {Kreidberg}, L. and {Le Bouquin}, J.-B. and {Lutz}, D. and {Mang}, F. and {Marleau}, G.-D. and {Millour}, F. and {More}, N. and {Nowak}, M. and {Ott}, T. and {Otten}, G. and {Paumard}, T. and {Rabien}, S. and {Rau}, C. and {Ribeiro}, D.~C. and {Sadun Bordoni}, M. and {Sauter}, J. and {Shangguan}, J. and {Shimizu}, T.~T. and {Sykes}, C. and {Soulain}, A. and {Spezzano}, S. and {Straubmeier}, C. and {Stolker}, T. and {Sturm}, E. and {Subroweit}, M. and {Tacconi}, L.~J. and {van Dishoeck}, E.~F. and {Vigan}, A. and {Widmann}, F. and {Wieprecht}, E. and {Winterhalder}, T.~O. and {Woillez}, J.},
        title = "{The cool brown dwarf Gliese 229 B is a close binary}",
      journal = {\nat},
         year = 2024,
        month = oct,
       volume = {634},
       number = {8036},
        pages = {1070-1074},
          doi = {10.1038/s41586-024-08064-x},
archivePrefix = {arXiv},
       eprint = {2410.11953},
 primaryClass = {astro-ph.SR},
       adsurl = {https://ui.adsabs.harvard.edu/abs/2024Natur.634.1070X}
}

@ARTICLE{xuan2026,
       author = {{Xuan}, Jerry W. and {Ruffio}, Jean-Baptiste and {Chachan}, Yayaati and {Ohno}, Kazumasa and {Kesseli}, Aurora and {Murray-Clay}, Ruth and {Lee}, Eve J. and {Moses}, Julianne I. and {Balmer}, William O. and {Baburaj}, Aneesh and {Blake}, Geoffrey A. and {Johnstone}, Doug and {Zhang}, Yapeng and {Knutson}, Heather A. and {Mawet}, Dimitri and {Beichman}, Charles and {Hodapp}, Klaus and {Perrin}, Marshall D. and {Konopacky}, Quinn and {Meyer}, Michael and {Bryden}, Geoffrey and {Greene}, Thomas P. and {Leisenring}, Jarron and {Ygouf}, Marie and {Benneke}, Bj{\"o}rn and {Inglis}, Julie and {Wallack}, Nicole L.},
        title = "{The Compositions of the HR 8799 Planets Reflect Accretion of Both Solids and Metal-enriched Gas}",
      journal = {\apj},
         year = 2026,
        month = mar,
       volume = {1000},
       number = {1},
          eid = {27},
        pages = {27},
          doi = {10.3847/1538-4357/ae448f},
archivePrefix = {arXiv},
       eprint = {2602.09422},
 primaryClass = {astro-ph.EP},
       adsurl = {https://ui.adsabs.harvard.edu/abs/2026ApJ..1000...27X}
}

@article{ruffio2026,
  author  = {Ruffio, Jean-Baptiste and Xuan, Jerry W. and Chachan, Yayaati and Kesseli, Aurora and Lee, Eve J. and Beichman, Charles and Hodapp, Klaus and Balmer, William O. and Konopacky, Quinn and Perrin, Marshall D. and Mawet, Dimitri and Knutson, Heather A. and Bryden, Geoffrey and Greene, Thomas P. and Johnstone, Doug and Leisenring, Jarron and Meyer, Michael and Ygouf, Marie},
  title   = {Jupiter-like uniform metal enrichment in a system of multiple giant exoplanets},
  journal = {Nature Astronomy},
  year    = {2026},
  date    = {2026-02-09},
  doi     = {10.1038/s41550-026-02783-z},
  url     = {https://doi.org/10.1038/s41550-026-02783-z},
  issn    = {2397-3366},
}

@ARTICLE{kammerer2025,
       author = {{Kammerer}, J. and {Winterhalder}, T.~O. and {Lacour}, S. and {Stolker}, T. and {Marleau}, G.-D. and {Balmer}, W.~O. and {Moore}, A.~F. and {Piscarreta}, L. and {Toci}, C. and {M{\'e}rand}, A. and {Nowak}, M. and {Rickman}, E.~L. and {Pueyo}, L. and {Pourr{\'e}}, N. and {Nasedkin}, E. and {Wang}, J.~J. and {Bourdarot}, G. and {Eisenhauer}, F. and {Henning}, Th. and {Garcia Lopez}, R. and {van Dishoeck}, E.~F. and {Forveille}, T. and {Monnier}, J.~D. and {Abuter}, R. and {Amorim}, A. and {Benisty}, M. and {Berger}, J.-P. and {Beust}, H. and {Blunt}, S. and {Boccaletti}, A. and {Bonnefoy}, M. and {Bonnet}, H. and {Bordoni}, M.~S. and {Brandner}, W. and {Cantalloube}, F. and {Caselli}, P. and {Ceva}, W. and {Charnay}, B. and {Chauvin}, G. and {Chavez}, A. and {Chomez}, A. and {Choquet}, E. and {Christiaens}, V. and {Cl{\'e}net}, Y. and {Coud{\'e} du Foresto}, V. and {Cridland}, A. and {Davies}, R. and {Dembet}, R. and {Dexter}, J. and {Drescher}, A. and {Duvert}, G. and {Eckart}, A. and {Fontanive}, C. and {F{\"o}rster Schreiber}, N.~M. and {Garcia}, P. and {Gendron}, E. and {Genzel}, R. and {Gillessen}, S. and {Girard}, J.~H. and {Grant}, S. and {Hagelberg}, J. and {Haubois}, X. and {Hei{\ss}el}, G. and {Hinkley}, S. and {Hippler}, S. and {Houll{\'e}}, M. and {Hubert}, Z. and {Jocou}, L. and {Keppler}, M. and {Kervella}, P. and {Kreidberg}, L. and {Kurtovic}, N.~T. and {Lagrange}, A.-M. and {Lapeyr{\`e}re}, V. and {Le Bouquin}, J.-B. and {Lutz}, D. and {Maire}, A.-L. and {Mang}, F. and {Matthews}, E.~C. and {Molli{\`e}re}, P. and {Mordasini}, C. and {Mouillet}, D. and {Ott}, T. and {Otten}, G.~P.~P.~L. and {Paladini}, C. and {Paumard}, T. and {Perraut}, K. and {Perrin}, G. and {Pfuhl}, O. and {Ribeiro}, D.~C. and {Rustamkulov}, Z. and {S{\'e}gransan}, D. and {Shangguan}, J. and {Shimizu}, T. and {Samland}, M. and {Sing}, D. and {Stadler}, J. and {Straub}, O. and {Straubmeier}, C. and {Sturm}, E. and {Tacconi}, L.~J. and {Udry}, S. and {Vigan}, A. and {Vincent}, F. and {von Fellenberg}, S.~D. and {Widmann}, F. and {Woillez}, J. and {Yazici}, S.},
        title = "{The ExoGRAVITY survey: A K-band spectral library of giant exoplanet and brown dwarf companions}",
      journal = {\aap},
         year = 2025,
        month = dec,
       volume = {704},
          eid = {A318},
        pages = {A318},
          doi = {10.1051/0004-6361/202556860},
archivePrefix = {arXiv},
       eprint = {2510.08691},
 primaryClass = {astro-ph.EP},
       adsurl = {https://ui.adsabs.harvard.edu/abs/2025A&A...704A.318K}
}

@ARTICLE{polyansky2018,
       author = {{Polyansky}, Oleg L. and {Kyuberis}, Aleksandra A. and {Zobov}, Nikolai F. and {Tennyson}, Jonathan and {Yurchenko}, Sergei N. and {Lodi}, Lorenzo},
        title = "{ExoMol molecular line lists XXX: a complete high-accuracy line list for water}",
      journal = {\mnras},
         year = 2018,
        month = oct,
       volume = {480},
       number = {2},
        pages = {2597-2608},
          doi = {10.1093/mnras/sty1877},
archivePrefix = {arXiv},
       eprint = {1807.04529},
 primaryClass = {astro-ph.EP},
       adsurl = {https://ui.adsabs.harvard.edu/abs/2018MNRAS.480.2597P}
}

@ARTICLE{hauschildt1999,
       author = {{Hauschildt}, Peter H. and {Allard}, France and {Baron}, E.},
        title = "{The NextGen Model Atmosphere Grid for 3000<=T$_{eff}$<=10,000 K}",
      journal = {\apj},
         year = 1999,
        month = feb,
       volume = {512},
       number = {1},
        pages = {377-385},
          doi = {10.1086/306745},
archivePrefix = {arXiv},
       eprint = {astro-ph/9807286},
 primaryClass = {astro-ph},
       adsurl = {https://ui.adsabs.harvard.edu/abs/1999ApJ...512..377H}
}

@ARTICLE{petrus2024,
       author = {{Petrus}, Simon and {Whiteford}, Niall and {Patapis}, Polychronis and {Biller}, Beth A. and {Skemer}, Andrew and {Hinkley}, Sasha and {Su{\'a}rez}, Genaro and {Palma-Bifani}, Paulina and {Morley}, Caroline V. and {Tremblin}, Pascal and {Charnay}, Benjamin and {Vos}, Johanna M. and {Wang}, Jason J. and {Stone}, Jordan M. and {Bonnefoy}, Micka{\"e}l and {Chauvin}, Ga{\"e}l and {Miles}, Brittany E. and {Carter}, Aarynn L. and {Lueber}, Anna and {Helling}, Christiane and {Sutlieff}, Ben J. and {Janson}, Markus and {Gonzales}, Eileen C. and {Hoch}, Kielan K.~W. and {Absil}, Olivier and {Balmer}, William O. and {Boccaletti}, Anthony and {Bonavita}, Mariangela and {Booth}, Mark and {Bowler}, Brendan P. and {Briesemeister}, Zackery W. and {Bryan}, Marta L. and {Calissendorff}, Per and {Cantalloube}, Faustine and {Chen}, Christine H. and {Choquet}, Elodie and {Christiaens}, Valentin and {Cugno}, Gabriele and {Currie}, Thayne and {Danielski}, Camilla and {De Furio}, Matthew and {Dupuy}, Trent J. and {Factor}, Samuel M. and {Faherty}, Jacqueline K. and {Fitzgerald}, Michael P. and {Fortney}, Jonathan J. and {Franson}, Kyle and {Girard}, Julien H. and {Grady}, Carol A. and {Henning}, Thomas and {Hines}, Dean C. and {Hood}, Callie E. and {Howe}, Alex R. and {Kalas}, Paul and {Kammerer}, Jens and {Kennedy}, Grant M. and {Kenworthy}, Matthew A. and {Kervella}, Pierre and {Kim}, Minjae and {Kitzmann}, Daniel and {Kraus}, Adam L. and {Kuzuhara}, Masayuki and {Lagage}, Pierre-Olivier and {Lagrange}, Anne-Marie and {Lawson}, Kellen and {Lazzoni}, Cecilia and {Leisenring}, Jarron M. and {Lew}, Ben W.~P. and {Liu}, Michael C. and {Liu}, Pengyu and {Llop-Sayson}, Jorge and {Lloyd}, James P. and {Macintosh}, Bruce and {M{\^a}lin}, Mathilde and {Manjavacas}, Elena and {Marino}, Sebasti{\'a}n and {Marley}, Mark S. and {Marois}, Christian and {Martinez}, Raquel A. and {Matthews}, Elisabeth C. and {Matthews}, Brenda C. and {Mawet}, Dimitri and {Mazoyer}, Johan and {McElwain}, Michael W. and {Metchev}, Stanimir and {Meyer}, Michael R. and {Millar-Blanchaer}, Maxwell A. and {Molli{\`e}re}, Paul and {Moran}, Sarah E. and {Mukherjee}, Sagnick and {Pantin}, Eric and {Perrin}, Marshall D. and {Pueyo}, Laurent and {Quanz}, Sascha P. and {Quirrenbach}, Andreas and {Ray}, Shrishmoy and {Rebollido}, Isabel and {Adams Redai}, Jea and {Ren}, Bin B. and {Rickman}, Emily and {Sallum}, Steph and {Samland}, Matthias and {Sargent}, Benjamin and {Schlieder}, Joshua E. and {Stapelfeldt}, Karl R. and {Tamura}, Motohide and {Tan}, Xianyu and {Theissen}, Christopher A. and {Uyama}, Taichi and {Vasist}, Malavika and {Vigan}, Arthur and {Wagner}, Kevin and {Ward-Duong}, Kimberly and {Wolff}, Schuyler G. and {Worthen}, Kadin and {Wyatt}, Mark C. and {Ygouf}, Marie and {Zurlo}, Alice and {Zhang}, Xi and {Zhang}, Keming and {Zhang}, Zhoujian and {Zhou}, Yifan},
        title = "{The JWST Early Release Science Program for Direct Observations of Exoplanetary Systems. V. Do Self-consistent Atmospheric Models Represent JWST Spectra? A Showcase with VHS 1256{\textendash}1257 b}",
      journal = {\apjl},
         year = 2024,
        month = may,
       volume = {966},
       number = {1},
          eid = {L11},
        pages = {L11},
          doi = {10.3847/2041-8213/ad3e7c},
archivePrefix = {arXiv},
       eprint = {2312.03852},
 primaryClass = {astro-ph.EP},
       adsurl = {https://ui.adsabs.harvard.edu/abs/2024ApJ...966L..11P}
}

@ARTICLE{balmer2025,
       author = {{Balmer}, William O. and {Kammerer}, Jens and {Pueyo}, Laurent and {Perrin}, Marshall D. and {Girard}, Julien H. and {Leisenring}, Jarron M. and {Lawson}, Kellen and {Dennen}, Henry and {van der Marel}, Roeland P. and {Beichman}, Charles A. and {Bryden}, Geoffrey and {Llop-Sayson}, Jorge and {Valenti}, Jeff A. and {Lothringer}, Joshua D. and {Lewis}, Nikole K. and {M{\^a}lin}, Mathilde and {Rebollido}, Isabel and {Rickman}, Emily and {Hoch}, Kielan K.~W. and {Soummer}, R{\'e}mi and {Clampin}, Mark and {Mountain}, C. Matt},
        title = "{JWST-TST High Contrast: Living on the Wedge, or, NIRCam Bar Coronagraphy Reveals CO$_{2}$ in the HR 8799 and 51 Eri Exoplanets' Atmospheres}",
      journal = {\aj},
         year = 2025,
        month = apr,
       volume = {169},
       number = {4},
          eid = {209},
        pages = {209},
          doi = {10.3847/1538-3881/adb1c6},
archivePrefix = {arXiv},
       eprint = {2503.13608},
 primaryClass = {astro-ph.EP},
       adsurl = {https://ui.adsabs.harvard.edu/abs/2025AJ....169..209B}
}

@INPROCEEDINGS{lodders2010,
       author = {{Lodders}, Katharina},
        title = "{Solar System Abundances of the Elements}",
    booktitle = {Principles and Perspectives in Cosmochemistry},
         year = 2010,
       editor = {{Goswami}, Aruna and {Reddy}, B. Eswar},
       series = {Astrophysics and Space Science Proceedings},
       volume = {16},
        month = jan,
        pages = {379},
          doi = {10.1007/978-3-642-10352-0_8},
archivePrefix = {arXiv},
       eprint = {1010.2746},
 primaryClass = {astro-ph.SR},
       adsurl = {https://ui.adsabs.harvard.edu/abs/2010ASSP...16..379L}
}

@ARTICLE{hauschildt1997,
       author = {{Hauschildt}, Peter H. and {Baron}, E. and {Allard}, France},
        title = "{Parallel Implementation of the PHOENIX Generalized Stellar Atmosphere Program}",
      journal = {\apj},
         year = 1997,
        month = jul,
       volume = {483},
       number = {1},
        pages = {390-398},
          doi = {10.1086/304233},
archivePrefix = {arXiv},
       eprint = {astro-ph/9607087},
 primaryClass = {astro-ph},
       adsurl = {https://ui.adsabs.harvard.edu/abs/1997ApJ...483..390H}
}

@ARTICLE{morley2024,
       author = {{Morley}, Caroline V. and {Mukherjee}, Sagnick and {Marley}, Mark S. and {Fortney}, Jonathan J. and {Visscher}, Channon and {Lupu}, Roxana and {Gharib-Nezhad}, Ehsan and {Thorngren}, Daniel and {Freedman}, Richard and {Batalha}, Natasha},
        title = "{The Sonora Substellar Atmosphere Models. III. Diamondback: Atmospheric Properties, Spectra, and Evolution for Warm Cloudy Substellar Objects}",
      journal = {\apj},
         year = 2024,
        month = nov,
       volume = {975},
       number = {1},
          eid = {59},
        pages = {59},
          doi = {10.3847/1538-4357/ad71d5},
archivePrefix = {arXiv},
       eprint = {2402.00758},
 primaryClass = {astro-ph.SR},
       adsurl = {https://ui.adsabs.harvard.edu/abs/2024ApJ...975...59M}
}

@ARTICLE{ackerman2001,
       author = {{Ackerman}, Andrew S. and {Marley}, Mark S.},
        title = "{Precipitating Condensation Clouds in Substellar Atmospheres}",
      journal = {\apj},
         year = 2001,
        month = aug,
       volume = {556},
       number = {2},
        pages = {872-884},
          doi = {10.1086/321540},
archivePrefix = {arXiv},
       eprint = {astro-ph/0103423},
 primaryClass = {astro-ph},
       adsurl = {https://ui.adsabs.harvard.edu/abs/2001ApJ...556..872A}
}

@ARTICLE{marley1999,
       author = {{Marley}, Mark S. and {McKay}, Christopher P.},
        title = "{Thermal Structure of Uranus' Atmosphere}",
      journal = {\icarus},
         year = 1999,
        month = apr,
       volume = {138},
       number = {2},
        pages = {268-286},
          doi = {10.1006/icar.1998.6071},
       adsurl = {https://ui.adsabs.harvard.edu/abs/1999Icar..138..268M}
}

@ARTICLE{davis2025,
       author = {{Davis}, C. Evan and {Fortney}, Jonathan J. and {Iyer}, Aishwarya and {Mukherjee}, Sagnick and {Morley}, Caroline V. and {Marley}, Mark S. and {Line}, Michael and {Muirhead}, Philip S.},
        title = "{The Sonora Substellar Atmosphere Models VI. Red Diamondback: Extending Diamondback with SPHINX for Brown Dwarf Early Evolution}",
      journal = {arXiv e-prints},
         year = 2025,
        month = oct,
          eid = {arXiv:2510.08694},
        pages = {arXiv:2510.08694},
          doi = {10.48550/arXiv.2510.08694},
archivePrefix = {arXiv},
       eprint = {2510.08694},
 primaryClass = {astro-ph.SR},
       adsurl = {https://ui.adsabs.harvard.edu/abs/2025arXiv251008694D}
}

@ARTICLE{karadili2021,
       author = {{Karalidi}, Theodora and {Marley}, Mark and {Fortney}, Jonathan J. and {Morley}, Caroline and {Saumon}, Didier and {Lupu}, Roxana and {Visscher}, Channon and {Freedman}, Richard},
        title = "{The Sonora Substellar Atmosphere Models. II. Cholla: A Grid of Cloud-free, Solar Metallicity Models in Chemical Disequilibrium for the JWST Era}",
      journal = {\apj},
         year = 2021,
        month = dec,
       volume = {923},
       number = {2},
          eid = {269},
        pages = {269},
          doi = {10.3847/1538-4357/ac3140},
archivePrefix = {arXiv},
       eprint = {2110.11824},
 primaryClass = {astro-ph.EP},
       adsurl = {https://ui.adsabs.harvard.edu/abs/2021ApJ...923..269K}
}

@ARTICLE{marleau2014,
       author = {{Marleau}, G.-D. and {Cumming}, A.},
        title = "{Constraining the initial entropy of directly detected exoplanets}",
      journal = {\mnras},
         year = 2014,
        month = jan,
       volume = {437},
       number = {2},
        pages = {1378-1399},
          doi = {10.1093/mnras/stt1967},
archivePrefix = {arXiv},
       eprint = {1302.1517},
 primaryClass = {astro-ph.EP},
       adsurl = {https://ui.adsabs.harvard.edu/abs/2014MNRAS.437.1378M}
}

@ARTICLE{marley2007,
       author = {{Marley}, Mark S. and {Fortney}, Jonathan J. and {Hubickyj}, Olenka and {Bodenheimer}, Peter and {Lissauer}, Jack J.},
        title = "{On the Luminosity of Young Jupiters}",
      journal = {\apj},
         year = 2007,
        month = jan,
       volume = {655},
       number = {1},
        pages = {541-549},
          doi = {10.1086/509759},
archivePrefix = {arXiv},
       eprint = {astro-ph/0609739},
 primaryClass = {astro-ph},
       adsurl = {https://ui.adsabs.harvard.edu/abs/2007ApJ...655..541M}
}

@ARTICLE{bildsten1997,
       author = {{Bildsten}, Lars and {Brown}, Edward F. and {Matzner}, Christopher D. and {Ushomirsky}, Greg},
        title = "{Lithium Depletion in Fully Convective Pre-Main-Sequence Stars}",
      journal = {\apj},
         year = 1997,
        month = jun,
       volume = {482},
       number = {1},
        pages = {442-447},
          doi = {10.1086/304151},
archivePrefix = {arXiv},
       eprint = {astro-ph/9612155},
 primaryClass = {astro-ph},
       adsurl = {https://ui.adsabs.harvard.edu/abs/1997ApJ...482..442B}
}

@ARTICLE{madurowicz2025,
       author = {{Madurowicz}, Alexander and {Ruffio}, Jean-Baptiste and {Macintosh}, Bruce and {Perrin}, Marshall and {Konopacky}, Quinn M. and {Baburaj}, Aneesh and {Hoch}, Kielan},
        title = "{Direct Spectroscopy of 51 Eridani b with JWST NIRSpec}",
      journal = {\aj},
         year = 2025,
        month = dec,
       volume = {170},
       number = {6},
          eid = {326},
        pages = {326},
          doi = {10.3847/1538-3881/ae1028},
archivePrefix = {arXiv},
       eprint = {2510.08327},
 primaryClass = {astro-ph.EP},
       adsurl = {https://ui.adsabs.harvard.edu/abs/2025AJ....170..326M}
}

@ARTICLE{morley2012,
       author = {{Morley}, Caroline V. and {Fortney}, Jonathan J. and {Marley}, Mark S. and {Visscher}, Channon and {Saumon}, Didier and {Leggett}, S.~K.},
        title = "{Neglected Clouds in T and Y Dwarf Atmospheres}",
      journal = {\apj},
         year = 2012,
        month = sep,
       volume = {756},
       number = {2},
          eid = {172},
        pages = {172},
          doi = {10.1088/0004-637X/756/2/172},
archivePrefix = {arXiv},
       eprint = {1206.4313},
 primaryClass = {astro-ph.SR},
       adsurl = {https://ui.adsabs.harvard.edu/abs/2012ApJ...756..172M}
}

@ARTICLE{Mukherjee2024,
       author = {{Mukherjee}, Sagnick and {Fortney}, Jonathan J. and {Morley}, Caroline V. and {Batalha}, Natasha E. and {Marley}, Mark S. and {Karalidi}, Theodora and {Visscher}, Channon and {Lupu}, Roxana and {Freedman}, Richard and {Gharib-Nezhad}, Ehsan},
        title = "{The Sonora Substellar Atmosphere Models. IV. Elf Owl: Atmospheric Mixing and Chemical Disequilibrium with Varying Metallicity and C/O Ratios}",
      journal = {\apj},
         year = 2024,
        month = mar,
       volume = {963},
       number = {1},
          eid = {73},
        pages = {73},
          doi = {10.3847/1538-4357/ad18c2},
       adsurl = {https://ui.adsabs.harvard.edu/abs/2024ApJ...963...73M}
}

@ARTICLE{wogan2025,
       author = {{Wogan}, Nicholas F. and {Mang}, James and {Batalha}, Natasha E. and {Zahnle}, Kevin and {Mukherjee}, Sagnick and {Visscher}, Channon and {Fortney}, Jonathan J. and {Marley}, Mark S. and {Morley}, Caroline V.},
        title = "{The Sonora Substellar Atmosphere Models. V. A Correction to the Disequilibrium Abundance of CO$_{2}$ for Sonora Elf Owl}",
      journal = {Research Notes of the American Astronomical Society},
         year = 2025,
        month = may,
       volume = {9},
       number = {5},
          eid = {108},
        pages = {108},
          doi = {10.3847/2515-5172/add407},
archivePrefix = {arXiv},
       eprint = {2505.03994},
 primaryClass = {astro-ph.EP},
       adsurl = {https://ui.adsabs.harvard.edu/abs/2025RNAAS...9..108W}
}

@ARTICLE{hoch2025,
       author = {{Hoch}, K.~K.~W. and {Rowland}, M. and {Petrus}, S. and {Nasedkin}, E. and {Ingebretsen}, C. and {Kammerer}, J. and {Perrin}, M. and {D'Orazi}, V. and {Balmer}, W.~O. and {Barman}, T. and {Bonnefoy}, M. and {Chauvin}, G. and {Chen}, C. and {De Rosa}, R.~J. and {Girard}, J. and {Gonzales}, E. and {Kenworthy}, M. and {Konopacky}, Q.~M. and {Macintosh}, B. and {Moran}, S.~E. and {Morley}, C.~V. and {Palma-Bifani}, P. and {Pueyo}, L. and {Ren}, B. and {Rickman}, E. and {Ruffio}, J. -B. and {Theissen}, C.~A. and {Ward-Duong}, K. and {Zhang}, Y.},
        title = "{Silicate clouds and a circumplanetary disk in the YSES-1 exoplanet system}",
      journal = {\nat},
         year = 2025,
        month = jul,
       volume = {643},
       number = {8073},
        pages = {938-942},
          doi = {10.1038/s41586-025-09174-w},
archivePrefix = {arXiv},
       eprint = {2507.18861},
 primaryClass = {astro-ph.EP},
       adsurl = {https://ui.adsabs.harvard.edu/abs/2025Natur.643..938H}
}

@ARTICLE{xuan2024,
       author = {{Xuan}, Jerry W. and {Perrin}, Marshall D. and {Mawet}, Dimitri and {Knutson}, Heather A. and {Mukherjee}, Sagnick and {Zhang}, Yapeng and {Hoch}, Kielan K.~W. and {Wang}, Jason J. and {Inglis}, Julie and {Wallack}, Nicole L. and {Ruffio}, Jean-Baptiste},
        title = "{Atmospheric Abundances and Bulk Properties of the Binary Brown Dwarf Gliese 229Bab from JWST/MIRI Spectroscopy}",
      journal = {\apjl},
         year = 2024,
        month = dec,
       volume = {977},
       number = {2},
          eid = {L32},
        pages = {L32},
          doi = {10.3847/2041-8213/ad92f9},
archivePrefix = {arXiv},
       eprint = {2411.10571},
 primaryClass = {astro-ph.SR},
       adsurl = {https://ui.adsabs.harvard.edu/abs/2024ApJ...977L..32X}
}

@ARTICLE{mukherjee2025,
       author = {{Mukherjee}, Sagnick and {Fortney}, Jonathan J. and {Wogan}, Nicholas F. and {Sing}, David K. and {Ohno}, Kazumasa},
        title = "{Effects of Planetary Parameters on Disequilibrium Chemistry in Irradiated Planetary Atmospheres: From Gas Giants to Sub-Neptunes}",
      journal = {\apj},
         year = 2025,
        month = jun,
       volume = {985},
       number = {2},
          eid = {209},
        pages = {209},
          doi = {10.3847/1538-4357/adc7b3},
archivePrefix = {arXiv},
       eprint = {2410.17169},
 primaryClass = {astro-ph.EP},
       adsurl = {https://ui.adsabs.harvard.edu/abs/2025ApJ...985..209M}
}

@ARTICLE{nasedkin2025,
       author = {{Nasedkin}, Evert and {Schrader}, Merle and {Vos}, Johanna M. and {Biller}, Beth and {Burningham}, Ben and {Cowan}, Nicolas B. and {Faherty}, Jacqueline and {Gonzales}, Eileen and {Lam}, Madeline B. and {McCarthy}, Allison M. and {Muirhead}, Philip S. and {O'Toole}, Cian and {Plummer}, Michael K. and {Su{\'a}rez}, Genaro and {Tan}, Xianyu and {Visscher}, Channon and {Whiteford}, Niall and {Zhou}, Yifan},
        title = "{The JWST Weather Report: retrieving temperature variations, auroral heating, and static cloud coverage on SIMP-0136}",
      journal = {arXiv e-prints},
         year = 2025,
        month = jul,
          eid = {arXiv:2507.07772},
        pages = {arXiv:2507.07772},
          doi = {10.48550/arXiv.2507.07772},
archivePrefix = {arXiv},
       eprint = {2507.07772},
 primaryClass = {astro-ph.EP},
       adsurl = {https://ui.adsabs.harvard.edu/abs/2025arXiv250707772N}
}

@ARTICLE{ZJ2025,
       author = {{Zhang}, Zhoujian and {Molli{\`e}re}, Paul and {Fortney}, Jonathan J. and {Marley}, Mark S.},
        title = "{ELemental Abundances of Planets and Brown Dwarfs Imaged around Stars (ELPIS). II. The Jupiter-like Inhomogeneous Atmosphere of the First Directly Imaged Planetary-mass Companion 2MASS 1207 b}",
      journal = {\aj},
         year = 2025,
        month = aug,
       volume = {170},
       number = {2},
          eid = {64},
        pages = {64},
          doi = {10.3847/1538-3881/addfcb},
archivePrefix = {arXiv},
       eprint = {2502.18559},
 primaryClass = {astro-ph.EP},
       adsurl = {https://ui.adsabs.harvard.edu/abs/2025AJ....170...64Z}
}

@ARTICLE{luhman2023,
       author = {{Luhman}, K.~L. and {Tremblin}, P. and {Birkmann}, S.~M. and {Manjavacas}, E. and {Valenti}, J. and {Alves de Oliveira}, C. and {Beck}, T.~L. and {Giardino}, G. and {L{\"u}tzgendorf}, N. and {Rauscher}, B.~J. and {Sirianni}, M.},
        title = "{JWST/NIRSpec Observations of the Planetary Mass Companion TWA 27B}",
      journal = {\apjl},
         year = 2023,
        month = jun,
       volume = {949},
       number = {2},
          eid = {L36},
        pages = {L36},
          doi = {10.3847/2041-8213/acd635},
archivePrefix = {arXiv},
       eprint = {2305.18603},
 primaryClass = {astro-ph.EP},
       adsurl = {https://ui.adsabs.harvard.edu/abs/2023ApJ...949L..36L}
}

@ARTICLE{lew2024,
       author = {{Lew}, Ben W.~P. and {Roellig}, Thomas and {Batalha}, Natasha E. and {Line}, Michael and {Greene}, Thomas and {Murkherjee}, Sagnick and {Freedman}, Richard and {Meyer}, Michael and {Beichman}, Charles and {Alves de Oliveira}, Catarina and {De Furio}, Matthew and {Johnstone}, Doug and {Greenbaum}, Alexandra Z. and {Marley}, Mark and {Fortney}, Jonathan J. and {Young}, Erick T. and {Leisenring}, Jarron and {Boyer}, Martha and {Hodapp}, Klaus and {Misselt}, Karl and {Stansberry}, John and {Rieke}, Marcia},
        title = "{High-precision Atmospheric Characterization of a Y Dwarf with JWST NIRSpec G395H Spectroscopy: Isotopologue, C/O Ratio, Metallicity, and the Abundances of Six Molecular Species}",
      journal = {\aj},
         year = 2024,
        month = may,
       volume = {167},
       number = {5},
          eid = {237},
        pages = {237},
          doi = {10.3847/1538-3881/ad3425},
archivePrefix = {arXiv},
       eprint = {2402.05900},
 primaryClass = {astro-ph.EP},
       adsurl = {https://ui.adsabs.harvard.edu/abs/2024AJ....167..237L}
}

@ARTICLE{Yurchenko2017,
       author = {{Yurchenko}, Sergei N. and {Amundsen}, David S. and {Tennyson}, Jonathan and {Waldmann}, Ingo P.},
        title = "{A hybrid line list for CH$_{4}$ and hot methane continuum}",
      journal = {\aap},
         year = 2017,
        month = sep,
       volume = {605},
          eid = {A95},
        pages = {A95},
          doi = {10.1051/0004-6361/201731026},
archivePrefix = {arXiv},
       eprint = {1706.05724},
 primaryClass = {astro-ph.EP},
       adsurl = {https://ui.adsabs.harvard.edu/abs/2017A&A...605A..95Y}
}

@ARTICLE{li2015,
       author = {{Li}, Gang and {Gordon}, Iouli E. and {Rothman}, Laurence S. and {Tan}, Yan and {Hu}, Shui-Ming and {Kassi}, Samir and {Campargue}, Alain and {Medvedev}, Emile S.},
        title = "{Rovibrational Line Lists for Nine Isotopologues of the CO Molecule in the X $^{1}${\ensuremath{\Sigma}}$^{+}$ Ground Electronic State}",
      journal = {\apjs},
         year = 2015,
        month = jan,
       volume = {216},
       number = {1},
          eid = {15},
        pages = {15},
          doi = {10.1088/0067-0049/216/1/15},
       adsurl = {https://ui.adsabs.harvard.edu/abs/2015ApJS..216...15L}
}

@ARTICLE{hitemp2010,
       author = {{Rothman}, L.~S. and {Gordon}, I.~E. and {Barber}, R.~J. and {Dothe}, H. and {Gamache}, R.~R. and {Goldman}, A. and {Perevalov}, V.~I. and {Tashkun}, S.~A. and {Tennyson}, J.},
        title = "{HITEMP, the high-temperature molecular spectroscopic database}",
      journal = {\jqsrt},
         year = 2010,
        month = oct,
       volume = {111},
        pages = {2139-2150},
          doi = {10.1016/j.jqsrt.2010.05.001},
       adsurl = {https://ui.adsabs.harvard.edu/abs/2010JQSRT.111.2139R}
}

@ARTICLE{miles2023,
       author = {{Miles}, Brittany E. and {Biller}, Beth A. and {Patapis}, Polychronis and {Worthen}, Kadin and {Rickman}, Emily and {Hoch}, Kielan K.~W. and {Skemer}, Andrew and {Perrin}, Marshall D. and {Whiteford}, Niall and {Chen}, Christine H. and {Sargent}, B. and {Mukherjee}, Sagnick and {Morley}, Caroline V. and {Moran}, Sarah E. and {Bonnefoy}, Mickael and {Petrus}, Simon and {Carter}, Aarynn L. and {Choquet}, Elodie and {Hinkley}, Sasha and {Ward-Duong}, Kimberly and {Leisenring}, Jarron M. and {Millar-Blanchaer}, Maxwell A. and {Pueyo}, Laurent and {Ray}, Shrishmoy and {Sallum}, Steph and {Stapelfeldt}, Karl R. and {Stone}, Jordan M. and {Wang}, Jason J. and {Absil}, Olivier and {Balmer}, William O. and {Boccaletti}, Anthony and {Bonavita}, Mariangela and {Booth}, Mark and {Bowler}, Brendan P. and {Chauvin}, Gael and {Christiaens}, Valentin and {Currie}, Thayne and {Danielski}, Camilla and {Fortney}, Jonathan J. and {Girard}, Julien H. and {Grady}, Carol A. and {Greenbaum}, Alexandra Z. and {Henning}, Thomas and {Hines}, Dean C. and {Janson}, Markus and {Kalas}, Paul and {Kammerer}, Jens and {Kennedy}, Grant M. and {Kenworthy}, Matthew A. and {Kervella}, Pierre and {Lagage}, Pierre-Olivier and {Lew}, Ben W.~P. and {Liu}, Michael C. and {Macintosh}, Bruce and {Marino}, Sebastian and {Marley}, Mark S. and {Marois}, Christian and {Matthews}, Elisabeth C. and {Matthews}, Brenda C. and {Mawet}, Dimitri and {McElwain}, Michael W. and {Metchev}, Stanimir and {Meyer}, Michael R. and {Molliere}, Paul and {Pantin}, Eric and {Quirrenbach}, Andreas and {Rebollido}, Isabel and {Ren}, Bin B. and {Schneider}, Glenn and {Vasist}, Malavika and {Wyatt}, Mark C. and {Zhou}, Yifan and {Briesemeister}, Zackery W. and {Bryan}, Marta L. and {Calissendorff}, Per and {Cantalloube}, Faustine and {Cugno}, Gabriele and {De Furio}, Matthew and {Dupuy}, Trent J. and {Factor}, Samuel M. and {Faherty}, Jacqueline K. and {Fitzgerald}, Michael P. and {Franson}, Kyle and {Gonzales}, Eileen C. and {Hood}, Callie E. and {Howe}, Alex R. and {Kraus}, Adam L. and {Kuzuhara}, Masayuki and {Lagrange}, Anne-Marie and {Lawson}, Kellen and {Lazzoni}, Cecilia and {Liu}, Pengyu and {Llop-Sayson}, Jorge and {Lloyd}, James P. and {Martinez}, Raquel A. and {Mazoyer}, Johan and {Quanz}, Sascha P. and {Redai}, Jea Adams and {Samland}, Matthias and {Schlieder}, Joshua E. and {Tamura}, Motohide and {Tan}, Xianyu and {Uyama}, Taichi and {Vigan}, Arthur and {Vos}, Johanna M. and {Wagner}, Kevin and {Wolff}, Schuyler G. and {Ygouf}, Marie and {Zhang}, Xi and {Zhang}, Keming and {Zhang}, Zhoujian},
        title = "{The JWST Early-release Science Program for Direct Observations of Exoplanetary Systems II: A 1 to 20 {\ensuremath{\mu}}m Spectrum of the Planetary-mass Companion VHS 1256-1257 b}",
      journal = {\apjl},
         year = 2023,
        month = mar,
       volume = {946},
       number = {1},
          eid = {L6},
        pages = {L6},
          doi = {10.3847/2041-8213/acb04a},
archivePrefix = {arXiv},
       eprint = {2209.00620},
 primaryClass = {astro-ph.EP},
       adsurl = {https://ui.adsabs.harvard.edu/abs/2023ApJ...946L...6M}
}

@ARTICLE{tan2022,
       author = {{Tan}, Yan and {Skinner}, Frances M. and {Samuels}, Shanelle and {Hargreaves}, Robert J. and {Hashemi}, Robab and {Gordon}, Iouli E.},
        title = "{H$_{2}$, He, and CO$_{2}$ Pressure-induced Parameters for the HITRAN Database. II. Line Lists of CO$_{2}$, N$_{2}$O, CO, SO$_{2}$, OH, OCS, H$_{2}$CO, HCN, PH$_{3}$, H$_{2}$S, and GeH$_{4}$}",
      journal = {\apjs},
         year = 2022,
        month = oct,
       volume = {262},
       number = {2},
          eid = {40},
        pages = {40},
          doi = {10.3847/1538-4365/ac83a6},
       adsurl = {https://ui.adsabs.harvard.edu/abs/2022ApJS..262...40T}
}

@ARTICLE{tennyson2020,
       author = {{Tennyson}, Jonathan and {Yurchenko}, Sergei N. and {Al-Refaie}, Ahmed F. and {Clark}, Victoria H.~J. and {Chubb}, Katy L. and {Conway}, Eamon K. and {Dewan}, Akhil and {Gorman}, Maire N. and {Hill}, Christian and {Lynas-Gray}, A.~E. and {Mellor}, Thomas and {McKemmish}, Laura K. and {Owens}, Alec and {Polyansky}, Oleg L. and {Semenov}, Mikhail and {Somogyi}, Wilfrid and {Tinetti}, Giovanna and {Upadhyay}, Apoorva and {Waldmann}, Ingo and {Wang}, Yixin and {Wright}, Samuel and {Yurchenko}, Olga P.},
        title = "{The 2020 release of the ExoMol database: Molecular line lists for exoplanet and other hot atmospheres}",
      journal = {\jqsrt},
         year = 2020,
        month = nov,
       volume = {255},
          eid = {107228},
        pages = {107228},
          doi = {10.1016/j.jqsrt.2020.107228},
archivePrefix = {arXiv},
       eprint = {2007.13022},
 primaryClass = {astro-ph.SR},
       adsurl = {https://ui.adsabs.harvard.edu/abs/2020JQSRT.25507228T}
}

@ARTICLE{asplund2009,
       author = {{Asplund}, Martin and {Grevesse}, Nicolas and {Sauval}, A. Jacques and {Scott}, Pat},
        title = "{The Chemical Composition of the Sun}",
      journal = {\araa},
         year = 2009,
        month = sep,
       volume = {47},
       number = {1},
        pages = {481-522},
          doi = {10.1146/annurev.astro.46.060407.145222},
archivePrefix = {arXiv},
       eprint = {0909.0948},
 primaryClass = {astro-ph.SR},
       adsurl = {https://ui.adsabs.harvard.edu/abs/2009ARA&A..47..481A}
}

@ARTICLE{Ruffio2023,
       author = {{Ruffio}, Jean-Baptiste and {Perrin}, Marshall D. and {Hoch}, Kielan K.~W. and {Kammerer}, Jens and {Konopacky}, Quinn M. and {Pueyo}, Laurent and {Madurowicz}, Alex and {Rickman}, Emily and {Theissen}, Christopher A. and {Agrawal}, Shubh and {Greenbaum}, Alexandra Z. and {Miles}, Brittany E. and {Barman}, Travis S. and {Balmer}, William O. and {Llop-Sayson}, Jorge and {Girard}, Julien H. and {Rebollido}, Isabel and {Soummer}, R{\'e}mi and {Allen}, Natalie H. and {Anderson}, Jay and {Beichman}, Charles A. and {Bellini}, Andrea and {Bryden}, Geoffrey and {Espinoza}, N{\'e}stor and {Glidden}, Ana and {Huang}, Jingcheng and {Lewis}, Nikole K. and {Libralato}, Mattia and {Louie}, Dana R. and {Sohn}, Sangmo Tony and {Seager}, Sara and {van der Marel}, Roeland P. and {Wakeford}, Hannah R. and {Watkins}, Laura L. and {Ygouf}, Marie and {Mountain}, C. Matt},
        title = "{JWST-TST High Contrast: Achieving Direct Spectroscopy of Faint Substellar Companions Next to Bright Stars with the NIRSpec Integral Field Unit}",
      journal = {\aj},
         year = 2024,
        month = aug,
       volume = {168},
       number = {2},
          eid = {73},
        pages = {73},
          doi = {10.3847/1538-3881/ad5281},
archivePrefix = {arXiv},
       eprint = {2310.09902},
 primaryClass = {astro-ph.EP},
       adsurl = {https://ui.adsabs.harvard.edu/abs/2024AJ....168...73R}
}

@ARTICLE{marley2021,
       author = {{Marley}, Mark S. and {Saumon}, Didier and {Visscher}, Channon and {Lupu}, Roxana and {Freedman}, Richard and {Morley}, Caroline and {Fortney}, Jonathan J. and {Seay}, Christopher and {Smith}, Adam J.~R.~W. and {Teal}, D.~J. and {Wang}, Ruoyan},
        title = "{The Sonora Brown Dwarf Atmosphere and Evolution Models. I. Model Description and Application to Cloudless Atmospheres in Rainout Chemical Equilibrium}",
      journal = {\apj},
         year = 2021,
        month = oct,
       volume = {920},
       number = {2},
          eid = {85},
        pages = {85},
          doi = {10.3847/1538-4357/ac141d},
archivePrefix = {arXiv},
       eprint = {2107.07434},
 primaryClass = {astro-ph.SR},
       adsurl = {https://ui.adsabs.harvard.edu/abs/2021ApJ...920...85M}
}

@ARTICLE{brandt2021,
       author = {{Brandt}, G. Mirek and {Dupuy}, Trent J. and {Li}, Yiting and {Chen}, Minghan and {Brandt}, Timothy D. and {Wong}, Tin Long Sunny and {Currie}, Thayne and {Bowler}, Brendan P. and {Liu}, Michael C. and {Best}, William M.~J. and {Phillips}, Mark W.},
        title = "{Improved Dynamical Masses for Six Brown Dwarf Companions Using Hipparcos and Gaia EDR3}",
      journal = {\aj},
         year = 2021,
        month = dec,
       volume = {162},
       number = {6},
          eid = {301},
        pages = {301},
          doi = {10.3847/1538-3881/ac273e},
archivePrefix = {arXiv},
       eprint = {2109.07525},
 primaryClass = {astro-ph.SR},
       adsurl = {https://ui.adsabs.harvard.edu/abs/2021AJ....162..301B}
}

@ARTICLE{allard2012,
       author = {{Allard}, F. and {Homeier}, D. and {Freytag}, B.},
        title = "{Models of very-low-mass stars, brown dwarfs and exoplanets}",
      journal = {Philosophical Transactions of the Royal Society of London Series A},
         year = 2012,
        month = jun,
       volume = {370},
       number = {1968},
        pages = {2765-2777},
          doi = {10.1098/rsta.2011.0269},
archivePrefix = {arXiv},
       eprint = {1112.3591},
 primaryClass = {astro-ph.SR},
       adsurl = {https://ui.adsabs.harvard.edu/abs/2012RSPTA.370.2765A}
}

@ARTICLE{yurchenko2014,
       author = {{Yurchenko}, Sergei N. and {Tennyson}, Jonathan},
        title = "{ExoMol line lists - IV. The rotation-vibration spectrum of methane up to 1500 K}",
      journal = {\mnras},
         year = 2014,
        month = may,
       volume = {440},
       number = {2},
        pages = {1649-1661},
          doi = {10.1093/mnras/stu326},
archivePrefix = {arXiv},
       eprint = {1401.4852},
 primaryClass = {astro-ph.EP},
       adsurl = {https://ui.adsabs.harvard.edu/abs/2014MNRAS.440.1649Y}
}

@MISC{foreman-mackey2013,
       author = {{Foreman-Mackey}, Daniel and {Conley}, Alex and {Meierjurgen Farr}, Will and {Hogg}, David W. and {Lang}, Dustin and {Marshall}, Phil and {Price-Whelan}, Adrian and {Sanders}, Jeremy and {Zuntz}, Joe},
        title = "{emcee: The MCMC Hammer}",
         year = 2013,
        month = mar,
          eid = {ascl:1303.002},
        pages = {ascl:1303.002},
archivePrefix = {ascl},
       eprint = {1303.002},
       adsurl = {https://ui.adsabs.harvard.edu/abs/2013ascl.soft03002F}
}

@ARTICLE{goodman2010,
       author = {{Goodman}, Jonathan and {Weare}, Jonathan},
        title = "{Ensemble samplers with affine invariance}",
      journal = {Communications in Applied Mathematics and Computational Science},
         year = 2010,
        month = jan,
       volume = {5},
       number = {1},
        pages = {65-80},
          doi = {10.2140/camcos.2010.5.65},
       adsurl = {https://ui.adsabs.harvard.edu/abs/2010CAMCS...5...65G}
}

@ARTICLE{theissen2022,
       author = {{Theissen}, Christopher A. and {Konopacky}, Quinn M. and {Lu}, Jessica R. and {Kim}, Dongwon and {Zhang}, Stella Y. and {Hsu}, Chih-Chun and {Chu}, Laurie and {Wei}, Lingfeng},
        title = "{The 3D Kinematics of the Orion Nebula Cluster: NIRSPEC-AO Radial Velocities of the Core Population}",
      journal = {\apj},
         year = 2022,
        month = feb,
       volume = {926},
       number = {2},
          eid = {141},
        pages = {141},
          doi = {10.3847/1538-4357/ac3252},
archivePrefix = {arXiv},
       eprint = {2105.05871},
 primaryClass = {astro-ph.GA},
       adsurl = {https://ui.adsabs.harvard.edu/abs/2022ApJ...926..141T}
}

@ARTICLE{hsu2021,
       author = {{Hsu}, Chih-Chun and {Burgasser}, Adam J. and {Theissen}, Christopher A. and {Gelino}, Christopher R. and {Birky}, Jessica L. and {Diamant}, Sharon J.~M. and {Bardalez Gagliuffi}, Daniella C. and {Aganze}, Christian and {Blake}, Cullen H. and {Faherty}, Jacqueline K.},
        title = "{The Brown Dwarf Kinematics Project (BDKP). V. Radial and Rotational Velocities of T Dwarfs from Keck/NIRSPEC High-resolution Spectroscopy}",
      journal = {\apjs},
         year = 2021,
        month = dec,
       volume = {257},
       number = {2},
          eid = {45},
        pages = {45},
          doi = {10.3847/1538-4365/ac1c7d},
archivePrefix = {arXiv},
       eprint = {2107.01222},
 primaryClass = {astro-ph.SR},
       adsurl = {https://ui.adsabs.harvard.edu/abs/2021ApJS..257...45H}
}

@ARTICLE{burgasser2016,
       author = {{Burgasser}, Adam J. and {Lopez}, Mike A. and {Mamajek}, Eric E. and {Gagn{\'e}}, Jonathan and {Faherty}, Jacqueline K. and {Tallis}, Melisa and {Choban}, Caleb and {Tamiya}, Tomoki and {Escala}, Ivanna and {Aganze}, Christian},
        title = "{The First Brown Dwarf/Planetary-mass Object in the 32 Orionis Group}",
      journal = {\apj},
         year = 2016,
        month = mar,
       volume = {820},
       number = {1},
          eid = {32},
        pages = {32},
          doi = {10.3847/0004-637X/820/1/32},
archivePrefix = {arXiv},
       eprint = {1602.03022},
 primaryClass = {astro-ph.SR},
       adsurl = {https://ui.adsabs.harvard.edu/abs/2016ApJ...820...32B}
}

@ARTICLE{blake2010,
       author = {{Blake}, Cullen H. and {Charbonneau}, David and {White}, Russel J.},
        title = "{The NIRSPEC Ultracool Dwarf Radial Velocity Survey}",
      journal = {\apj},
         year = 2010,
        month = nov,
       volume = {723},
       number = {1},
        pages = {684-706},
          doi = {10.1088/0004-637X/723/1/684},
archivePrefix = {arXiv},
       eprint = {1008.3874},
 primaryClass = {astro-ph.SR},
       adsurl = {https://ui.adsabs.harvard.edu/abs/2010ApJ...723..684B}
}

@ARTICLE{allard2001,
       author = {{Allard}, France and {Hauschildt}, Peter H. and {Alexander}, David R. and {Tamanai}, Akemi and {Schweitzer}, Andreas},
        title = "{The Limiting Effects of Dust in Brown Dwarf Model Atmospheres}",
      journal = {\apj},
         year = 2001,
        month = jul,
       volume = {556},
       number = {1},
        pages = {357-372},
          doi = {10.1086/321547},
archivePrefix = {arXiv},
       eprint = {astro-ph/0104256},
 primaryClass = {astro-ph},
       adsurl = {https://ui.adsabs.harvard.edu/abs/2001ApJ...556..357A}
}

@ARTICLE{hargreaves2020,
       author = {{Hargreaves}, Robert J. and {Gordon}, Iouli E. and {Rey}, Michael and {Nikitin}, Andrei V. and {Tyuterev}, Vladimir G. and {Kochanov}, Roman V. and {Rothman}, Laurence S.},
        title = "{An Accurate, Extensive, and Practical Line List of Methane for the HITEMP Database}",
      journal = {\apjs},
         year = 2020,
        month = apr,
       volume = {247},
       number = {2},
          eid = {55},
        pages = {55},
          doi = {10.3847/1538-4365/ab7a1a},
archivePrefix = {arXiv},
       eprint = {2001.05037},
 primaryClass = {astro-ph.EP},
       adsurl = {https://ui.adsabs.harvard.edu/abs/2020ApJS..247...55H}
}

@ARTICLE{barman2011,
       author = {{Barman}, Travis S. and {Macintosh}, Bruce and {Konopacky}, Quinn M. and {Marois}, Christian},
        title = "{Clouds and Chemistry in the Atmosphere of Extrasolar Planet HR8799b}",
      journal = {\apj},
         year = 2011,
        month = may,
       volume = {733},
       number = {1},
          eid = {65},
        pages = {65},
          doi = {10.1088/0004-637X/733/1/65},
archivePrefix = {arXiv},
       eprint = {1103.3895},
 primaryClass = {astro-ph.EP},
       adsurl = {https://ui.adsabs.harvard.edu/abs/2011ApJ...733...65B}
}

@ARTICLE{kammerer2021,
       author = {{Kammerer}, J. and {Lacour}, S. and {Stolker}, T. and {Molli{\`e}re}, P. and {Sing}, D.~K. and {Nasedkin}, E. and {Kervella}, P. and {Wang}, J.~J. and {Ward-Duong}, K. and {Nowak}, M. and {Abuter}, R. and {Amorim}, A. and {Asensio-Torres}, R. and {Baub{\"o}ck}, M. and {Benisty}, M. and {Berger}, J. -P. and {Beust}, H. and {Blunt}, S. and {Boccaletti}, A. and {Bohn}, A. and {Bolzer}, M. -L. and {Bonnefoy}, M. and {Bonnet}, H. and {Brandner}, W. and {Cantalloube}, F. and {Caselli}, P. and {Charnay}, B. and {Chauvin}, G. and {Choquet}, E. and {Christiaens}, V. and {Cl{\'e}net}, Y. and {Coud{\'e} du Foresto}, V. and {Cridland}, A. and {Dembet}, R. and {Dexter}, J. and {de Zeeuw}, P.~T. and {Drescher}, A. and {Duvert}, G. and {Eckart}, A. and {Eisenhauer}, F. and {Gao}, F. and {Garcia}, P. and {Garcia Lopez}, R. and {Gendron}, E. and {Genzel}, R. and {Gillessen}, S. and {Girard}, J. and {Haubois}, X. and {Hei{\ss}el}, G. and {Henning}, T. and {Hinkley}, S. and {Hippler}, S. and {Horrobin}, M. and {Houll{\'e}}, M. and {Hubert}, Z. and {Jocou}, L. and {Keppler}, M. and {Kreidberg}, L. and {Lagrange}, A. -M. and {Lapeyr{\`e}re}, V. and {Le Bouquin}, J. -B. and {L{\'e}na}, P. and {Lutz}, D. and {Maire}, A. -L. and {M{\'e}rand}, A. and {Monnier}, J.~D. and {Mouillet}, D. and {M{\"u}ller}, A. and {Ott}, T. and {Otten}, G.~P.~P.~L. and {Paladini}, C. and {Paumard}, T. and {Perraut}, K. and {Perrin}, G. and {Pfuhl}, O. and {Pueyo}, L. and {Rameau}, J. and {Rodet}, L. and {Rousset}, G. and {Rustamkulov}, Z. and {Shangguan}, J. and {Shimizu}, T. and {Stadler}, J. and {Straub}, O. and {Straubmeier}, C. and {Sturm}, E. and {Tacconi}, L.~J. and {van Dishoeck}, E.~F. and {Vigan}, A. and {Vincent}, F. and {von Fellenberg}, S.~D. and {Widmann}, F. and {Wieprecht}, E. and {Wiezorrek}, E. and {Woillez}, J. and {Yazici}, S.},
        title = "{GRAVITY K-band spectroscopy of HD 206893 B. Brown dwarf or exoplanet}",
      journal = {\aap},
         year = 2021,
        month = aug,
       volume = {652},
          eid = {A57},
        pages = {A57},
          doi = {10.1051/0004-6361/202140749},
archivePrefix = {arXiv},
       eprint = {2106.08249},
 primaryClass = {astro-ph.EP},
       adsurl = {https://ui.adsabs.harvard.edu/abs/2021A&A...652A..57K}
}

@ARTICLE{burningham2017,
       author = {{Burningham}, Ben and {Marley}, M.~S. and {Line}, M.~R. and {Lupu}, R. and {Visscher}, C. and {Morley}, C.~V. and {Saumon}, D. and {Freedman}, R.},
        title = "{Retrieval of atmospheric properties of cloudy L dwarfs}",
      journal = {\mnras},
         year = 2017,
        month = sep,
       volume = {470},
       number = {1},
        pages = {1177-1197},
          doi = {10.1093/mnras/stx1246},
archivePrefix = {arXiv},
       eprint = {1701.01257},
 primaryClass = {astro-ph.SR},
       adsurl = {https://ui.adsabs.harvard.edu/abs/2017MNRAS.470.1177B}
}

@ARTICLE{zhang2021,
       author = {{Zhang}, Yapeng and {Snellen}, Ignas A.~G. and {Bohn}, Alexander J. and {Molli{\`e}re}, Paul and {Ginski}, Christian and {Hoeijmakers}, H. Jens and {Kenworthy}, Matthew A. and {Mamajek}, Eric E. and {Meshkat}, Tiffany and {Reggiani}, Maddalena and {Snik}, Frans},
        title = "{The $^{13}$CO-rich atmosphere of a young accreting super-Jupiter}",
      journal = {\nat},
         year = 2021,
        month = jul,
       volume = {595},
       number = {7867},
        pages = {370-372},
          doi = {10.1038/s41586-021-03616-x},
archivePrefix = {arXiv},
       eprint = {2107.06297},
 primaryClass = {astro-ph.EP},
       adsurl = {https://ui.adsabs.harvard.edu/abs/2021Natur.595..370Z}
}

@ARTICLE{oberg2011,
       author = {{{\"O}berg}, Karin I. and {Murray-Clay}, Ruth and {Bergin}, Edwin A.},
        title = "{The Effects of Snowlines on C/O in Planetary Atmospheres}",
      journal = {\apjl},
         year = 2011,
        month = dec,
       volume = {743},
       number = {1},
          eid = {L16},
        pages = {L16},
          doi = {10.1088/2041-8205/743/1/L16},
archivePrefix = {arXiv},
       eprint = {1110.5567},
 primaryClass = {astro-ph.GA},
       adsurl = {https://ui.adsabs.harvard.edu/abs/2011ApJ...743L..16O}
}

@ARTICLE{wilcomb2020,
       author = {{Wilcomb}, Kielan K. and {Konopacky}, Quinn M. and {Barman}, Travis S. and {Theissen}, Christopher A. and {Ruffio}, Jean-Baptiste and {Brock}, Laci and {Macintosh}, Bruce and {Marois}, Christian},
        title = "{Moderate-resolution K-band Spectroscopy of Substellar Companion {\ensuremath{\kappa}} Andromedae b}",
      journal = {\aj},
         year = 2020,
        month = nov,
       volume = {160},
       number = {5},
          eid = {207},
        pages = {207},
          doi = {10.3847/1538-3881/abb9b1},
archivePrefix = {arXiv},
       eprint = {2009.08959},
 primaryClass = {astro-ph.EP},
       adsurl = {https://ui.adsabs.harvard.edu/abs/2020AJ....160..207W}
}

@ARTICLE{ruffio2019,
       author = {{Ruffio}, Jean-Baptiste and {Macintosh}, Bruce and {Konopacky}, Quinn M. and {Barman}, Travis and {De Rosa}, Robert J. and {Wang}, Jason J. and {Wilcomb}, Kielan K. and {Czekala}, Ian and {Marois}, Christian},
        title = "{Radial Velocity Measurements of HR 8799 b and c with Medium Resolution Spectroscopy}",
      journal = {\aj},
         year = 2019,
        month = nov,
       volume = {158},
       number = {5},
          eid = {200},
        pages = {200},
          doi = {10.3847/1538-3881/ab4594},
archivePrefix = {arXiv},
       eprint = {1909.07571},
 primaryClass = {astro-ph.EP},
       adsurl = {https://ui.adsabs.harvard.edu/abs/2019AJ....158..200R}
}

@ARTICLE{brandtHGCA2021,
       author = {{Brandt}, Timothy D.},
        title = "{The Hipparcos-Gaia Catalog of Accelerations: Gaia EDR3 Edition}",
      journal = {ApJS},
         year = 2021,
        month = jun,
       volume = {254},
       number = {2},
          eid = {42},
        pages = {42},
          doi = {10.3847/1538-4365/abf93c},
archivePrefix = {arXiv},
       eprint = {2105.11662},
 primaryClass = {astro-ph.GA},
       adsurl = {https://ui.adsabs.harvard.edu/abs/2021ApJS..254...42B}
}

@ARTICLE{bowler2020astrometry,
       author = {{Bowler}, Brendan P. and {Blunt}, Sarah C. and {Nielsen}, Eric L.},
        title = "{Population-level Eccentricity Distributions of Imaged Exoplanets and Brown Dwarf Companions: Dynamical Evidence for Distinct Formation Channels}",
      journal = {\aj},
         year = 2020,
        month = feb,
       volume = {159},
       number = {2},
          eid = {63},
        pages = {63},
          doi = {10.3847/1538-3881/ab5b11},
archivePrefix = {arXiv},
       eprint = {1911.10569},
 primaryClass = {astro-ph.EP},
       adsurl = {https://ui.adsabs.harvard.edu/abs/2020AJ....159...63B}
}

@ARTICLE{wenger2000,
       author = {{Wenger}, M. and {Ochsenbein}, F. and {Egret}, D. and {Dubois}, P. and
         {Bonnarel}, F. and {Borde}, S. and {Genova}, F. and {Jasniewicz}, G. and
         {Lalo{\"e}}, S. and {Lesteven}, S. and {Monier}, R.},
        title = "{The SIMBAD astronomical database. The CDS reference database for astronomical objects}",
      journal = {\aaps},
         year = "2000",
        month = "Apr",
       volume = {143},
        pages = {9-22},
          doi = {10.1051/aas:2000332},
archivePrefix = {arXiv},
       eprint = {astro-ph/0002110},
 primaryClass = {astro-ph},
       adsurl = {https://ui.adsabs.harvard.edu/abs/2000A&AS..143....9W}
}

@ARTICLE{ochsenbein2000,
       author = {{Ochsenbein}, F. and {Bauer}, P. and {Marcout}, J.},
        title = "{The VizieR database of astronomical catalogues}",
      journal = {\aaps},
         year = "2000",
        month = "Apr",
       volume = {143},
        pages = {23-32},
          doi = {10.1051/aas:2000169},
archivePrefix = {arXiv},
       eprint = {astro-ph/0002122},
 primaryClass = {astro-ph},
       adsurl = {https://ui.adsabs.harvard.edu/abs/2000A&AS..143...23O}
}

@ARTICLE{rickman2024,
       author = {{Rickman}, Emily L. and {Ceva}, Will and {Matthews}, Elisabeth C. and {S{\'e}gransan}, Damien and {Bowler}, Brendan and {Forveille}, Thierry and {Franson}, Kyle and {Hagelberg}, Janis and {Udry}, St{\'e}phane and {Vigan}, Arthur},
        title = "{The discovery of two new benchmark likely brown dwarfs with precise dynamical masses at the stellar-substellar boundary}",
      journal = {arXiv e-prints},
         year = 2024,
        month = jan,
          eid = {arXiv:2401.10058},
        pages = {arXiv:2401.10058},
          doi = {10.48550/arXiv.2401.10058},
archivePrefix = {arXiv},
       eprint = {2401.10058},
 primaryClass = {astro-ph.SR},
       adsurl = {https://ui.adsabs.harvard.edu/abs/2024arXiv240110058R}
}

@ARTICLE{rickman2020,
       author = {{Rickman}, E.~L. and {S{\'e}gransan}, D. and {Hagelberg}, J. and {Beuzit}, J. -L. and {Cheetham}, A. and {Delisle}, J. -B. and {Forveille}, T. and {Udry}, S.},
        title = "{Spectral and atmospheric characterisation of a new benchmark brown dwarf HD 13724 B}",
      journal = {\aap},
         year = 2020,
        month = mar,
       volume = {635},
          eid = {A203},
        pages = {A203},
          doi = {10.1051/0004-6361/202037524},
archivePrefix = {arXiv},
       eprint = {2002.08319},
 primaryClass = {astro-ph.EP},
       adsurl = {https://ui.adsabs.harvard.edu/abs/2020A&A...635A.203R}
}

@ARTICLE{hoch2024,
       author = {{Hoch}, Kielan K.~W. and {Theissen}, Christopher A. and {Barman}, Travis S. and {Perrin}, Marshall D. and {Ruffio}, Jean-Baptiste and {Rickman}, Emily and {Konopacky}, Quinn M. and {Manjavacas}, Elena and {Balmer}, William O. and {Pueyo}, Laurent and {Kammerer}, Jens and {van der Marel}, Roeland P. and {Lewis}, Nikole K. and {Girard}, Julien H. and {Seager}, Sara and {Clampin}, Mark and {Mountain}, C. Matt},
        title = "{JWST-TST High Contrast: Spectroscopic Characterization of the Benchmark Brown Dwarf HD 19467 B with the NIRSpec Integral Field Spectrograph}",
      journal = {\aj},
         year = 2024,
        month = oct,
       volume = {168},
       number = {4},
          eid = {187},
        pages = {187},
          doi = {10.3847/1538-3881/ad6cd3},
archivePrefix = {arXiv},
       eprint = {2408.03830},
 primaryClass = {astro-ph.SR},
       adsurl = {https://ui.adsabs.harvard.edu/abs/2024AJ....168..187H}
}

@ARTICLE{rickman2019,
       author = {{Rickman}, E.~L. and {S{\'e}gransan}, D. and {Marmier}, M. and {Udry}, S. and {Bouchy}, F. and {Lovis}, C. and {Mayor}, M. and {Pepe}, F. and {Queloz}, D. and {Santos}, N.~C. and {Allart}, R. and {Bonvin}, V. and {Bratschi}, P. and {Cersullo}, F. and {Chazelas}, B. and {Choplin}, A. and {Conod}, U. and {Deline}, A. and {Delisle}, J. -B. and {Dos Santos}, L.~A. and {Figueira}, P. and {Giles}, H.~A.~C. and {Girard}, M. and {Lavie}, B. and {Martin}, D. and {Motalebi}, F. and {Nielsen}, L.~D. and {Osborn}, H. and {Ottoni}, G. and {Raimbault}, M. and {Rey}, J. and {Roger}, T. and {Seidel}, J.~V. and {Stalport}, M. and {Su{\'a}rez Mascare{\~n}o}, A. and {Triaud}, A. and {Turner}, O. and {Weber}, L. and {Wyttenbach}, A.},
        title = "{The CORALIE survey for southern extrasolar planets. XVIII. Three new massive planets and two low-mass brown dwarfs at greater than 5 AU separation}",
      journal = {\aap},
         year = 2019,
        month = may,
       volume = {625},
          eid = {A71},
        pages = {A71},
          doi = {10.1051/0004-6361/201935356},
archivePrefix = {arXiv},
       eprint = {1904.01573},
 primaryClass = {astro-ph.EP},
       adsurl = {https://ui.adsabs.harvard.edu/abs/2019A&A...625A..71R}
}

@ARTICLE{2025arXiv251206083L,
       author = {{Li}, Yaguang and {Liu}, Michael C. and {Dupuy}, Trent J. and {Huber}, Daniel and {Zhang}, Jingwen and {Hey}, Daniel and {Costa}, R.~R. and {Reersted Larsen}, Jens and {Ong}, J.~M. Joel and {Basu}, Sarbani and {Metcalfe}, Travis S. and {Zhou}, Yixiao and {van Saders}, Jennifer and {Bedding}, Timothy R. and {Hon}, Marc and {Kjeldsen}, Hans and {Campante}, Tiago L. and {Monteiro}, M{\'a}rio J.~P.~F.~G. and {Sloth Lundkvist}, Mia and {Lykke Winther}, Mark and {Chontos}, Ashley and {Saunders}, Nicholas and {Carmichael}, Theron W. and {Bouchez}, Antonin and {Alvarez}, Carlos and {Walker}, Sam and {Sepulveda}, Aldo G. and {Isaacson}, Howard and {Howard}, Andrew W. and {Gibson}, Steven R. and {Halverson}, Samuel and {Rider}, Kodi and {Roy}, Arpita and {Baker}, Ashley D. and {Edelstein}, Jerry and {Smith}, Chris and {Fulton}, Benjamin J. and {Walawender}, Josh},
        title = "{A Test of Substellar Evolutionary Models with High-Precision Ages from Asteroseismology and Gyrochronology for the Benchmark System HR 7672AB}",
      journal = {arXiv e-prints},
         year = 2025,
        month = dec,
          eid = {arXiv:2512.06083},
        pages = {arXiv:2512.06083},
archivePrefix = {arXiv},
       eprint = {2512.06083},
 primaryClass = {astro-ph.SR},
       adsurl = {https://ui.adsabs.harvard.edu/abs/2025arXiv251206083L}
}

@ARTICLE{2023ApJ...956...99B,
       author = {{Balmer}, William O. and {Pueyo}, Laurent and {Stolker}, Tomas and {Reggiani}, Henrique and {Maire}, A.-L. and {Lacour}, S. and {Molli{\`e}re}, P. and {Nowak}, M. and {Sing}, D. and {Pourr{\'e}}, N. and {Blunt}, S. and {Wang}, J.~J. and {Rickman}, E. and {Kammerer}, J. and {Henning}, Th. and {Ward-Duong}, K. and {Abuter}, R. and {Amorim}, A. and {Asensio-Torres}, R. and {Benisty}, M. and {Berger}, J.-P. and {Beust}, H. and {Boccaletti}, A. and {Bohn}, A. and {Bonnefoy}, M. and {Bonnet}, H. and {Bourdarot}, G. and {Brandner}, W. and {Cantalloube}, F. and {Caselli}, P. and {Charnay}, B. and {Chauvin}, G. and {Chavez}, A. and {Choquet}, E. and {Christiaens}, V. and {Cl{\'e}net}, Y. and {Coud{\'e} Du Foresto}, V. and {Cridland}, A. and {Dembet}, R. and {Dexter}, J. and {Drescher}, A. and {Duvert}, G. and {Eckart}, A. and {Eisenhauer}, F. and {Gao}, F. and {Garcia}, P. and {Garcia Lopez}, R. and {Gendron}, E. and {Genzel}, R. and {Gillessen}, S. and {Girard}, J.~H. and {Haubois}, X. and {Hei{\ss}el}, G. and {Hinkley}, S. and {Hippler}, S. and {Horrobin}, M. and {Houll{\'e}}, M. and {Hubert}, Z. and {Jocou}, L. and {Keppler}, M. and {Kervella}, P. and {Kreidberg}, L. and {Lagrange}, A.-M. and {Lapeyr{\`e}re}, V. and {Le Bouquin}, J.-B. and {L{\'e}na}, P. and {Lutz}, D. and {Monnier}, J.~D. and {Mouillet}, D. and {Nasedkin}, E. and {Ott}, T. and {Otten}, G.~P.~P.~L. and {Paladini}, C. and {Paumard}, T. and {Perraut}, K. and {Perrin}, G. and {Pfuhl}, O. and {Rameau}, J. and {Rodet}, L. and {Rousset}, G. and {Rustamkulov}, Z. and {Shangguan}, J. and {Shimizu}, T. and {Stadler}, J. and {Straub}, O. and {Straubmeier}, C. and {Sturm}, E. and {Tacconi}, L.~J. and {van Dishoeck}, E.~F. and {Vigan}, A. and {Vincent}, F. and {von Fellenberg}, S.~D. and {Widmann}, F. and {Wieprecht}, E. and {Wiezorrek}, E. and {Winterhalder}, T. and {Woillez}, J. and {Yazici}, S. and {Young}, A. and {Gravity Collaboration}},
        title = "{VLTI/GRAVITY Observations and Characterization of the Brown Dwarf Companion HD 72946 B}",
      journal = {\apj},
         year = 2023,
        month = oct,
       volume = {956},
       number = {2},
          eid = {99},
        pages = {99},
          doi = {10.3847/1538-4357/acf761},
archivePrefix = {arXiv},
       eprint = {2309.04403},
 primaryClass = {astro-ph.SR},
       adsurl = {https://ui.adsabs.harvard.edu/abs/2023ApJ...956...99B}
}

@ARTICLE{2023AJ....165...39F,
       author = {{Franson}, Kyle and {Bowler}, Brendan P. and {Bonavita}, Mariangela and {Brandt}, Timothy D. and {Chen}, Minghan and {Samland}, Matthias and {Zhang}, Zhoujian and {Lueber}, Anna and {Heng}, Kevin and {Kitzmann}, Daniel and {Wolf}, Trevor and {Jones}, Brandon A. and {Tran}, Quang H. and {Bardalez Gagliuffi}, Daniella C. and {Biller}, Beth and {Chilcote}, Jeffrey and {Crepp}, Justin R. and {Dupuy}, Trent J. and {Faherty}, Jacqueline and {Fontanive}, Cl{\'e}mence and {Groff}, Tyler D. and {Gratton}, Raffaele and {Guyon}, Olivier and {Jensen-Clem}, Rebecca and {Jovanovic}, Nemanja and {Kasdin}, N. Jeremy and {Lozi}, Julien and {Magnier}, Eugene A. and {Mu{\v{z}}i{\'c}}, Koraljka and {Sanghi}, Aniket and {Theissen}, Christopher A.},
        title = "{Astrometric Accelerations as Dynamical Beacons: Discovery and Characterization of HIP 21152 B, the First T-dwarf Companion in the Hyades}",
      journal = {\aj},
         year = 2023,
        month = feb,
       volume = {165},
       number = {2},
          eid = {39},
        pages = {39},
          doi = {10.3847/1538-3881/aca408},
archivePrefix = {arXiv},
       eprint = {2211.09840},
 primaryClass = {astro-ph.SR},
       adsurl = {https://ui.adsabs.harvard.edu/abs/2023AJ....165...39F}
}

@article{molliere_petitradtrans_2019,
	title = {{petitRADTRANS}: {A} {Python} radiative transfer package for exoplanet characterization and retrieval},
	volume = {627},
	issn = {0004-6361, 1432-0746},
	shorttitle = {{petitRADTRANS}},
	url = {https://www.aanda.org/10.1051/0004-6361/201935470},
	doi = {10.1051/0004-6361/201935470},
	language = {en},
	urldate = {2020-11-05},
	journal = {A\&A},
	author = {Mollière, P. and Wardenier, J. P. and van Boekel, R. and Henning, Th. and Molaverdikhani, K. and Snellen, I. A. G.},
	month = jul,
	year = {2019},
	pages = {A67},
}

@article{nasedkin_atmospheric_2024, 
    doi = {10.21105/joss.05875}, 
    url = {https://doi.org/10.21105/joss.05875}, 
    year = {2024}, 
    publisher = {The Open Journal}, 
    volume = {9}, 
    number = {96}, 
    pages = {5875}, 
    author = {Evert Nasedkin and Paul Mollière and Doriann Blain}, 
    title = {Atmospheric Retrievals with petitRADTRANS}, journal = {Journal of Open Source Software} 
}

@ARTICLE{exomol_ch4MM_2024,
author = {{Guest}, Elizabeth R. and {Tennyson}, Jonathan and {Yurchenko}, Sergei N.},
title = "{Predicting the rotational dependence of line broadening using machine learning}",
journal = {Journal of Molecular Spectroscopy},
year = 2024,
month = mar,
volume = {401},
eid = {111901},
pages = {111901},
doi = {10.1016/j.jms.2024.111901},
adsurl = {https://ui.adsabs.harvard.edu/abs/2024JMoSp.40111901G}
}

@Article{ExoMol_H2O,
author = {O. L. Polyansky and A. A. Kyuberis   and N. F. Zobov and J. Tennyson and  S. N. Yurchenko and  L. Lodi},
title = {{ExoMol molecular line lists XXX: a complete high-accuracy  line list for water}},
journal = {\mnras},
doi={10.1093/mnras/sty1877},
Volume={480},
Pages={2597-2608},
year = {2018}}

@Article{ExoMol_NH3,
author = {P. A. Coles and and Sergei N. Yurchenko and Jonathan Tennyson},
title = {{ExoMol molecular line lists XXXV: a rotation-vibration line list for hot ammonia}},
journal = {\mnras},
volume={490},
issue = {4},
Pages={4638--4647},
doi={10.1093/mnras/stz2778},
year = {2019}}

@article{ExoMol_CO2,
author = {Sergei N. Yurchenko and Thomas M. Mellor and Richard S. Freedman and Jonathan Tennyson},
title = {{ExoMol line lists – XXXIX. Ro-vibrational molecular line list for CO$_2$}},
journal = {\mnras},
    volume = {496},
    number = {4},
    pages = {5282-5291},
year = {2020},
    doi = {10.1093/mnras/staa1874}
}

@Article{ExoMol_H2S, 
author = {A. A. A. Azzam  and S. N. Yurchenko and J. Tennyson and O. V. Naumenko},
title = {{ExoMol line lists XVI: A Hot Line List for H$_{2}$S}},
journal = {\mnras},
volume = {460},
year  = {2016},
doi= {10.1093/mnras/stw1133},
pages = {4063-4074}}

@article{ExoMol_HCN,
author = {R J Barber and J K Strange and C Hill and O L  Polyansky and G Ch Mellau and S N Yurchenko and Jonathan Tennyson},
title = {{ExoMol line lists -- III. An improved hot rotation-vibration line list for HCN and HNC}},
journal = {\mnras},
volume = {437}, 
pages = {1828-1835},
year = {2014},
doi = {10.1093/mnras/stt2011}
}

@article{allard_k-h_2016,
	title = {K-{H}₂ line shapes for the spectra of cool brown dwarfs},
	volume = {589},
	doi = {10.1051/0004-6361/201628270},
	journal = {A\&A},
	author = {Allard, N. F. and Spiegelman, F. and Kielkopf, J. F.},
	year = {2016},
	pages = {A21},
}

@article{allard_new_2019,
	title = {New study of the line profiles of sodium perturbed by {H}₂},
	volume = {628},
	doi = {10.1051/0004-6361/201935593},
	journal = {A\&A},
	author = {Allard, N. F. and Spiegelman, F. and Leininger, T. and Mollière, P.},
	year = {2019},
	pages = {A120},
}

@ARTICLE{wende_feh_2010,
       author = {{Wende}, S. and {Reiners}, A. and {Seifahrt}, A. and {Bernath}, P.~F.},
        title = "{CRIRES spectroscopy and empirical line-by-line identification of FeH molecular absorption in an M dwarf}",
      journal = {\aap},
         year = 2010,
        month = nov,
       volume = {523},
          eid = {A58},
        pages = {A58},
          doi = {10.1051/0004-6361/201015220},
archivePrefix = {arXiv},
       eprint = {1007.4116},
 primaryClass = {astro-ph.SR},
       adsurl = {https://ui.adsabs.harvard.edu/abs/2010A&A...523A..58W}
}

@ARTICLE{exomol_ph3,
       author = {{Sousa-Silva}, Clara and {Al-Refaie}, Ahmed F. and {Tennyson}, Jonathan and {Yurchenko}, Sergei N.},
        title = "{ExoMol line lists - VII. The rotation-vibration spectrum of phosphine up to 1500 K}",
      journal = {\mnras},
         year = 2015,
        month = jan,
       volume = {446},
       number = {3},
        pages = {2337-2347},
          doi = {10.1093/mnras/stu2246},
archivePrefix = {arXiv},
       eprint = {1410.2917},
 primaryClass = {astro-ph.EP},
       adsurl = {https://ui.adsabs.harvard.edu/abs/2015MNRAS.446.2337S}
}

@article{feroz_importance_2013,
       author = {{Feroz}, Farhan and {Hobson}, Michael P. and {Cameron}, Ewan and {Pettitt}, Anthony N.},
        title = "{Importance Nested Sampling and the MultiNest Algorithm}",
      journal = {The Open Journal of Astrophysics},
         year = 2019,
        month = nov,
       volume = {2},
       number = {1},
          eid = {10},
        pages = {10},
          doi = {10.21105/astro.1306.2144},
archivePrefix = {arXiv},
       eprint = {1306.2144},
 primaryClass = {astro-ph.IM},
       adsurl = {https://ui.adsabs.harvard.edu/abs/2019OJAp....2E..10F}
}

@article{feroz_multinest_2009,
	title = {{MULTINEST}: an efficient and robust {Bayesian} inference tool for cosmology and particle physics},
	volume = {398},
	doi = {10.1111/j.1365-2966.2009.14548.x},
	number = {4},
	journal = {\mnras},
	author = {Feroz, F. and Hobson, M. P. and Bridges, M.},
	month = oct,
	year = {2009},
	pages = {1601--1614},
}

@ARTICLE{zhang_elpis_2023,
       author = {{Zhang}, Zhoujian and {Molli{\`e}re}, Paul and {Hawkins}, Keith and {Manea}, Catherine and {Fortney}, Jonathan J. and {Morley}, Caroline V. and {Skemer}, Andrew and {Marley}, Mark S. and {Bowler}, Brendan P. and {Carter}, Aarynn L. and {Franson}, Kyle and {Maas}, Zachary G. and {Sneden}, Christopher},
        title = "{ELemental abundances of Planets and brown dwarfs Imaged around Stars (ELPIS). I. Potential Metal Enrichment of the Exoplanet AF Lep b and a Novel Retrieval Approach for Cloudy Self-luminous Atmospheres}",
      journal = {\aj},
         year = 2023,
        month = nov,
       volume = {166},
       number = {5},
          eid = {198},
        pages = {198},
          doi = {10.3847/1538-3881/acf768},
archivePrefix = {arXiv},
       eprint = {2309.02488},
 primaryClass = {astro-ph.EP},
       adsurl = {https://ui.adsabs.harvard.edu/abs/2023AJ....166..198Z}
}

@INPROCEEDINGS{skilling_nestedsampling_2004,
       author = {{Skilling}, John},
        title = "{Nested Sampling}",
    booktitle = {Bayesian Inference and Maximum Entropy Methods in Science and Engineering: 24th International Workshop on Bayesian Inference and Maximum Entropy Methods in Science and Engineering},
         year = 2004,
       editor = {{Fischer}, Rainer and {Preuss}, Roland and {Toussaint}, Udo Von},
       series = {American Institute of Physics Conference Series},
       volume = {735},
        month = nov,
    publisher = {AIP},
        pages = {395-405},
          doi = {10.1063/1.1835238},
       adsurl = {https://ui.adsabs.harvard.edu/abs/2004AIPC..735..395S}
}

@article{buchner_x-ray_2014,
	title = {X-ray spectral modelling of the {AGN} obscuring region in the {CDFS}: {Bayesian} model selection and catalogue},
	volume = {564},
	doi = {10.1051/0004-6361/201322971},
	journal = {\aap},
	author = {Buchner, J. and Georgakakis, A. and Nandra, K. and Hsu, L. and Rangel, C. and Brightman, M. and Merloni, A. and Salvato, M. and Donley, J. and Kocevski, D.},
	month = apr,
	year = {2014},
	pages = {A125},
}

@article{rothman_hitemp_2010,
	title = {{HITEMP}, the high-temperature molecular spectroscopic database},
	volume = {111},
	doi = {10.1016/j.jqsrt.2010.05.001},
	journal = {\jqsrt},
	author = {Rothman, L. S. and Gordon, I. E. and Barber, R. J. and Dothe, H. and Gamache, R. R. and Goldman, A. and Perevalov, V. I. and Tashkun, S. A. and Tennyson, J.},
	month = oct,
	year = {2010},
	pages = {2139--2150},
}

@ARTICLE{nasedkin_hcicovariance_2023,
       author = {{Nasedkin}, E. and {Molli{\`e}re}, P. and {Wang}, J. and {Cantalloube}, F. and {Kreidberg}, L. and {Pueyo}, L. and {Stolker}, T. and {Vigan}, A.},
        title = "{Impacts of high-contrast image processing on atmospheric retrievals}",
      journal = {\aap},
         year = 2023,
        month = oct,
       volume = {678},
          eid = {A41},
        pages = {A41},
          doi = {10.1051/0004-6361/202346585},
archivePrefix = {arXiv},
       eprint = {2308.01343},
 primaryClass = {astro-ph.EP},
       adsurl = {https://ui.adsabs.harvard.edu/abs/2023A&A...678A..41N}
}

@INPROCEEDINGS{SVO,
       author = {{Rodrigo}, C. and {Solano}, E.},
        title = "{The SVO Filter Profile Service}",
    booktitle = {XIV.0 Scientific Meeting (virtual) of the Spanish Astronomical Society},
         year = 2020,
        month = jul,
          eid = {182},
        pages = {182},
       adsurl = {https://ui.adsabs.harvard.edu/abs/2020sea..confE.182R}
}

@ARTICLE{deregt_vhs1256_2026,
       author = {{de Regt}, S. and {Whiteford}, N. and {Miles}, B.~E. and {Gandhi}, S. and {Gonz{\'a}lez Picos}, D. and {Snellen}, I.~A.~G.},
        title = "{Native-resolution retrievals of VHS 1256-1257 b spanning the JWST/NIRSpec wavelength range: Chemical composition of a partially cloudy atmosphere}",
      journal = {arXiv e-prints},
         year = 2026,
        month = jul,
          eid = {arXiv:2607.00952},
        pages = {arXiv:2607.00952},
          doi = {10.48550/arXiv.2607.00952},
archivePrefix = {arXiv},
       eprint = {2607.00952},
 primaryClass = {astro-ph.EP},
       adsurl = {https://ui.adsabs.harvard.edu/abs/2026arXiv260700952D}
}

@ARTICLE{faherty_methane_2024,
       author = {{Faherty}, Jacqueline K. and {Burningham}, Ben and {Gagn{\'e}}, Jonathan and {Su{\'a}rez}, Genaro and {Vos}, Johanna M. and {Alejandro Merchan}, Sherelyn and {Morley}, Caroline V. and {Rowland}, Melanie and {Lacy}, Brianna and {Kiman}, Rocio and {Caselden}, Dan and {Kirkpatrick}, J. Davy and {Meisner}, Aaron and {Schneider}, Adam C. and {Kuchner}, Marc Jason and {Bardalez Gagliuffi}, Daniella Carolina and {Beichman}, Charles and {Eisenhardt}, Peter and {Gelino}, Christopher R. and {Gharib-Nezhad}, Ehsan and {Gonzales}, Eileen and {Marocco}, Federico and {Rothermich}, Austin James and {Whiteford}, Niall},
        title = "{Methane emission from a cool brown dwarf}",
      journal = {\nat},
         year = 2024,
        month = apr,
       volume = {628},
       number = {8008},
        pages = {511-514},
          doi = {10.1038/s41586-024-07190-w},
archivePrefix = {arXiv},
       eprint = {2404.10977},
 primaryClass = {astro-ph.SR},
       adsurl = {https://ui.adsabs.harvard.edu/abs/2024Natur.628..511F}
}

@ARTICLE{smith_heating_2026,
       author = {{Smith}, Kayla J. and {Marley}, Mark S. and {Koskinen}, Tommi T. and {Mang}, James},
        title = "{Understanding the Energy Input Required for Methane Emission on CWISEP J193518.59-154620.3: A Comprehensive Analysis}",
      journal = {arXiv e-prints},
         year = 2026,
        month = aug,
          eid = {arXiv:2608.17016},
        pages = {arXiv:2608.17016},
          doi = {10.48550/arXiv.2608.17016},
archivePrefix = {arXiv},
       eprint = {2608.17016},
 primaryClass = {astro-ph.EP},
       adsurl = {https://ui.adsabs.harvard.edu/abs/2026arXiv260817016S}
}

@ARTICLE{nowak_betapic_2020,
       author = {{GRAVITY Collaboration} and {Nowak}, M. and {Lacour}, S. and {Molli{\`e}re}, P. and {Wang}, J. and {Charnay}, B. and {van Dishoeck}, E.~F. and {Abuter}, R. and {Amorim}, A. and {Berger}, J.~P. and {Beust}, H. and {Bonnefoy}, M. and {Bonnet}, H. and {Brandner}, W. and {Buron}, A. and {Cantalloube}, F. and {Collin}, C. and {Chapron}, F. and {Cl{\'e}net}, Y. and {Coud{\'e} Du Foresto}, V. and {de Zeeuw}, P.~T. and {Dembet}, R. and {Dexter}, J. and {Duvert}, G. and {Eckart}, A. and {Eisenhauer}, F. and {F{\"o}rster Schreiber}, N.~M. and {F{\'e}dou}, P. and {Garcia Lopez}, R. and {Gao}, F. and {Gendron}, E. and {Genzel}, R. and {Gillessen}, S. and {Hau{\ss}mann}, F. and {Henning}, T. and {Hippler}, S. and {Hubert}, Z. and {Jocou}, L. and {Kervella}, P. and {Lagrange}, A.-M. and {Lapeyr{\`e}re}, V. and {Le Bouquin}, J.-B. and {L{\'e}na}, P. and {Maire}, A.-L. and {Ott}, T. and {Paumard}, T. and {Paladini}, C. and {Perraut}, K. and {Perrin}, G. and {Pueyo}, L. and {Pfuhl}, O. and {Rabien}, S. and {Rau}, C. and {Rodr{\'\i}guez-Coira}, G. and {Rousset}, G. and {Scheithauer}, S. and {Shangguan}, J. and {Straub}, O. and {Straubmeier}, C. and {Sturm}, E. and {Tacconi}, L.~J. and {Vincent}, F. and {Widmann}, F. and {Wieprecht}, E. and {Wiezorrek}, E. and {Woillez}, J. and {Yazici}, S. and {Ziegler}, D.},
        title = "{Peering into the formation history of {\ensuremath{\beta}} Pictoris b with VLTI/GRAVITY long-baseline interferometry}",
      journal = {\aap},
         year = 2020,
        month = jan,
       volume = {633},
          eid = {A110},
        pages = {A110},
          doi = {10.1051/0004-6361/201936898},
archivePrefix = {arXiv},
       eprint = {1912.04651},
 primaryClass = {astro-ph.EP},
       adsurl = {https://ui.adsabs.harvard.edu/abs/2020A&A...633A.110G}
}

@ARTICLE{balmer_aflepgrav_2025,
       author = {{Balmer}, William O. and {Franson}, Kyle and {Chomez}, Antoine and {Pueyo}, Laurent and {Stolker}, Tomas and {Lacour}, Sylvestre and {Nowak}, Mathias and {Nasedkin}, Evert and {Bonse}, Markus J. and {Thorngren}, Daniel and {Palma-Bifani}, Paulina and {Molli{\`e}re}, Paul and {Wang}, Jason J. and {Zhang}, Zhoujian and {Chavez}, Amanda and {Kammerer}, Jens and {Blunt}, Sarah and {Bowler}, Brendan P. and {Bonnefoy}, Mickael and {Brandner}, Wolfgang and {Charnay}, Benjamin and {Chauvin}, Gael and {Henning}, Th. and {Lagrange}, A.-M. and {Pourr{\'e}}, Nicolas and {Rickman}, Emily and {De Rosa}, Robert and {Vigan}, Arthur and {Winterhalder}, Thomas},
        title = "{VLTI/GRAVITY Observations of AF Lep b: Preference for Circular Orbits, Cloudy Atmospheres, and a Moderately Enhanced Metallicity}",
      journal = {\aj},
         year = 2025,
        month = jan,
       volume = {169},
       number = {1},
          eid = {30},
        pages = {30},
          doi = {10.3847/1538-3881/ad9265},
archivePrefix = {arXiv},
       eprint = {2411.05917},
 primaryClass = {astro-ph.EP},
       adsurl = {https://ui.adsabs.harvard.edu/abs/2025AJ....169...30B}
}

@ARTICLE{brandt2020,
       author = {{Brandt}, Timothy D. and {Dupuy}, Trent J. and {Bowler}, Brendan P. and {Bardalez Gagliuffi}, Daniella C. and {Faherty}, Jacqueline and {Brandt}, G. Mirek and {Michalik}, Daniel},
        title = "{A Dynamical Mass of 70 {\ensuremath{\pm}} 5 M$_{Jup}$ for Gliese 229B, the First T Dwarf}",
      journal = {\aj},
         year = 2020,
        month = oct,
       volume = {160},
       number = {4},
          eid = {196},
        pages = {196},
          doi = {10.3847/1538-3881/abb45e},
archivePrefix = {arXiv},
       eprint = {1910.01652},
 primaryClass = {astro-ph.SR},
       adsurl = {https://ui.adsabs.harvard.edu/abs/2020AJ....160..196B}
}

@ARTICLE{dupuy2014,
       author = {{Dupuy}, Trent J. and {Liu}, Michael C. and {Ireland}, Michael J.},
        title = "{New Evidence for a Substellar Luminosity Problem: Dynamical Mass for the Brown Dwarf Binary Gl 417BC}",
      journal = {\apj},
         year = 2014,
        month = aug,
       volume = {790},
       number = {2},
          eid = {133},
        pages = {133},
          doi = {10.1088/0004-637X/790/2/133},
archivePrefix = {arXiv},
       eprint = {1406.1184},
 primaryClass = {astro-ph.SR},
       adsurl = {https://ui.adsabs.harvard.edu/abs/2014ApJ...790..133D}
}

@ARTICLE{sahlmann2020,
       author = {{Sahlmann}, J. and {Burgasser}, A.~J. and {Bardalez Gagliuffi}, D.~C. and {Lazorenko}, P.~F. and {S{\'e}gransan}, D. and {Zapatero Osorio}, M.~R. and {Blake}, C.~H. and {Gelino}, C.~R. and {Mart{\'\i}n}, E.~L. and {Bouy}, H.},
        title = "{Astrometric orbits of spectral binary brown dwarfs - I. Massive T dwarf companions to 2M1059-21 and 2M0805+48}",
      journal = {\mnras},
         year = 2020,
        month = jun,
       volume = {495},
       number = {1},
        pages = {1136-1147},
          doi = {10.1093/mnras/staa1235},
archivePrefix = {arXiv},
       eprint = {2004.14889},
 primaryClass = {astro-ph.SR},
       adsurl = {https://ui.adsabs.harvard.edu/abs/2020MNRAS.495.1136S}
}

@ARTICLE{calamari2022,
       author = {{Calamari}, Emily and {Faherty}, Jacqueline K. and {Burningham}, Ben and {Gonzales}, Eileen and {Bardalez-Gagliuffi}, Daniella and {Vos}, Johanna M. and {Gemma}, Marina and {Whiteford}, Niall and {Gaarn}, Josefine},
        title = "{An Atmospheric Retrieval of the Brown Dwarf Gliese 229B}",
      journal = {\apj},
         year = 2022,
        month = dec,
       volume = {940},
       number = {2},
          eid = {164},
        pages = {164},
          doi = {10.3847/1538-4357/ac9cc9},
archivePrefix = {arXiv},
       eprint = {2210.13614},
 primaryClass = {astro-ph.SR},
       adsurl = {https://ui.adsabs.harvard.edu/abs/2022ApJ...940..164C}
}

@article{Agrawal_2023,
doi = {10.3847/1538-3881/acd6a3},
url = {https://dx.doi.org/10.3847/1538-3881/acd6a3},
year = {2023},
month = {jun},
publisher = {The American Astronomical Society},
volume = {166},
number = {1},
pages = {15},
author = {Agrawal, Shubh and Ruffio, Jean-Baptiste and Konopacky, Quinn M. and Macintosh, Bruce and Mawet, Dimitri and Nielsen, Eric L. and Hoch, Kielan K. W. and Liu, Michael C. and Barman, Travis S. and Thompson, William and Greenbaum, Alexandra Z. and Marois, Christian and Patience, Jenny},
title = {Detecting Exoplanets Closer to Stars with Moderate Spectral Resolution Integral-field Spectroscopy},
journal = {The Astronomical Journal}
}

@article{Rauscher_2024,
doi = {10.1088/1538-3873/ad1b36},
url = {https://dx.doi.org/10.1088/1538-3873/ad1b36},
year = {2024},
month = {jan},
publisher = {The Astronomical Society of the Pacific},
volume = {136},
number = {1},
pages = {015001},
author = {Rauscher, Bernard J.},
title = {NSClean: An Algorithm for Removing Correlated Noise from JWST NIRSpec Images},
journal = {Publications of the Astronomical Society of the Pacific}
}

@inproceedings{Perrin2012,
author = {Marshall D. Perrin and R{\'e}mi Soummer and Erin M. Elliott and Matthew D. Lallo and Anand Sivaramakrishnan},
title = {{Simulating point spread functions for the James Webb Space Telescope with WebbPSF}},
volume = {8442},
booktitle = {Space Telescopes and Instrumentation 2012: Optical, Infrared, and Millimeter Wave},
editor = {Mark C. Clampin and Giovanni G. Fazio and Howard A. MacEwen and Jacobus M. Oschmann Jr.},
organization = {International Society for Optics and Photonics},
publisher = {SPIE},
pages = {84423D},
year = {2012},
doi = {10.1117/12.925230},
URL = {https://doi.org/10.1117/12.925230}
}

@inproceedings{Perrin2014,
author = {Marshall D. Perrin and Anand Sivaramakrishnan and Charles-Philippe Lajoie and Erin Elliott and Laurent Pueyo and Swara Ravindranath and Lo{\"i}c Albert},
title = {{Updated point spread function simulations for JWST with WebbPSF}},
volume = {9143},
booktitle = {Space Telescopes and Instrumentation 2014: Optical, Infrared, and Millimeter Wave},
editor = {Jacobus M. Oschmann Jr. and Mark Clampin and Giovanni G. Fazio and Howard A. MacEwen},
organization = {International Society for Optics and Photonics},
publisher = {SPIE},
pages = {91433X},
year = {2014},
doi = {10.1117/12.2056689},
URL = {https://doi.org/10.1117/12.2056689}
}

@dataset{mukherjee_2023,
  author       = {Mukherjee, Sagnick and
                  Fortney, Jonathan and
                  Morley, Caroline and
                  Batalha, Natasha and
                  Marley, Mark and
                  Karalidi, Theodora and
                  Visscher, Channon and
                  Lupu, Roxana and
                  Freedman, Richard and
                  Gharib-Nezhad, Ehsan},
  title        = {The Sonora Substellar Atmosphere Models. IV. Elf
                   Owl: Atmospheric Mixing and Chemical
                   Disequilibrium with Varying Metallicity and C/O
                   Ratios (T- type Models)
                  },
  month        = dec,
  year         = 2023,
  publisher    = {Zenodo},
  doi          = {10.5281/zenodo.10385821},
  url          = {https://doi.org/10.5281/zenodo.10385821},
}

@Article{RuffioXuan2026,
author={Ruffio, Jean-Baptiste
and Xuan, Jerry W.
and Chachan, Yayaati
and Kesseli, Aurora
and Lee, Eve J.
and Beichman, Charles
and Hodapp, Klaus
and Balmer, William O.
and Konopacky, Quinn
and Perrin, Marshall D.
and Mawet, Dimitri
and Knutson, Heather A.
and Bryden, Geoffrey
and Greene, Thomas P.
and Johnstone, Doug
and Leisenring, Jarron
and Meyer, Michael
and Ygouf, Marie},
title={Jupiter-like uniform metal enrichment in a system of multiple giant exoplanets},
journal={Nature Astronomy},
year={2026},
month={Feb},
day={09},
issn={2397-3366},
doi={10.1038/s41550-026-02783-z},
url={https://doi.org/10.1038/s41550-026-02783-z}
}

@article{Greco_2016,
doi = {10.3847/1538-4357/833/2/134},
url = {https://doi.org/10.3847/1538-4357/833/2/134},
year = {2016},
month = {dec},
publisher = {The American Astronomical Society},
volume = {833},
number = {2},
pages = {134},
author = {Greco, Johnny P. and Brandt, Timothy D.},
title = {THE MEASUREMENT, TREATMENT, AND IMPACT OF SPECTRAL COVARIANCE AND BAYESIAN PRIORS IN INTEGRAL-FIELD SPECTROSCOPY OF EXOPLANETS},
journal = {The Astrophysical Journal}
}
\bibliographystyle{aasjournal}

\end{document}